\documentclass[11pt]{article}

\usepackage{cite}

\usepackage{graphicx}
\usepackage{latexsym}
\usepackage{amssymb}
\usepackage{tikz}
\usetikzlibrary{arrows.meta,shapes.misc,patterns,shapes.symbols,decorations.pathreplacing,decorations}

\usepackage[font=small]{caption}
\usepackage[font=footnotesize]{subcaption}
\usepackage{amsthm,amsmath}

\usepackage[margin=1in]{geometry}

\newcommand{\morabba}[1]{\,\begin{flushright}
\Rectsteel \\
\end{flushright}}

\newcommand{\G}{\,\mathcal{G}}
\newcommand{\GG}{\,\mathcal{G}^*}

\newcommand{\all}[2]{\,\begin{align}
#1 
\label{#2}
\end{align}
}

\makeatletter
\newcommand{\vast}{\bBigg@{4}}
\newcommand{\Vast}{\bBigg@{5}}

\makeatother

\begin{document}

\title{Inferring Structural Characteristics of Networks with Strong and Weak Ties from Fixed-Choice Surveys}

\author{Naghmeh Momeni and Michael~G.~Rabbat
\thanks{N. Momeni and M. Rabbat are with the Department
of Electrical and Computer Engineering, McGill University, Montreal, QC, Canada, H3A 0E9. Email: naghmeh.momenitaramsari@mail.mcgill.ca, michael.rabbat@mcgill.ca. \newline
\indent A preliminary version of this work was presented at the 2016 IEEE Workshop on Statistical Signal Processing~\cite{momeni2016}. \newline
\indent This work was supported by the Natural Sciences and Engineering Research Council of Canada through grants RGPIN-2012-341596 and RGPIN-2017-06266.}
}

\maketitle

\pgfdeclaredecoration{sl}{initial}{
        \state{initial}[width=\pgfdecoratedpathlength-1sp]{
            \pgfmoveto{\pgfpointorigin}
        }
        \state{final}{
            \pgflineto{\pgfpointorigin}
        }
    }

\tikzset{strong bidir/.style={->,ultra thick, shorten >=2mm, shorten <=2mm, 
            decoration={sl,raise=1mm},decorate}}
            
    \tikzset{weak bidir/.style={dashed,->,ultra thick,shorten >=2mm, shorten <=2mm, 
            decoration={sl,raise=1mm},decorate}}

    \tikzset{strong dir/.style = {->,ultra thick, shorten >=2 mm, shorten <=2mm}}
       \tikzset{weak dir/.style = {dashed,->,ultra thick, shorten >=2 mm, shorten <=2mm}}

\begin{abstract}
Knowing the structure of an offline social network facilitates a variety of analyses, including studying the rate at which infectious diseases may spread and identifying a subset of actors to immunize in order to reduce, as much as possible, the rate of spread. Offline social network topologies are typically estimated by surveying actors and asking them to list their neighbours. While identifying close friends and family (i.e., strong ties) can typically be done reliably, listing all of one's acquaintances (i.e., weak ties) is subject to error due to respondent fatigue. This issue is commonly circumvented through the use of so-called ``fixed choice''€ surveys where respondents are asked to name a fixed, small number of their weak ties (e.g., two or ten). Of course, the resulting \emph{crude} observed network will omit many ties, and using this crude network to infer properties of the network, such as its degree distribution or clustering coefficient, will lead to biased estimates. This paper develops estimators, based on the method of moments, for a number of network characteristics including those related to the first and second moments of the degree distribution as well as the network size, using fixed-choice survey data. Experiments with simulated data illustrate that the proposed estimators perform well across a variety of network topologies and measurement scenarios, and the resulting estimates are significantly more accurate than those obtained directly using the crude observed network, which are commonly used in the literature. We also describe a variation of the Jackknife procedure that can be used to obtain an estimate of the estimator variance.
\end{abstract}

\section{Introduction} \label{sec:intro}
  
  \subsection{Network Sampling}

Network science has quickly spread into diverse disciplines because it offers versatile and powerful tools to quantify the structure of interactions and connections. For social networks, for instance, the diffusion of information~\cite{bakshy2012role,yang2010modeling,moreno2004dynamics} and infectious disease~\cite{RevModPhys}, awareness~\cite{casterline2001diffusion}, and health behaviors~\cite{kohler2007social,latkin2003norms} are studied. The structural properties of the underlying social networks are central in these studies. Thus we need to observe and measure these properties. Like most large-scale systems, for practical considerations we need to find efficient ways of inferring these properties from a limited set of observations. This task is the focus of the network inference literature. 
Different sampling methods in the literature are suited for different practical requirements~\cite{kolaczyk2014statistical}.  Examples include: traceroute sampling~\cite{achlioptas2005bias,clauset2005accuracy,viger}, which is typically used for sampling the Internet; respondent-driven sampling methods~\cite{salganik2004sampling}, which are typically used for sampling social networks connecting hidden populations that are difficult to find and interview; crawling methods, other random-walk methods~\cite{gjoka2010}, and forest fire sampling~\cite{leskovec2006sampling}, which are typically used for the web and online social networks; and random node and link sampling~\cite{kurant2011walking,ribeiro2010estimating}.   

In this paper we focus on sampling offline social networks. 
We consider two features that are specific to social network research and that demand special consideration for network sampling. The first one involves degree truncation introduced in the measurement process, which we discuss more in Section~\ref{FCD}. The second one involves heterogeneity of link weights, which we discuss more in Section~\ref{SW}. After introducing these two features and pointing out the absence of theoretical results on inference methods for offline social networks, we focus on incorporating them into the mathematical treatment of the sampling procedure.  We introduce a setup to incorporate both of these features. We then focus on the problem of inference, which is the main contribution of this paper.

\subsection{Fixed Choice Design}\label{FCD}

  Most of the sampling methods for social networks can  be mathematically formulated as variants of snowball sampling. Snowball sampling consists of sampling an initial set of nodes and their incident links, then sampling their neighbors and their incident links, and so on. It is equivalent to running a breadth first search from the initial set of nodes, and is typically stopped at a given depth, so that not all links are traversed. Ideally, the sampling would proceed until new nodes and links are no longer encountered, so the entire network is sampled. This is impractical in most settings, and as we will discuss, even more so in offline social networks. 

In practice, information about offline social networks are typically obtained through personal interviews and surveys. In this context, each person is referred to as an \emph{ego}, and their 1-hop neighbors in the graph are called \emph{alters}. A zero-wave snowball sample would consist of simply selecting a set of interviewees and asking them to list their alters. This is called an \emph{ego-centric design}. For practical considerations of time and cost, the majority of social network data is ego-centric~\cite{marsden2011survey,marsden1990network,perkins2015social}.  Even this simple and economical design introduces challenges, such as imperfect recollections and other memory issues. A serious practical problem is respondent fatigue, which imposes limits on the interview time and the amount of information expected from respondents. The conventional way of approaching this problem is to employ the so-called \emph{fixed-choice design}, which amounts to imposing limits on the number of alters that each respondent is asked to list. There are numerous examples of classic and recent social networks studies that employ a fixed-choice design~\cite{wellman1979community,fischer1982dwell,
behrman2002social,helleringer2007sexual,christakis2007spread,
christakis2013social}. 

Interestingly, the social network studies that focus on diffusion of information, awareness, innovation and health behaviors directly use the crude, degree-truncated version of the network as the topology on which the diffusion processes take place. As pointed out recently in~\cite{JP}, the behavior of diffusion processes on the original networks and their degree-truncated variants can differ significantly. Thus, inferring the properties of the original network form the sampled data constitutes a significant step towards improving the results in the social network literature. The only prior works in the literature on inference of network properties from fixed-choice survey data do not differentiate between different types of social ties~\cite{babaku,hoff2013}.

\subsection{Strong and Weak Ties} \label{SW}
The second property of social networks that a sampling procedure should take into account is the heterogeneity of link weights. In social network studies, a conventional simplification is to divide social ties into strong and weak, and different questions in a survey specifically aim to elucidate different types of ties. For example, some survey questions target within-household and intimate relationships (such as secret sharing and intimate advice seeking), and others questions target between-household and weaker relations (conversations, interactions, etc.)~\cite{banerjee2013,perkins2015social}. Or in the context of student friendship networks, some ties pertain to within-school friendship bonds and some pertain to between-school ties~\cite{harris2009national}. Dividing the ties into two distinct categories is a first step towards a closer correspondence to actual survey data.

Strong and weak ties have different levels of impact in different phenomena, such as diffusion of information, providing social support, adopting health behaviors, and cooperation and trust~\cite{granovetter1973strength,friedkin1982information,
krackhardt1992strength,karsai2014time}. The distinct role of strong and weak social  ties has been also studied in online social networks~\cite{burke2013using}. 

The recent study~\cite{babaku} discusses modeling and inference for fixed-choice designs in the simplified scenario where there is only one type of tie. 
In actual studies, the nature of links are heterogeneous and the first approximation would be to dichotomize them. Moreover, typical surveys used to infer structural properties of offline social networks incorporate multiple questions, while the method of~\cite{babaku} effectively assumes that only one fixed-choice question has been posed.
While accounting for different types of ties (e.g., strong and weak) is a significant step towards making the approach more useful to sociologists, it also gives rise to a number of challenges. In particular, the model with strong and weak ties has more parameters to be estimated and requires carefully accounting for the interactions (e.g., correlations) between these parameters. Developing an inference methodology to address these challenges constitutes one of the main contributions of this manuscript, as discussed next.

\subsection{Contribution and Paper Organization}
In this paper, we study the problem of inferring network characteristics from surveys employing fixed-choice design questions. We focus on the case of networks with two distinct types of links (strong and weak). We propose an inference method to estimate network properties based on observing the sampled version of it, and we also describe a method to estimate the variance of the proposed estimators.

The rest of this manuscript is structured as follows. Section~\ref{sec:setup} presents the sampling setting, taking into account both features discussed above. Section~\ref{inf} formulates the inference problem and presents methods for estimating structural properties of the network from fixed-choice survey data. Then Section~\ref{sec:results} illustrates the performance of the proposed inference methodology via simulations, and compares the results with those of the crude version of the network (without accounting for the bias introduced by fixed-choice observations).

\section{Problem Formulation} \label{sec:setup}

Consider the following sampling setup. The \emph{original network}, whose properties we want to estimate, is denoted by $\mathcal{G}$. This original network has $N$ nodes, where $N$ is an unknown parameter to be estimated. The network is undirected, and links are of one or two types: weak and strong.
Thus, for each node in $\G$ we can define two distinct degrees, pertaining to the number of its strong links and weak links.

The sampling process starts with selecting a set of respondents (referred to as \emph{seeds}) denoted by $S_0$ with cardinality $|S_0| = n_0$.
Each seed is asked to name all of its strong neighbors and also $B$ of its weak neighbors, where $B$ is a given positive integer (see, e.g.,~\cite{perkins2015social}). That is, we assume that the problem of imperfect recollections can be neglected for the case of strong ties. Moreover, since the number of weak ties is typically large, we assume that the imposed limit is applied only to weak ties. We also assume that $B$ is much smaller than the smallest weak degree in the network, so that every node has at least $B$ weak ties to name. This is reasonable since typical fixed-choice designs use values for $B$ that are less than ten~\cite{wellman1979community,fischer1982dwell,
behrman2002social,helleringer2007sexual,christakis2007spread,
christakis2013social}.

The alters that each seed names might themselves belong to $S_0$. Let $S_1^s$ and $S_1^w$ denote the sets of non-seed strong and weak alters named by any seed, respectively. Note that $S_0$, $S_1^s$, and $S_1^w$ are not disjoint, and it is possible that some node may appear in all three sets; that is, a node may be a seed, it may be named as a strong tie of another seed, and it may be named as a weak tie of yet another seed. We denote the cardinality of $S_1^s$ by $n_1^s$ and the cardinality of $S_1^w$ by $n_1^w$. We refer to  the  subgraph of $\mathcal{G}$ constructed  from the seeds and the responses as the \emph{sampled network}, and we denote this sampled network by $\mathcal{G}^*$. 
 
\begin{figure}
\centering
\includegraphics[scale=0.5]{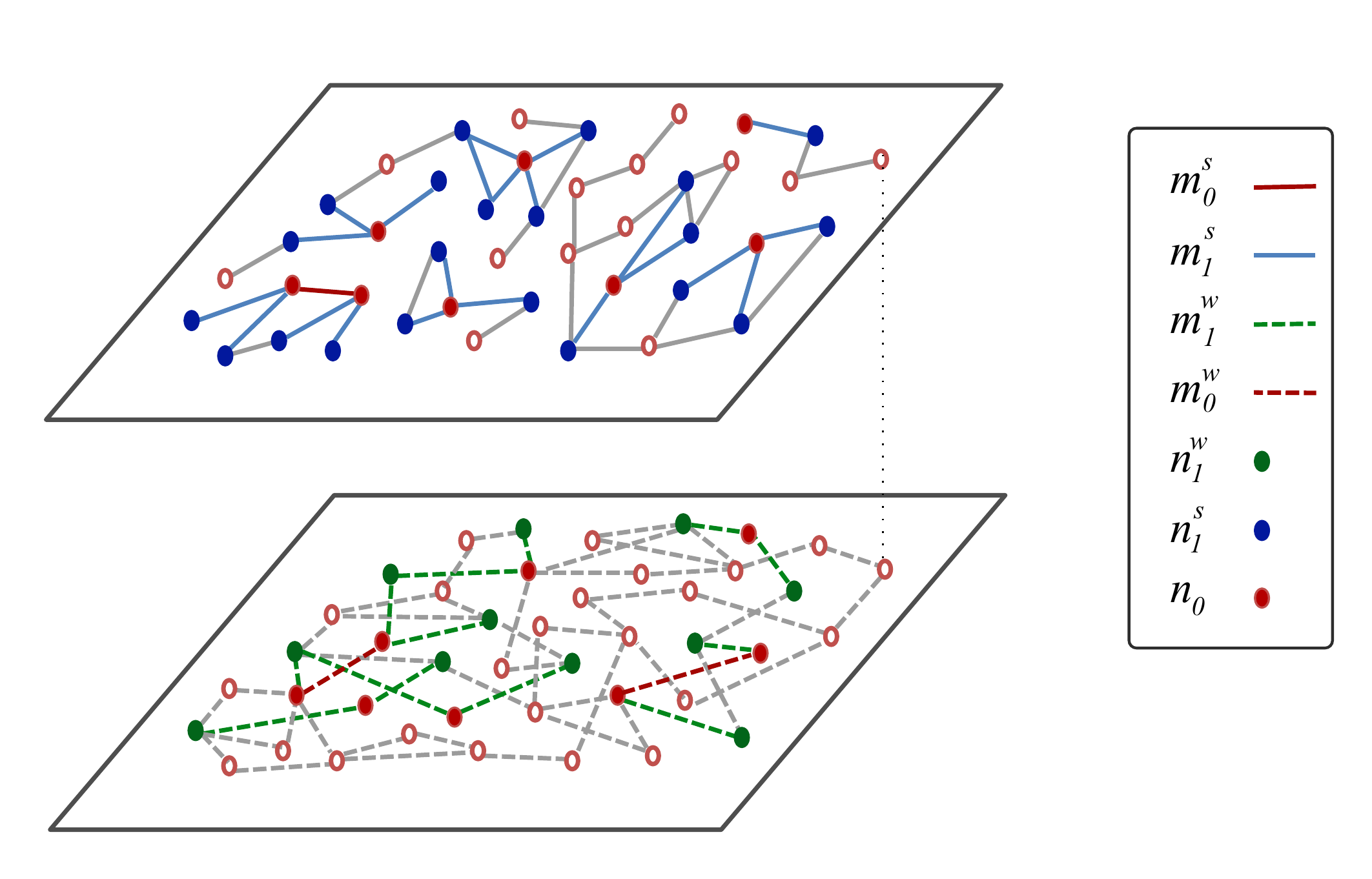}
\caption{A schematic illustration of the sampling setup.  The upper layer represents the strong ties and the lower one represents the weak ties. The set of nodes in two layers are the same and the links in two layers are exclusive. The seeds are depicted in red. In this example $B=2$. Gray links and hollow nodes exist in $\G$  but are not observed in $\GG$. The observables shown in the legend are equal to the number of corresponding nodes/links (as intorduced in Section~\ref{inf}).}\label{setup}
\end{figure}

Figure~\ref{setup} shows a schematic illustrating the sampling process. As can be seen, some of the sampled links connect two seeds, and others connect a seed to a non-seed. We denote the number of strong and weak links with both ends in $S_0$ by $m_0^s$ and $m_0^w$, respectively. 
We denote the number of links between $S_0$ and  $S_1^s$ by $m_1^s$ and, similarly, the number of links between $S_0$ and  $S_1^w$ by $m_1^w$. Table~\ref{notation} provides a summary of the notation used throughout paper.

\begin{table*}[]
\centering
\caption{Notation used for statistics of the original (unknown) and sampled (observed) networks.}
\label{my-label}
\resizebox{\textwidth}{!}{
\begin{tabular}{llll}
\multicolumn{2}{c}{\textbf{Original network (unknown)}}                                        & \multicolumn{2}{c}{\textbf{Sampled Network (observed)}}                                  \\ \hline
Variable       & \multicolumn{1}{l|}{Definition}                                               & Variable              & Definition                                                       \\ \hline
$\G$           & \multicolumn{1}{l|}{Original graph}                                           & $\GG$                 & Sampled graph (observed)                                         \\
$N$            & \multicolumn{1}{l|}{Number of nodes}                                          & $n_0$                 & number of seeds                                                  \\
$q$            & \multicolumn{1}{l|}{Sampling probablity}                                      & $n_1^s$               & number of non-seed strong alters named by seeds                  \\
$K_s$          & \multicolumn{1}{l|}{Average strong degree}                                    & $n_1^w$               & number of non-seed weak alters named by seeds                    \\
$K_w$          & \multicolumn{1}{l|}{Average weak degree}                                      & $m_0^s$               & number of strong links between seeds                             \\
$K_{ss}$       & \multicolumn{1}{l|}{Second moment of strong degrees}                          & $m_1^s$               & number of strong links between seeds and non-seeds               \\
$K_{ww}$       & \multicolumn{1}{l|}{Second moment of weak degrees}                            & $m_0^w$               & number of weak links between seeds                               \\
$K_{sw}$       & \multicolumn{1}{l|}{Cross-production moment of degrees}                       & $m_1^w$               & number of weak links between seeds and non-seeds                 \\
$T_{s^3}$      & \multicolumn{1}{l|}{Number of triangles with three strong links}              & $T^{\ast}_{s^3}$      & Number of observed triangles with three strong links             \\
$T_{s^2w}$     & \multicolumn{1}{l|}{Number of triangles with two strong and one weak links}   & $T^{\ast}_{s^2w}$     & Number of observed triangles with two strong and one weak links  \\
$T_{sw^2}$     & \multicolumn{1}{l|}{Number of triangles with one strong and two weak links}   & $T^{\ast}_{sw^2}$     & Number of observed triangles with one strong and two weak links  \\
$T_{w^3}$      & \multicolumn{1}{l|}{Number of triangles with three weak links}                & $T^{\ast}_{w^3}$      & Number of observed triangles with three weak links               \\
$\tau_{ss}$    & \multicolumn{1}{l|}{Number of total triads with two strong links}             & $\lambda^{\ast}_{ss}$ & Number of observed open triads with two strong links             \\
$\tau_{sw}$    & \multicolumn{1}{l|}{Number of total triads with one strong and one weak link} & $\lambda^{\ast}_{sw}$ & Number of observed open triads with one strong and one weak link \\
$\tau_{ww}$    & \multicolumn{1}{l|}{Number of total triads with two weak links}               & $\lambda^{\ast}_{ww}$ & Number of observed open triads with two weak links               \\
$\lambda_{ss}$ & \multicolumn{1}{l|}{Number of open triads with two strong links}              &                       &                                                                  \\
$\lambda_{sw}$ & \multicolumn{1}{l|}{Number of open triads with one strong and one weak link}  &                       &                                                                  \\
$\lambda_{ww}$ & \multicolumn{1}{l|}{Number of open triads with two weak links}                &                       &                                                                  \\
$CC$           & \multicolumn{1}{l|}{Clustering coefficient}                                   &                       &                                                                  \\
               &                                                                               &                       &                                                                 
\end{tabular}
}
\label{notation}
\end{table*}

Our objective is to infer properties of $\G$ given the observed subgraph $\GG$. We model the heterogeneity of links in the original graph with a two-layer network with the same set of nodes and two binary-valued adjacency matrices $A^s=[a_{ij}^s]$ and $A^w=[a_{ij}^w]$ representing strong and weak links, respectively. We denote the strong degree of node $i$ by $k_i^s = \sum_{j=1}^N a_{ij}^s$ and its weak degree by $k_i^w = \sum_{j=1}^N a_{ij}^w$. 

Specifically, the parameters to be estimated are the number of nodes in the original network $N$, the average strong and weak degrees $K_s = \frac{1}{n} \sum_{i=1}^N k_i^s$ and $K_w = \frac{1}{n} \sum_{i=1}^N k_i^w$, the second moments of the degrees (or equivalently, the variance of the degree distributions and the correlation between strong and weak degrees) $K_{ss}=\frac{1}{n} \sum_{i=1}^N (k_i^s)^2$, $K_{ww}=\frac{1}{n} \sum_{i=1}^N (k_i^w)^2$, $K_{sw}=\frac{1}{n} \sum_{i=1}^N k_i^s k_i^w$, as well as the number of triads and triangles of different types (see Sec.~\ref{sec:second-moments}), and the clustering coefficient~\cite{newman2010}.

\section{Inference Methodology} \label{inf}
  
 We use the method of moments to perform inference. We need a generative model for the observables so that their expected values can be written as a function of the desired variables. Then the method of moments proceeds by finding the least squares fit between the observables and their expected values.

  We model the selection of seeds as an i.i.d.~Bernoulli process, in which each node in the network is chosen as a seed independently with probability $q$, which is unknown. Since we seek a non-parametric framework, we also assume that the weak neighbors named by each seed are chosen uniformly at random from all of the weak neighbors of the seed. Let $X_i$ be a Bernoulli random variable with probability $q$ associated with each node $i=1,\dots,N$. If node $i$ is a seed (i.e., $i \in S_0$ is surveyed) then $X_i=1$, and otherwise $X_i=0$. 
  
Extensions to the more general case where weak neighbors are not chosen uniformly at random, but rather are chosen according to some other distribution (e.g., proportional to the neighbor's degree) may be of interest, but we leave this to future work. Likewise, it may be of interest to relax the assumption that seeds accurately report all of their strong ties (e.g., to account for forgetting one or two). Indeed, if strong ties are inadvertently omitted, then the estimates produced by the procedure described below will be biased, since they don't account for this source of error. In practice, it would be impractical to assume statistics about the number of strong ties omitted, and it would also need to be estimated. We also leave this extension to future work.
  
  With this model and notation, our next step is to find expressions for the expected values of observed statistics $n_0$, $m_0^s$, $m_1^s$, $n_1^s$, $m_0^w$, $m_1^w$, $n_1^w$. Our approach to inferring the desired parameters will proceed in two stages, which we describe below. The first stage only involves estimating the first moments of the node degrees, and the second stage involves estimating the second moments.
  
\subsection{First Moments}

  The number of nodes that are seeds can be written as $n_0=\sum_{i=1}^N X_{i}$ and therefore we have 
\begin{equation}
\mathbb{E}[n_0]=Nq.\label{e1}
\end{equation}
Similarly, $m_0^s$ and $m_1^s$ can be written as 
$m_0^s=\frac{1}{2}\sum_{i=1}^N\sum_{j=1}^NX_iX_ja_{ij}^s$ and $m_1^s=\frac{1}{2}\sum_{i=1}^N\sum_{j=1}^NX_i(1-X_j)a_{ij}^s$, respectively. Therefore we have
\begin{equation}
\mathbb{E}[m_0^s]=\frac{1}{2}q^2\sum_{i=1}^Nk_i^s=\frac{1}{2}q^2NK_s\label{e2}
\end{equation}
and
\begin{equation}
\mathbb{E}[m_1^s]=q(1-q)\sum_{i=1}^Nk_i^s=q(1-q)NK_s,\label{e3}
\end{equation}
where $K_s = \frac{1}{N}\sum_{i=1}^N \sum_{j=1}^N a_{ij}^s$ is the (unknown) average strong degree. 

Let $M_i^s$ be a binary variable equal to 1 if and only if node $i$ is named as a strong neighbor by at least one seed. The total number of nodes in $S_1^s$ is equal to $\sum_{i=1}^N(1-X_i)M_i^s$. By approximating the strong degree $\sum_{j=1}^N a_{ij}^s$ of node $i$ by $K_s$, we have
\begin{equation}
\mathbb{E}[n_1^s]=\sum_{i=1}^N(1-q)(1-(1-q)^{k_i^s})\simeq N(1-q)\big(1-(1-q)^{K_s}\big).\label{e4}
\end{equation}
We discuss when this assumption is reasonable and investigate its consequences further in Section~\ref{sec:approximation-justification} below.

Note that if two seeds are connected with a strong link, each of them names the other one as an alter. If they are connected with a weak link, the two events corresponding to each one naming the other are assumed to be independent. We model the event that node $i$ names node $j$ as a weak neighbor as a Bernoulli variable $W_{ij}$ that is equal to 1 with probability $\frac{B}{k_i^w}$ and 0 otherwise. So the total number of weak links connecting any two seeds is equal to $\frac{1}{2}\sum_{i=1}^N \sum_{j=1}^N X_iX_ja_{ij}^w(W_{ij}+W_{ji}-W_{ij}W_{ji})$, and its expected valued can be approximated by
\begin{align}
\mathbb{E}[m_0^w]&=\frac{1}{2}\sum_{i=1}^N\sum_{j\in \mathcal{N}_i^w}q^2\left(\frac{B}{k_i^w}+\frac{B}{k_j^w}-\frac{B^2}{k_i^wk_j^w}\right) \nonumber \\
&=\frac{1}{2}q^2B\left(2N-B\sum_{i=1}^N\sum_{j\in \mathcal{N}_i^w} \frac{1}{k_i^wk_j^w}\right) \nonumber \\
&\simeq \frac{1}{2}q^2BN\left(2-\frac{B}{K_w}\right),\label{e5}
\end{align}
where $\mathcal{N}_i^w$ denotes the set of weak neighbors of node $i$. Similarly, $m_1^w=\sum_{i=1}^N\sum_{j=1}^NX_i(1-X_j)a_{ij}^wW_{ij}$ and
\begin{equation}
\mathbb{E}[m_1^w]=q(1-q)NB.\label{e6}
\end{equation}

  If we write down the expected value of $n_1^w$, the second moment of the degrees in the weak layer  ($K_{ww}$) appears. 
  As we will discuss below, $K_{ww}$ can be estimated along with the other second moments by studying the number of triangles and triads in the observed graph. Therefore, at this step of inference it is reasonable  to  disregard  $n_1^w$ from the analysis. 

   We have six non-linear equations and four unknowns, $N,q,K_w,K_s$. There are different possible ways to approach solving this system of equations. 
   One is using the generalized method of moments which minimizes the weighted squared errors of all six equations. This is infeasible because the three equations, \eqref{e1},~\eqref{e5}, and~\eqref{e6}, admit a closed-form solution; the errors of these equations become  zero, leading to the divergence of their corresponding weights.  
  To avoid such divergence, we  proceed as follows. First, using  Equations~\eqref{e1},~\eqref{e5}, and~\eqref{e6}, we solve directly for $\widehat{N}$, $\widehat{q}$, and $\widehat{K}_w$: 
  \begin{align}
  &\widehat{N}=\frac{Bn_0^2}{B n_0-m_1^w}\\
  &\widehat{q}=\frac{B n_0-m_1^w}{B n_0}\\
   &\widehat{K}_w=\frac{B(B n_0-m_1^w)}{2B n_0-2m_1^w-m_0^w}.
\end{align} 
Then, we substitute the estimated values into~\eqref{e2}, \eqref{e3}, and \eqref{e4} and estimate $K_s$ by solving the least squares problem,
\all{
\min_{K_s} 
\Bigg[
\left( m_0^s-\mathbb{E}[m_0^s] \right)^2
 +
 \left( m_1^s-\mathbb{E}[m_1^s]\right)^2
+
\left(n_1^s-\mathbb{E}[n_1^s] \right)^2
\Bigg]
.
}{opt}
where the three expectations are replaced with the expressions from \eqref{e2}, \eqref{e3}, and \eqref{e4}.

\subsection{Second Moments}
\label{sec:second-moments}

Next we proceed to estimate the second moments of the degrees. Note that the existence of the moments of the degree distribution is not an issue here. 
Networks of interest in this work have finite  degree moments. 
Diverging moments occur in heavy-tailed degree distributions (e.g., power-law) \emph{only} for infinite network size. Moreover, as mentioned above, degree distributions of offline social networks are generally much less skewed than online social networks (see~\cite{dunbar}, for example), since humans typically have limited time and capacity to maintain strong and weak ties. 
Thus, for the networks of interest in this work we can safely assume that the degree moments exist and are finite. 

In the following, we use the term \emph{triad} to refer to a three-node motif consisting of one node (the ego) and two of its neighbors. The neighbors can be connected (a \emph{closed} triad) or not (an \emph{open} triad).  For example, a triangle in the original network $\G$ comprises three closed triads since any of the three nodes can be selected as the ego. 

\begin{figure}
\centering

    \begin{subfigure}{0.12\textwidth} 
     \begin{tikzpicture}   

    \draw[fill=blue!5,line width=0.35mm] (0.85,1.3) circle (6pt)node[](1){} ;
    \draw[fill=blue!5,line width=0.35mm] (0,0) circle (6pt)node[](2){} ;
    \draw[fill=blue!5,line width=0.35mm] (1.7,0) circle (6pt)node[](3){} ;

    \foreach \x/\y in {1/2,1/3}
        \path[ultra thick](\x) edge(\y); 
    \foreach \x/\y in {}
    	\path[arrows={[scale=0.8]}] (\x) edge [strong bidir] (\y); 
	\foreach \x/\y in {}
   		\path[arrows={[scale=0.8]}] (\x) edge [weak dir] (\y); 
    \foreach \x/\y in {}
    	\path[arrows={[scale=0.8]}] (\x) edge [weak bidir] (\y); 

    \end{tikzpicture}
    \caption{}\label{tss}
    \end{subfigure}~~
    \begin{subfigure}{0.12\textwidth}
	\begin{tikzpicture}   
    \draw[fill=blue!5,line width=0.35mm] (0.85,1.3) circle (6pt)node[](1){} ;
    \draw[fill=blue!5,line width=0.35mm] (0,0) circle (6pt)node[](2){} ;
    \draw[fill=blue!5,line width=0.35mm] (1.7,0) circle (6pt)node[](3){} ;

    \foreach \x/\y in {1/2}
        \path[ultra thick](\x) edge(\y); 
    \foreach \x/\y in {1/3}
        \path[dashed,ultra thick](\x) edge(\y); 
	\foreach \x/\y in {}
   		\path[arrows={[scale=0.8]}] (\x) edge [weak dir] (\y); 
    \foreach \x/\y in {}
    	\path[arrows={[scale=0.8]}] (\x) edge [weak bidir] (\y); 

    \end{tikzpicture} \caption{}\label{tsw}
    \end{subfigure}~~
    \begin{subfigure}{0.12\textwidth}
    \begin{tikzpicture}   

    \draw[fill=blue!5,line width=0.35mm] (0.85,1.3) circle (6pt)node[](1){} ;
    \draw[fill=blue!5,line width=0.35mm] (0,0) circle (6pt)node[](2){} ;
    \draw[fill=blue!5,line width=0.35mm] (1.7,0) circle (6pt)node[](3){} ;

    \foreach \x/\y in {1/2,1/3}
        \path[dashed,ultra thick](\x) edge(\y); 
    \foreach \x/\y in {}
    	\path[arrows={[scale=0.8]}] (\x) edge [strong bidir] (\y); 
	\foreach \x/\y in {}
   		\path[arrows={[scale=0.8]}] (\x) edge [weak dir] (\y); 
    \foreach \x/\y in {}
    	\path[arrows={[scale=0.8]}] (\x) edge [weak bidir] (\y); 

    \end{tikzpicture} 
    \caption{}\label{tww}
     \end{subfigure}
     \caption{Different compositions of open triads. Solid lines denote strong ties, and dashed lines denote weak ties.}\label{triads_3}
\end{figure}
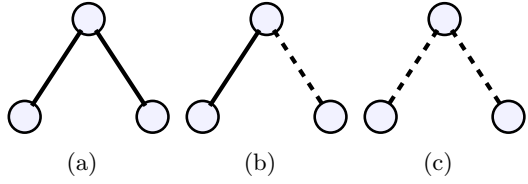

  To  employ  the method of moments for estimating $K_{ss}$, $K_{ww}$, and $K_{sw}$ and the clustering coefficient, we should again find variables that can be written as a function of the desired quantities (here, the second moments).  
The variables in $\G$   that can be written as a function of second moments are the number of different types of triads (closed and open). Due to link heterogeneity, we can have triads with different compositions, as illustrated in Figure~\ref{triads_3}.

Let $\tau_{ss}, \tau_{sw}$, and $\tau_{ww}$ denote the total number of triads in the original network, $\G$, similar to the ones in Figures~\ref{tss}, \ref{tsw}, and \ref{tww} , respectively.  
Then
\begin{align}
\tau_{ss}&=\sum_{i=1}^N \binom{k_i^s}{2}\simeq\frac{1}{2}N(K_{ss}-K_s)\label{kss}
\\ \tau_{sw}&=\sum_{i=1}^N \binom{k_i^s}{1}\binom{k_i^w}{1}\simeq N(K_{sw})\label{ksw}
\\ \tau_{ww}&=\sum_{i=1}^N \binom{k_i^w}{2}\simeq\frac{1}{2}N(K_{ww}-K_w).\label{kww}
\end{align}  
Note that these triads can be closed or open; the link (present or absent) between the two non-ego nodes is not accounted for here. 

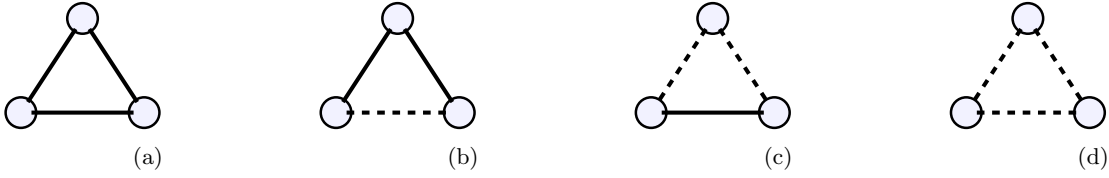
\begin{figure}
\centering
\resizebox{\columnwidth}{!}{
    \begin{subfigure}{0.24\columnwidth}
     \begin{tikzpicture}   

    \draw[fill=blue!5,line width=0.35mm] (0.85,1.3) circle (6pt)node[](1){} ;
    \draw[fill=blue!5,line width=0.35mm] (0,0) circle (6pt)node[](2){} ;
    \draw[fill=blue!5,line width=0.35mm] (1.7,0) circle (6pt)node[](3){} ;

    \foreach \x/\y in {1/2,1/3,2/3}
        \path[ultra thick](\x) edge(\y); 
    \foreach \x/\y in {}
    	\path[arrows={[scale=0.8]}] (\x) edge [strong bidir] (\y); 
	\foreach \x/\y in {}
   		\path[arrows={[scale=0.8]}] (\x) edge [weak dir] (\y); 
    \foreach \x/\y in {}
    	\path[arrows={[scale=0.8]}] (\x) edge [weak bidir] (\y); 

    \end{tikzpicture} 
    \caption{}\label{ts3}
    \end{subfigure}~~
    \begin{subfigure}{0.24\columnwidth}
 \begin{tikzpicture}   
    \draw[fill=blue!5,line width=0.35mm] (0.85,1.3) circle (6pt)node[](1){} ;
    \draw[fill=blue!5,line width=0.35mm] (0,0) circle (6pt)node[](2){} ;
    \draw[fill=blue!5,line width=0.35mm] (1.7,0) circle (6pt)node[](3){} ;

    \foreach \x/\y in {1/2,1/3}
        \path[ultra thick](\x) edge(\y); 
    \foreach \x/\y in {2/3}
    	\path[dashed,ultra thick] (\x) edge (\y); 
	\foreach \x/\y in {}
   		\path[arrows={[scale=0.8]}] (\x) edge [weak dir] (\y); 
    \foreach \x/\y in {}
    	\path[arrows={[scale=0.8]}] (\x) edge [weak bidir] (\y); 

    \end{tikzpicture} 
    \caption{}\label{ts2w}
    \end{subfigure}~~
    \begin{subfigure}{0.24\columnwidth}
    \begin{tikzpicture}   
    \draw[fill=blue!5,line width=0.35mm] (0.85,1.3) circle (6pt)node[](1){} ;
    \draw[fill=blue!5,line width=0.35mm] (0,0) circle (6pt)node[](2){} ;
    \draw[fill=blue!5,line width=0.35mm] (1.7,0) circle (6pt)node[](3){} ;

    \foreach \x/\y in {1/2,1/3}
        \path[dashed,ultra thick](\x) edge(\y); 
    \foreach \x/\y in {2/3}
    	\path[ultra thick] (\x) edge (\y); 
	\foreach \x/\y in {}
   		\path[arrows={[scale=0.8]}] (\x) edge [weak dir] (\y); 
    \foreach \x/\y in {}
    	\path[arrows={[scale=0.8]}] (\x) edge [weak bidir] (\y); 

    \end{tikzpicture} 
    \caption{}\label{tsw2}
     \end{subfigure}~~
     \begin{subfigure}{0.24\columnwidth}
    \begin{tikzpicture}   
    \draw[fill=blue!5,line width=0.35mm] (0.85,1.3) circle (6pt)node[](1){} ;
    \draw[fill=blue!5,line width=0.35mm] (0,0) circle (6pt)node[](2){} ;
    \draw[fill=blue!5,line width=0.35mm] (1.7,0) circle (6pt)node[](3){} ;

    \foreach \x/\y in {1/2,1/3,2/3}
        \path[dashed,ultra thick](\x) edge(\y); 
    \foreach \x/\y in {}
    	\path[arrows={[scale=0.8]}] (\x) edge [strong bidir] (\y); 
	\foreach \x/\y in {}
   		\path[arrows={[scale=0.8]}] (\x) edge [weak dir] (\y); 
    \foreach \x/\y in {}
    	\path[arrows={[scale=0.8]}] (\x) edge [weak bidir] (\y); 

    \end{tikzpicture} 
    \caption{}\label{tw3}
     \end{subfigure}}
     \caption{Different compositions of triangles. Solid links denote strong ties, and dashed links denote weak ties.}\label{triangles_4}
\end{figure}

There are four compositions of triangles, as illustrated in Figure~\ref{triangles_4}, based on the type of each link. The number of each of these triangles in $\G$ is denoted by $T_{s^3}, T_{s^2w}, T_{sw^2}$,and $T_{w^3}$. Recall that each triangle comprises three closed triads.  For instance, a triangle with three strong edges gets counted as three $ss$ triads, and a triangle with one strong edge and two weak edges corresponds to one $ww$ triad and two $sw$ triads. 

The total number of possible triads in $\G$ can be written as a function of the number of open triads and the number of triangles in the network:
\begin{equation}
\tau_{ss}=\lambda_{ss}+3T_{s^3}+T_{s^2w}
\label{ss}
\end{equation}
\begin{equation}
\tau_{sw}=\lambda_{sw}+2T_{s^2w}+2T_{sw^2}
\label{sw}
\end{equation}
\begin{equation}
\tau_{ww}=\lambda_{ww}+3T_{w^3}+T_{sw^2},
\label{ww}
\end{equation}
where $\lambda_{ss}, \lambda_{sw}$, and $\lambda_{ww}$ denote the number of different open triads (Figure~\ref{triads_3}) in $\G$. 

So based on Equations~\eqref{ss}, \eqref{sw}, and \eqref{ww}, in order to estimate the total number of triads  in $\G$, we need to separately estimate the number of triangles, as well as the number of open triads.  To this end, we need to find the expected values of the number of triangles and open triads in $\GG$ as a function of these values in $\G$ and the estimated variables in the first step of inference (that is, $N$, $q$, $K_{s}$, $K_w$).

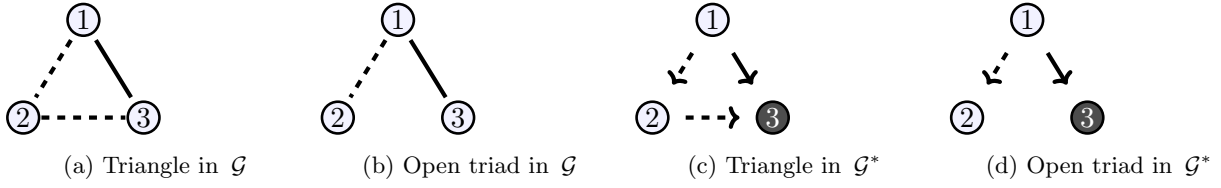
\begin{figure}
\centering
    \begin{subfigure}{0.24\textwidth}
     \begin{tikzpicture}   
    \draw[fill=blue!5,line width=0.35mm] (0.8,1.3) circle (6pt)node[](1){1} ;
    \draw[fill=blue!5,line width=0.35mm] (0,0) circle (6pt)node[](2){2} ;
    \draw[fill=blue!5,line width=0.35mm] (1.6,0) circle (6pt)node[](3){3} ;
   \foreach \x/\y in {1/3}
        \path[ultra thick](\x) edge(\y); 
   \foreach \x/\y in {1/2,2/3}
        \path[dashed,ultra thick](\x) edge(\y); 
    \foreach \x/\y in {}
    	\path[arrows={[scale=0.8]}] (\x) edge [strong bidir] (\y); 
	\foreach \x/\y in {}
   		\path[arrows={[scale=0.8]}] (\x) edge [weak dir] (\y); 
    \foreach \x/\y in {}
    	\path[arrows={[scale=0.8]}] (\x) edge [weak bidir] (\y); 

    \end{tikzpicture} 
    \caption{Triangle in $\G$}\label{trianG}
    \end{subfigure}~ 
    \begin{subfigure}{0.24\textwidth}
     \begin{tikzpicture}   
    \draw[fill=blue!5,line width=0.35mm] (0.8,1.3) circle (6pt)node[](1){1} ;
    \draw[fill=blue!5,line width=0.35mm] (0,0) circle (6pt)node[](2){2} ;
    \draw[fill=blue!5,line width=0.35mm] (1.6,0) circle (6pt)node[](3){3} ;
   \foreach \x/\y in {1/3}
        \path[ultra thick](\x) edge(\y); 
   \foreach \x/\y in {1/2}
        \path[dashed,ultra thick](\x) edge(\y); 
    \foreach \x/\y in {}
    	\path[arrows={[scale=0.8]}] (\x) edge [strong bidir] (\y); 
	\foreach \x/\y in {}
   		\path[arrows={[scale=0.8]}] (\x) edge [weak dir] (\y); 
    \foreach \x/\y in {}
    	\path[arrows={[scale=0.8]}] (\x) edge [weak bidir] (\y); 

    \end{tikzpicture} 
    \caption{Open triad in $\G$}\label{triadG}
    \end{subfigure}~
 \begin{subfigure}{0.24\textwidth}   
 \begin{tikzpicture}   
    
    \draw[fill=blue!5,line width=0.35mm] (0.8,1.3) circle (6pt)node[](1){1} ;
    \draw[fill=blue!5,line width=0.35mm] (0,0) circle (6pt)node[](2){2} ;
    \draw[fill=black!70,line width=0.35mm] (1.6,0) circle (6pt)node[](3){\color{white} 3} ;

    \foreach \x/\y in {1/3}
        \path[arrows={[scale=0.8]}] (\x) edge [strong dir] (\y); 
    \foreach \x/\y in {}
    	\path[arrows={[scale=0.8]}] (\x) edge [strong bidir] (\y); 
	\foreach \x/\y in {2/3,1/2}
   		\path[arrows={[scale=0.8]}] (\x) edge [weak dir] (\y); 
    \foreach \x/\y in {}
    	\path[arrows={[scale=0.8]}] (\x) edge [weak bidir] (\y); 

    \end{tikzpicture}
    \caption{Triangle in $\GG$}\label{trianGG}
  \end{subfigure}~
    \begin{subfigure}{0.24\textwidth}
    \begin{tikzpicture}   
    \draw[fill=blue!5,line width=0.35mm] (0.8,1.3) circle (6pt)node[](1){1} ;
    \draw[fill=blue!5,line width=0.35mm] (0,0) circle (6pt)node[](2){2} ;
    \draw[fill=black!70,line width=0.35mm] (1.6,0) circle (6pt)node[](3){\color{white} 3} ;

    \foreach \x/\y in {1/3}
        \path[arrows={[scale=0.8]}] (\x) edge [strong dir] (\y); 
    \foreach \x/\y in {}
    	\path[arrows={[scale=0.8]}] (\x) edge [strong bidir] (\y); 
	\foreach \x/\y in {1/2}
   		\path[arrows={[scale=0.8]}] (\x) edge [weak dir] (\y); 
    \foreach \x/\y in {}
    	\path[arrows={[scale=0.8]}] (\x) edge [weak bidir] (\y); a
    \end{tikzpicture}
    \caption{Open triad in $\GG$}\label{triadGG}
  \end{subfigure} 
\caption{Example of a triangle and an open triad in  $\G$  being observed in  $\GG$. The hollow nodes are chosen as respondents, and the solid node is not chosen. Solid lines  represent strong links. Dashed  lines represent weak links. Arrows indicate mentioning  the adjacent node in the interview.}\label{transform}
\end{figure}

Let us consider two illustrative examples. Consider  the triangle shown in Figure~\ref{trianG}. Depending on whether each of the three nodes are selected as seeds, and if so whether they name the other nodes, this triangle may or may not appear in $\GG$. One possible scenario in which the triangle can be observed in $\GG$ is illustrated in Figure~\ref{trianGG}. This event happens if:
\begin{enumerate}
\item Only nodes 1 and 2 are selected as seeds; 
\item Node 1 names node 2 (and not the reverse); and 
\item Node 2 names node 3 (the reverse cannot occur since node 3 is not a seed). 
\end{enumerate}
The probability of this event (all three points above occurring simultaneously) is $q^2(1-q)b_{11}b_{01}$, where $b_{11}$  denotes the probability that a seed (here, node 1) names one strong link and one weak link and similarly, $b_{01}$ denotes the probability that a seed (here, node 2) names no strong link and one weak link in the triad. These probabilities depend on the degrees of the seed. However, we approximate them for an arbitrary node $x$   by
\begin{equation}
b_{01}=\frac{\binom{k_x^w-2}{B-1}}{\binom{k_x^w}{B}}=\frac{B(k_x^w-B)}{k_x^w(k_i^w-1)}\simeq\frac{B(K_w-B)}{K_w(K_w-1)},
\end{equation}
and
\begin{equation}
b_{11}=\frac{\binom{k_x^w-1}{B-1}}{\binom{k_x^w}{B}}=\frac{B}{k_x^w}\simeq \frac{B}{K_w}.
\end{equation}

Similarly, we can define $b_{00},b_{10},b_{20}$, and $b_{02}$. Their approximated expressions are shown in Table~\ref{B_A} in the Appendix. There are 42 possible ways that a triangle in $\G$ can be observed in $\GG$, and these factors can be used as building blocks for calculating the probabilities pertaining to all 42 possible ways. We denote the probability of observing triangles in $\GG$ by $\{\rho_{j}; j=1,2,...,42\}$. All of the triangles and corresponding expressions for $\rho_j$ are presented in Figure~\ref{all_triangles} and Table~\ref{rho} in the Appendix.

The same triangle in Figure~\ref{trianG} can be observed as an open triad in $\GG$. One possible scenario is illustrated in Figure~\ref{triadGG}. The probability of this event is equal to $q^2(1-q)b_{11}b_{00}$. There are 31 possible ways an open triad can be observed in $\GG$ (see Figure~\ref{all_triads}). We denote the probability of observing triangles in $\G$ as open triads in $\GG$ by $\{\pi_i; i=1,2,...,31\}$ (Table~\ref{pi}).

Open triads in $\GG$ are not observed whenever at least one link in a triangle in $\G$ is absent (not named). They can be observed if an open triad with the same composition is preserved during the sampling process. Consider again the open triad in Figure~\ref{triadGG}. It can originate from the triangle in Figure~\ref{trianG} or from the open triad in Figure~\ref{triadG}. Note that in the latter case, node 2 in not connected to node 3 in $\G$. So the absence of this link in $\GG$ is not the result of node 2 not naming node 3 (unlike the case of triangle to triad). We can write the probability of observing this triad originating from the triad in Figure~\ref{triadG} as $q^2(1-q)b_{11}a_{00}$, where $a_{00}$ corresponds to node 2 not naming any strong or weak link while it  is connected to only one of them. For an arbitrary node $x$, $a_{00}$ can be approximated by
\begin{equation}
a_{00}=\frac{\binom{k_x^w-1}{B}}{\binom{k_x^w}{B}}=\frac{k_x^w-B}{k_x^w}\simeq 1-\frac{B}{K_w}.
\end{equation}
Similarly, we can define $a_{01}$ and $a_{10}$. The approximation of these quantities are presented in Table~\ref{B_A} in the Appendix. We denote the probability of observing open triads in $\G$ as open triads in $\GG$ by $\{\phi_i; i=1,2,...,31\}$ (Table~\ref{phi}).

Let us denote the expected number of different types of triangles and open triads in $\GG$ with the same notation introduced for the original network with the addition of $\ast$ superscrtipts. For the triangles we have
\begin{align}
&T^{\ast}_{s^3}=T_{s^3}\times \sum_{i=1}^2 {\rho_i} \label{Ts3star} \\ 
&T^{\ast}_{s^2w}=T_{s^2w}\times \sum_{i=3}^{7}{\rho_i}\label{Ts2wstar} \\
&T^{\ast}_{sw^2}=T_{sw^2}\times \sum_{i=8}^{17}{\rho_i}\label{Tsw2star}\\ 
&T^{\ast}_{w^3}=T_{w^3}\times \sum_{i=18}^{26}{\rho_i}.\label{Tw3star}
\end{align}
 For example, Equation~\eqref{Ts2wstar} states that  the   $s^2w$ triangles  in $\G$ can be observed in $\GG$ under 5 different events, depicted in Figure~\ref{all_triangles}, whose probabilities are listed in Table~\ref{rho}. A similar explanation and reasoning follows for the other triangular configurations. 
Also, the expected number of open triads in $\GG$ can be written as
\begin{align}
&\lambda^{\ast}_{ss}=3T_{s^3}\times \pi_3+T_{s^2w}\times\sum_{i=1}^4{\pi_i}+\lambda_{ss}\times\sum_{i=1}^4{\phi_i} \label{tssstar}
\\ &\lambda^{\ast}_{sw}=2 T_{s^2 w }\times \pi_6+ 2 T_{sw^2}\times\sum_{i=5}^{13} {\pi_i}+\lambda_{ss}\times\sum_{i=5}^{13} {\phi_i}\label{tswstar}
\\ &\lambda^{\ast}_{ww}=T_{sw^2}\times \pi_{14}+3T_{w^3}\times\sum_{i=14}^{24}{\pi_i}+\lambda_{ww}\times\sum_{i=14}^{24}{\phi_i}.\label{twwstar}
\end{align}

Note that the coefficients $a_{i,j}$ and $b_{i,j}$ in Table~\ref{B_A} are all functions of $B$ and $K_w$. Similarly, the coefficients $\rho_i$, $\pi_i$, and $\phi_i$ only depend on the values from Table~\ref{B_A} and $q$. Since the parameter $B$ is assumed to be known, given estimates of $q$ and $K_w$ we can approximate all of these coefficients. Then we can use the estimated values in conjunction with Equations \eqref{Ts3star}--\eqref{twwstar} to estimate the number of triangles and triads. Note that, given the coefficients, these are all linear equations in the unknown parameters, so estimation reduces to solving a system of linear equations. Finally, we use the estimated numbers of triangles and triads to estimate the degree correlations, $K_{ss}$, $K_{sw}$, and $K_{ww}$ via Equations \eqref{kss}, \eqref{ksw}, and \eqref{kww}, which are also linear equations in the unknowns.

 \subsection{Summary of the Inference Method}

The following steps summarize the entire proposed inference method.
\begin{enumerate}
\item{Estimate $\widehat{N}, \widehat{q}$, and $\widehat{K}_w$ (Equations~\eqref{e1}, \eqref{e2}, and \eqref{e3}).}
\item{Estimate the strong degree $\widehat{K}_s$ \eqref{opt}.}
\item{Plug in the estimated weak degree $\widehat{K}_w$ and sampling probability $\widehat{q}$ to estimate values for $\{\rho_j; j=1,2,...,26\}$ and $\{(\pi_i,\phi_i); i=1,2,...,24\}$ (Tables~\ref{rho}, \ref{pi}, \ref{phi}). }
\item{Count the number of observed triangles ($T^{\ast}_{s^3}$, $T^{\ast}_{s^2w}$, $T^{\ast}_{sw^2}=T_{sw^2}$, and $T^{\ast}_{w^3}=T_{w^3}$) and triads ($\lambda^{\ast}_{ss}$, $\lambda^{\ast}_{sw}$, and $\lambda^{\ast}_{ww}$) in $\GG$.}
\item{Estimate the number of different triangles in $\G$  (Equations~\eqref{Ts3star}, \eqref{Ts2wstar}, \eqref{Tsw2star}, \eqref{Tw3star}).}
\item{Estimate number of different open triads in $\G$ (Equations~\eqref{tssstar}, \eqref{tswstar}, and \eqref{twwstar}).}
\item{Estimate the total number of all triads in $\G$ (Equations~\eqref{ss}, \eqref{sw}, and \eqref{ww}).}
\item{Estimate $\widehat{K}_{ss}$, $\widehat{K}_{sw}$, and  $\widehat{K}_{ww}$ (Equations \eqref{kss}, \eqref{ksw}, and \eqref{kww}).}
\end{enumerate}

The computational complexity of this method is dominated by step~4, which involves counting all triangles and triads in the observed network. Typical studies of offline social networks focus on villages populations smaller than $10^4$. For observed networks of this size, running the entire inference procedure takes about one second on a contemporary laptop computer.

\subsection{The Average Degree Approximation}
\label{sec:approximation-justification}
Many steps of the development above involve approximating the individual node degrees $k_i^s$ and $k_i^w$ with the average values $K_s$ and $K_w$. Let us briefly describe why this is both  practically  and  theoretically  reasonable.  
First note that the less skewed the degree distribution is, the better the said approximation performs. 
  Although there is no social network study in which a full real-world offline social network has been observed, there are studies which provide the degree distribution. For example, see Figure~1 in~\cite{dunbar}. 
Offline social networks exhibit reasonably concentrated degree distributions, not heavy tailed. So it is expected that the adopted approximation is not a significant source of error. 

Let us also estimate the error theoretically. Consider a network whose weak and strong degree distributions are both Poisson; i.e., suppose $k_i^s$, $i=1,\dots,N$ are i.i.d.~Poisson random variables with mean $K_s$. It is straightforward to show that
\begin{equation}
\mathbb{E}\left[\sum_i(1-q)^{k_i^s}\right]=Ne^{-qK_s}.
\end{equation}
Evaluating the Taylor expansion of this expression at $q=0$, we find that the relative error is
\begin{equation*}
\frac{\mathbb{E}[\sum_i(1-q)^{k_i^s}]}{N(1-q)^{K_s}}=1+\frac{K_sq^2}{2}+\frac{K_sq^3}{3}+\frac{K_s(K_s+2)q^4}{8}+O(q^5).
\end{equation*}
Thus, the leading term in the error is proportional to $q^2$, which is reasonably small (note that in typical offline social networks, $K_s$ may be a few 10's while $q$ is significantly smaller than one). 

\begin{figure}
\centering
\includegraphics[width=0.5\textwidth]{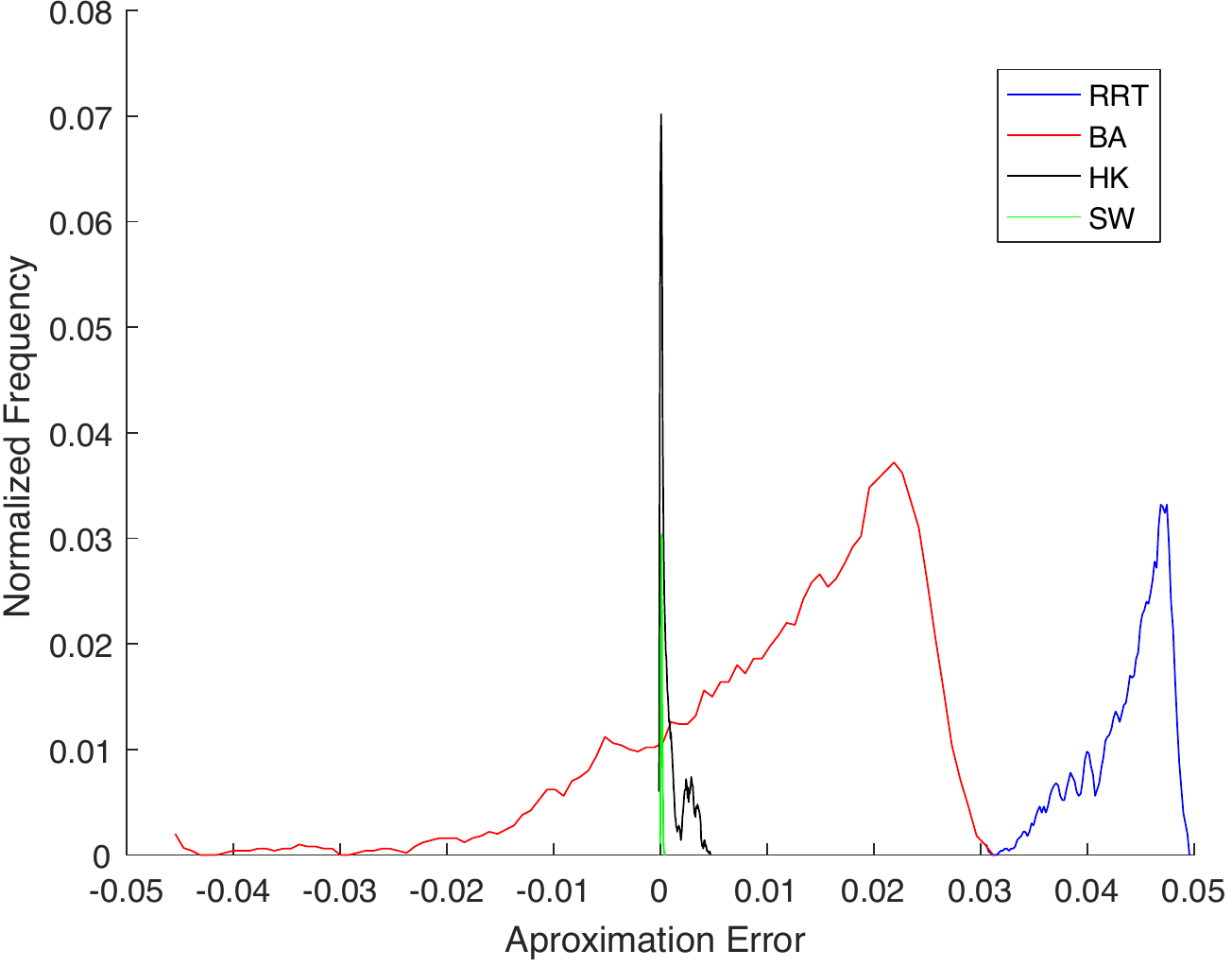}
\caption{
The distribution of relative error due to the approximation made in Equation~\eqref{e5} for the SW, RRT, BA, and HK families of networks. For each family we generated 1000 networks of size 1000, with parameters randomly generated. As mentioned in the text, BA and RRT are worst-case scenarios due to their extreme skew, yet the relative error is reasonably small for them. For the more realistic models of SW and HK, the error is  considerably smaller. 
}\label{j1}
\end{figure}

For Equation~\eqref{e5}, theoretical calculation cannot be performed in closed form even for Poisson networks.
 Thus, to verify its applicability, we tested it on a variety of network models with different properties. Figure~\ref{j1} presents the distributions of $\frac{\hat{Y}}{Y}$, where $Y$ is the sum ${\sum_i \sum_{j \in \mathcal{N}_i^w} \frac{1}{k_i^w k_j^w}}$, approximated in Equation~\eqref{e5} by the expression ${\widehat{Y}=\frac{N}{K_w}}$. 
 In the simulation we employ four distinct families of networks with different properties to investigate the robustness of the approximation. The four synthetic network models are: Small-world~(SW) \cite{newman1999scaling},   Barabasi-Albert~(BA) \cite{barabasi1999}, Random Recursive Trees~(RRT) \cite{RRT}, and the high-clustering scale-free model of Holme and Kim~(HK) \cite{HK}. The difference between BA and RRT is that in the BA model incoming nodes choose their neighbors preferentially (i.e., with degree-proportional probabilities), whereas in the RRT model they choose them uniformly at random. The BA and RRT models generate networks with unrealistically high-skewed degree distributions for offline social networks; we present them here as extreme worst-case scenarios. 
 
 We generate 1000 random networks from each family, with parameters randomly generated, and with sizes fixed at $N=1000$. The distribution of relative errors is presented in Figure~\ref{j1}. 
  It can be observed that the relative error for these networks is less than 6\%. For the HK and SW models, which may be considered more realistic models of offline social networks, the relative error is about 1\%. This demonstrates the reasonable accuracy of the approximations. 

 \section{Results and Discussions} \label{sec:results}

\subsection{Performance of the Proposed Estimators}

To verify the accuracy of the estimators, the ideal scenario would be to have data from real-world offline social networks that have been fully observed (i.e., ground truth), along with their sampled versions. We found no fully-observed real-world offline social network dataset available in the literature. Full observation is almost impossible due to practical and privacy considerations. Thus we use synthetic networks for evaluation. 

We verify the accuracy of the proposed estimators via Monte Carlo simulations over 500 synthetic networks. In each MC trial, we build a synthetic two-layer network. The set of nodes in the two layers is the same. One layer represents the strong links and the other represents weak links. All the synthetic networks are generated according to a modified Watts-Strogatz model, with the difference being that edges are randomly added instead of being randomly rewired~\cite{newman1999scaling}. We randomly sample model parameters such that the average degree of the weak layer falls between 100 and 200 and the average degree of the strong layer is between 10 and 20. 
These values are justified by the substantial literature in evolutionary psychology and neuroscience~\cite{miritello2013time,dunbar2015structure,dunbar2014social,
sutcliffe2012relationships,hill2008network,zhou2005discrete,
saramaki2014persistence} which suggests that the human brain has evolved to maintain approximately $150$ active social ties, with an `inner circle' of up to 20 members. 
The results are not sensitive to these precise values; increasing the average degree in the weak layer does not substantially change the results.

We apply the sampling process on this network. Then, we infer the desired variables and compare them to the true values. The number of weak links named by seeds is $B=10$. We first keep $q=0.1$ constant and increase $N$ to confirm that the performance improves as the network size increases. We then fix $N=4000$ and vary $q$ to study the effect of sampling proportion. The same simulations are repeated with $B=2$. Finally, we fix $N=4000$ and $q=0.1$ and vary $B$.

\begin{figure*}
        \centering
        \begin{subfigure}[b]{ 0.32\textwidth}
                \includegraphics[width=\textwidth]{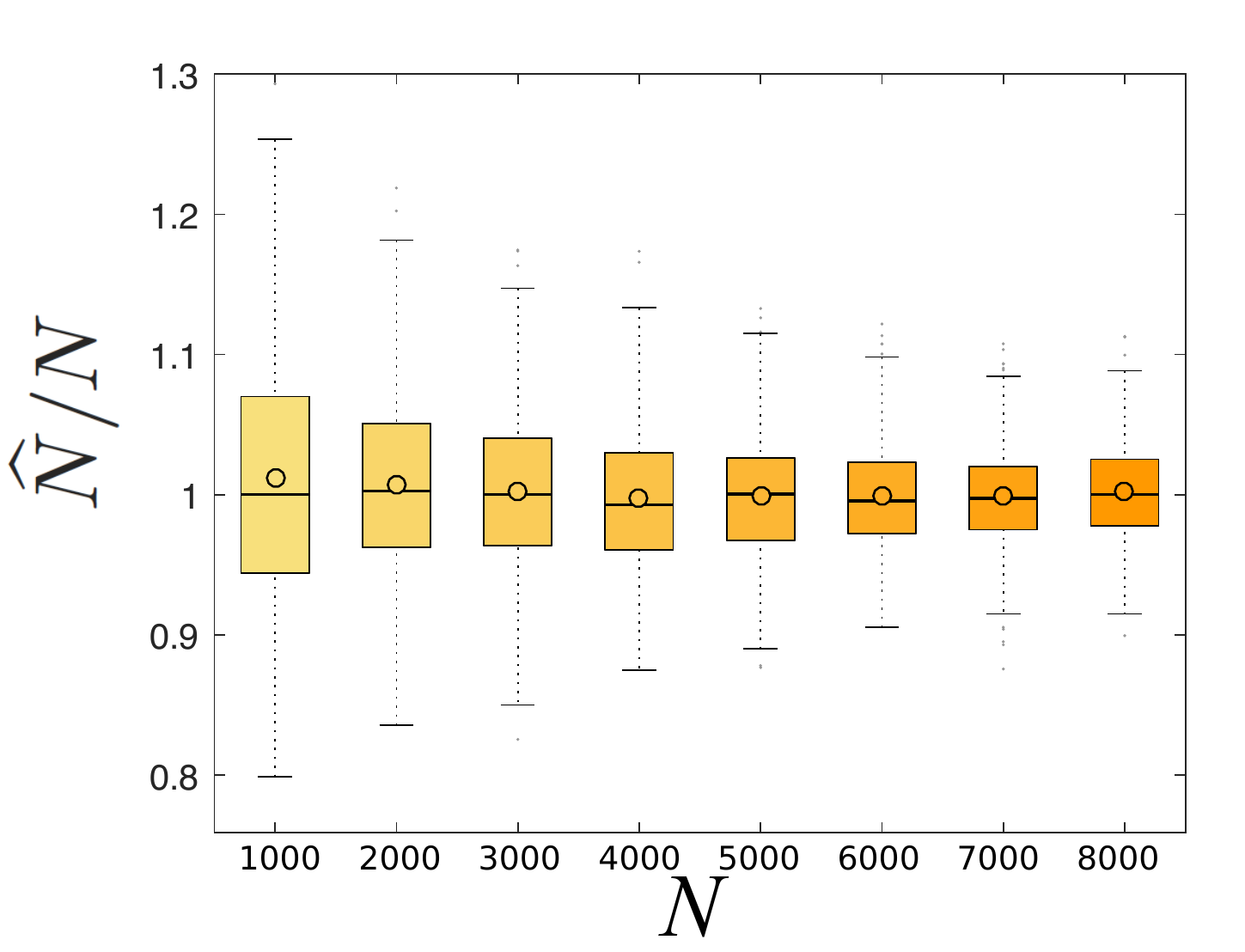}
                \caption{Number of Nodes, $q=0.1$}
                \label{N}
        \end{subfigure}%
        ~ 
        \begin{subfigure}[b]{ 0.32\textwidth}
                \includegraphics[width=\textwidth]{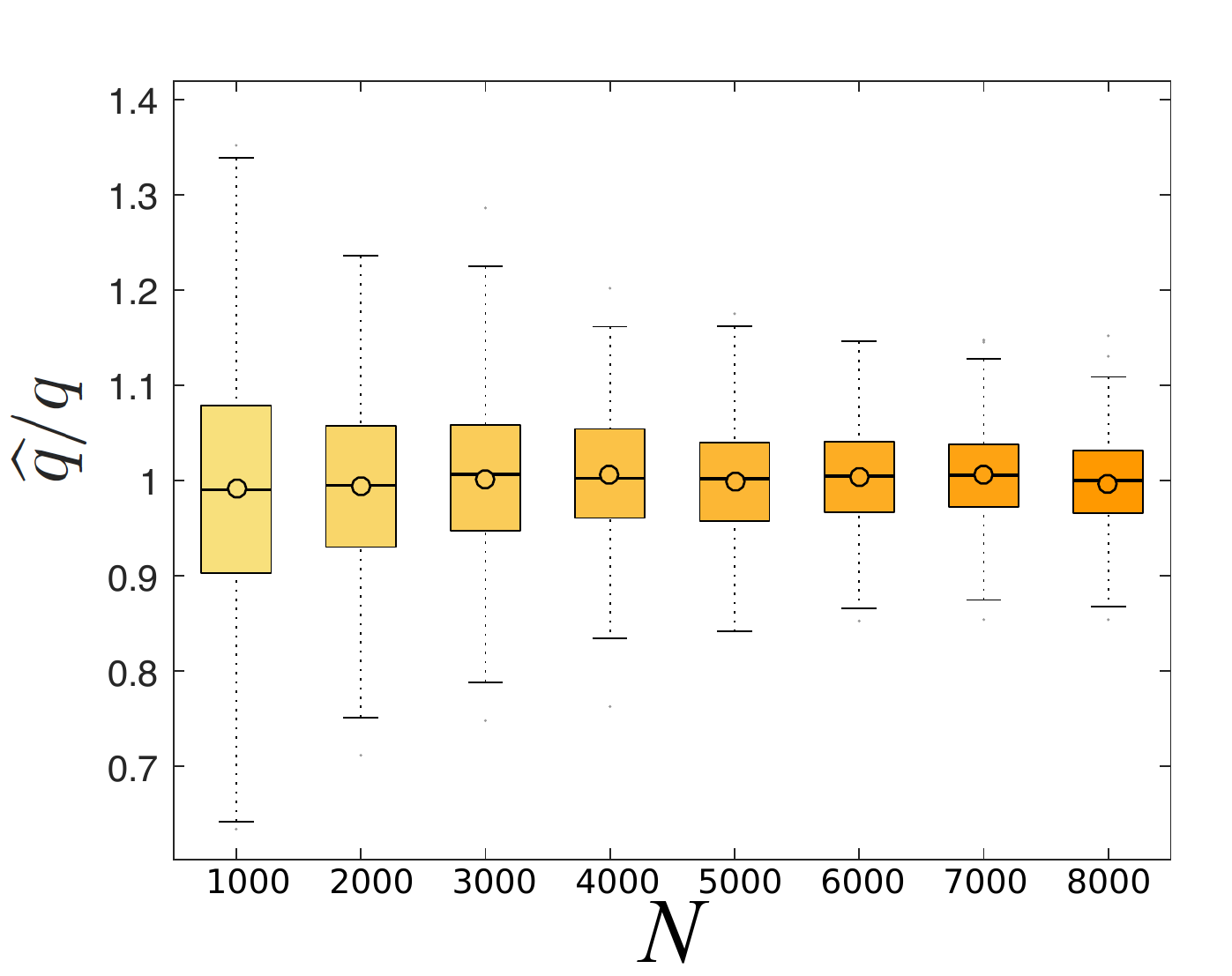}
                \caption{Sampling probability, $q=0.1$}
                
        \end{subfigure}
        ~
        \begin{subfigure}[b]{0.32\textwidth}
                \includegraphics[width=\textwidth]{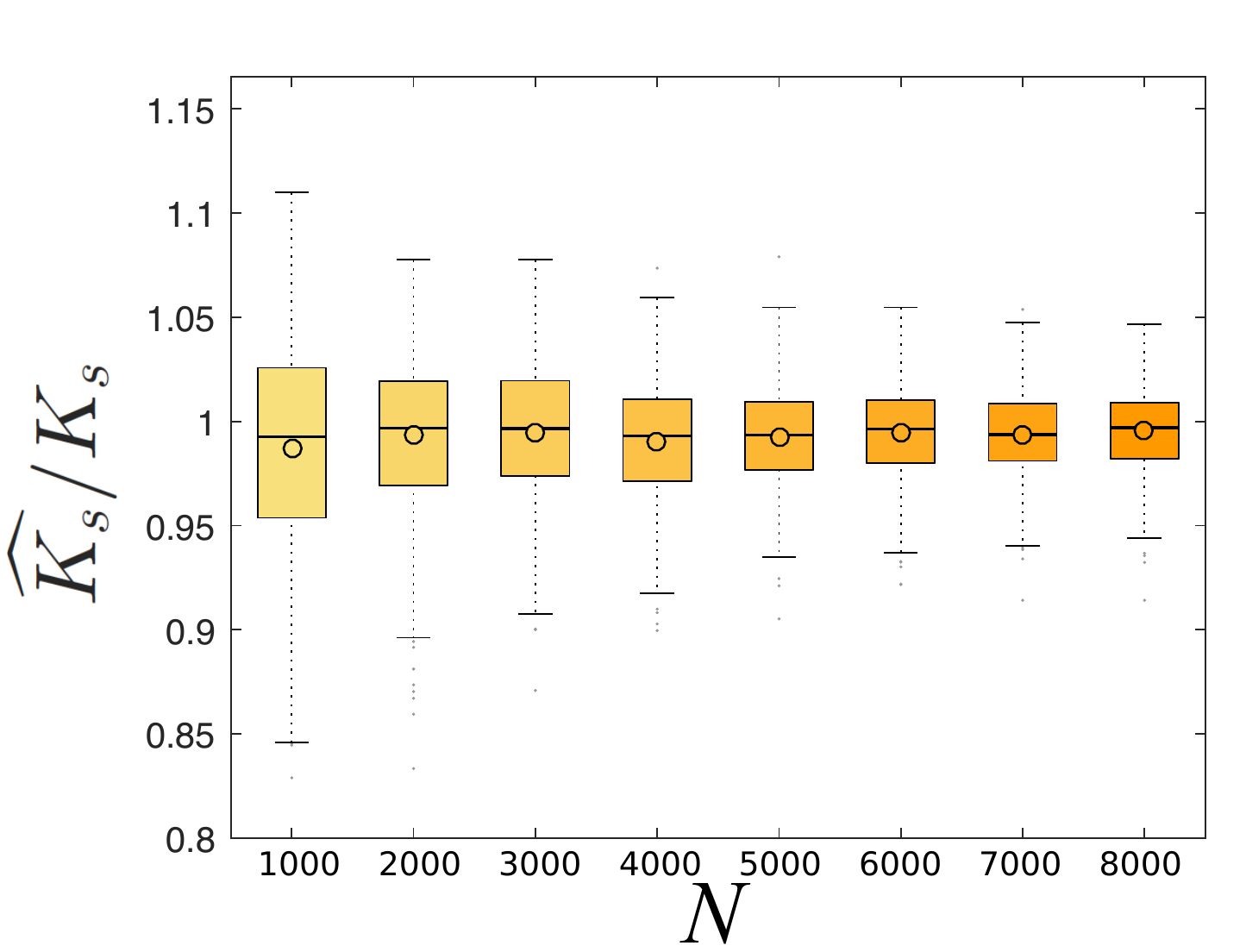}
                \caption{Average strong degree, $q=0.1$ }
                \label{ks_N}
        \end{subfigure}%
        \\
        
        \vspace{1em}
        
        \begin{subfigure}[b]{0.32\textwidth}
                \includegraphics[width=\textwidth]{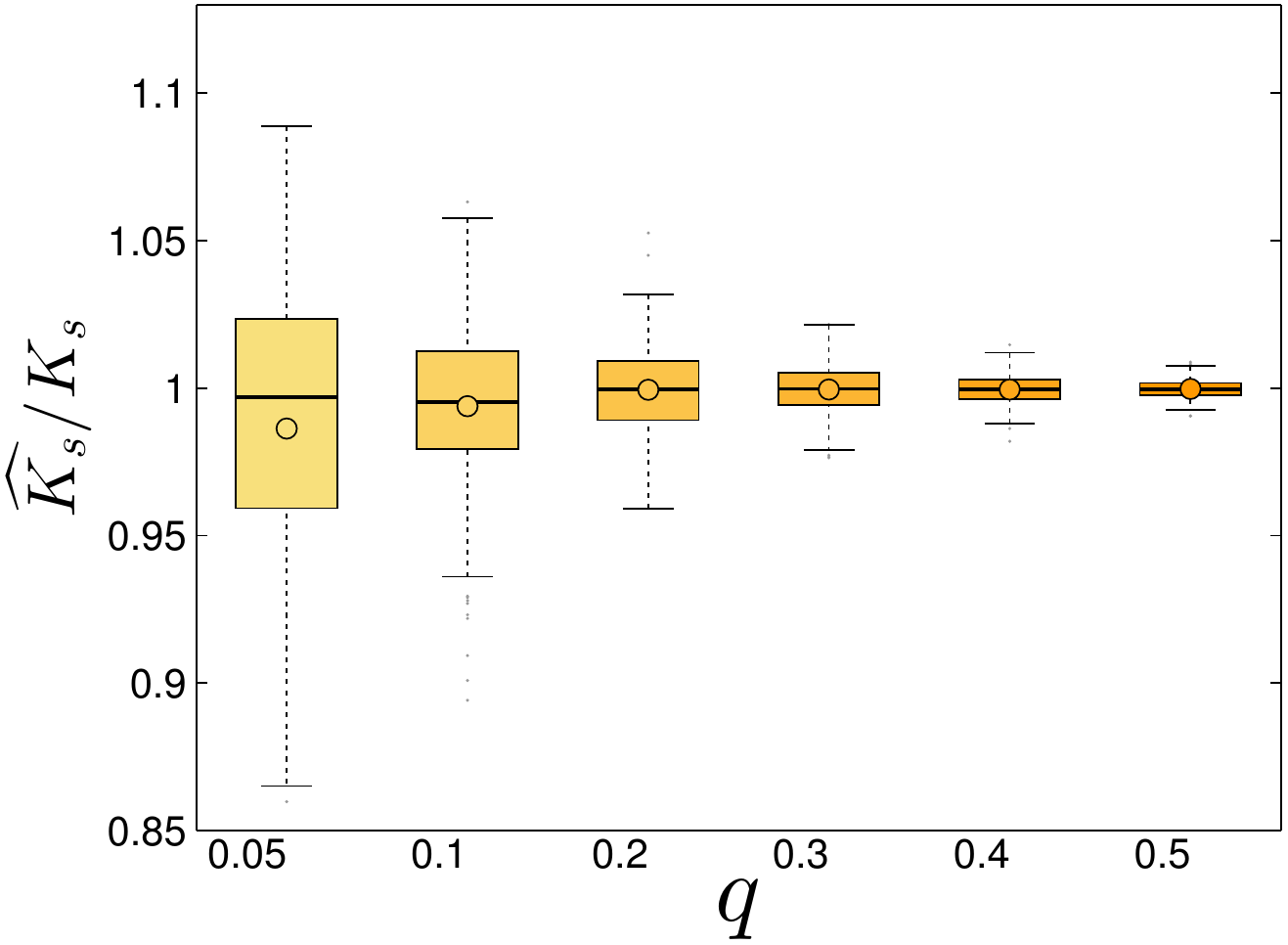}
                \caption{Average strong degree, $N=4000$ }
                \label{ks_q}
        \end{subfigure}
        ~
        \begin{subfigure}[b]{0.32\textwidth}
                \includegraphics[width=\textwidth]{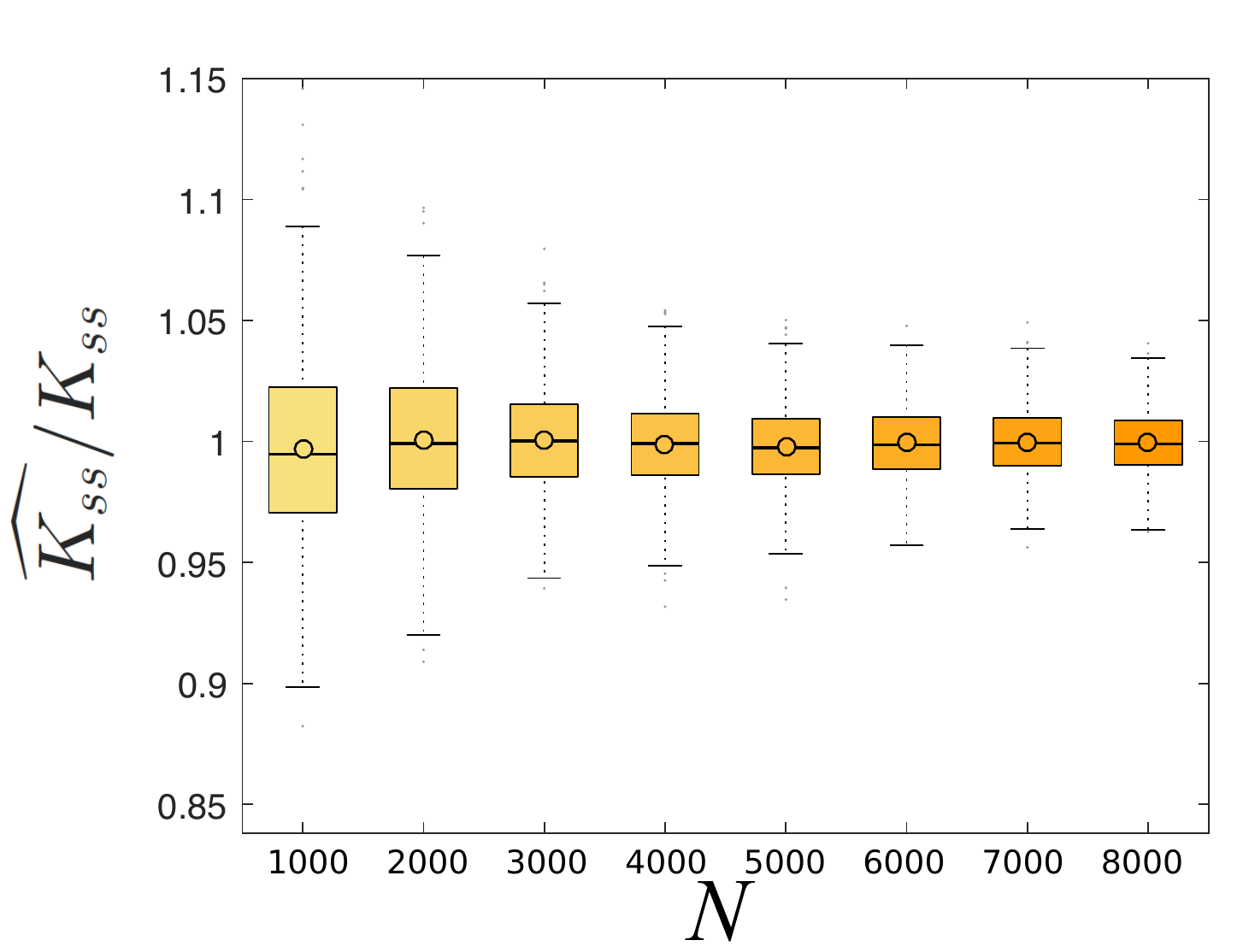}
                \caption{Second moment of strong degrees, $q=0.1$}
                \label{ks2_N}
        \end{subfigure}%
        ~ 
        \begin{subfigure}[b]{0.32\textwidth}
                \includegraphics[width=\textwidth]{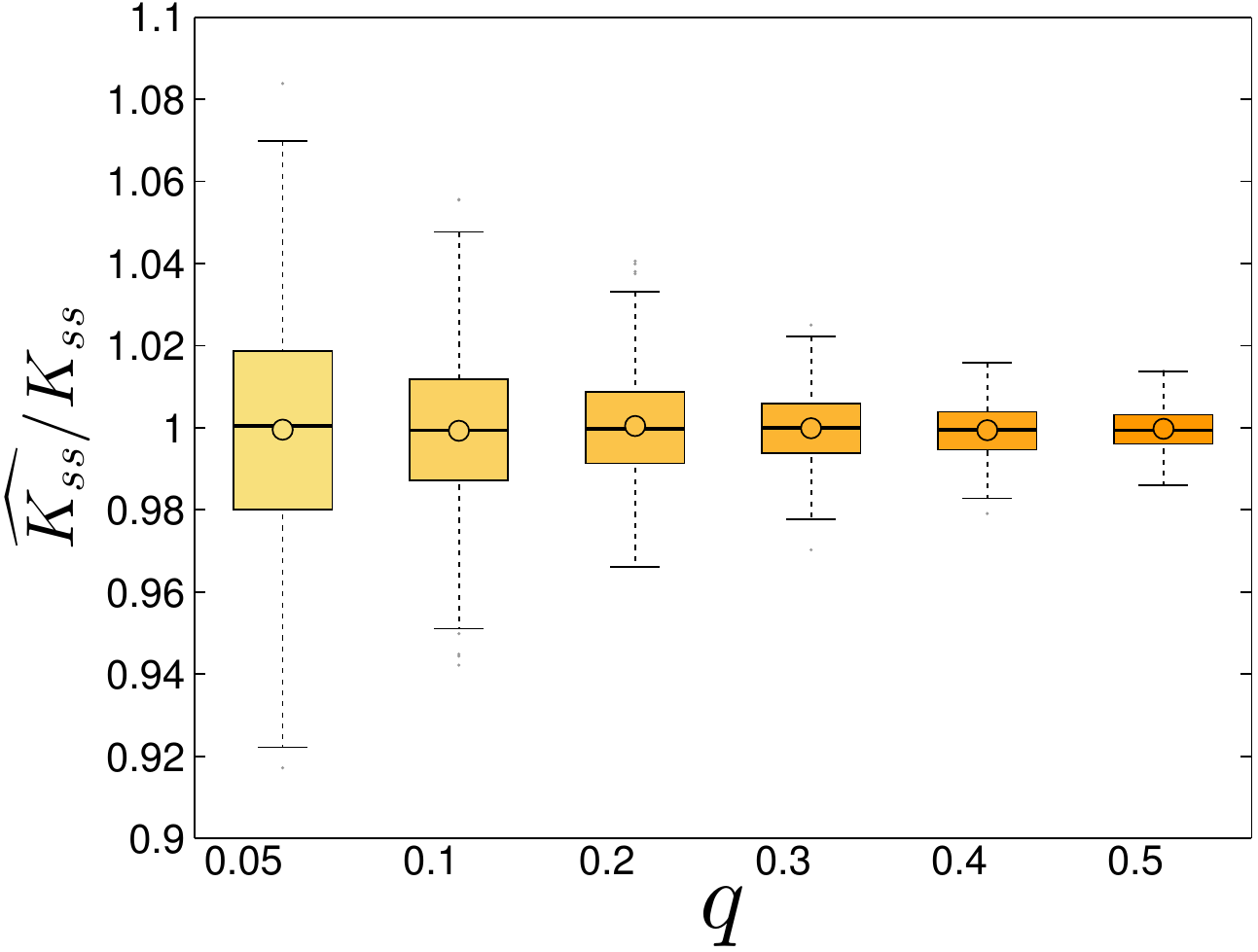}
                \caption{Second moment of strong degrees, $N=4000$}
                \label{ks2_q}
        \end{subfigure}
        \caption{\footnotesize{Distribution of ratio of estimated values and true values ($B=10$).} }\label{fig:results1}
\end{figure*}

Figure~\ref{fig:results1} shows the empirical distribution of the ratio of the estimated values to the true values for $N$, $q$, $K_s$, and $K_{ss}$ (all for $B=10$).  In all cases, the estimator does exhibit some bias for smaller sizes of networks, $N$, and samples, $q$.  As the number of nodes or the sample size increases the variability of the estimates decreases, as does the bias. The results when $B=2$ are similar to the case of $B=10$ (see the supplementary material); only the variability of the estimates is greater, but not significantly so. 

\begin{figure*}
        \centering
        \begin{subfigure}[b]{ 0.32\textwidth}
                \includegraphics[width=\textwidth]{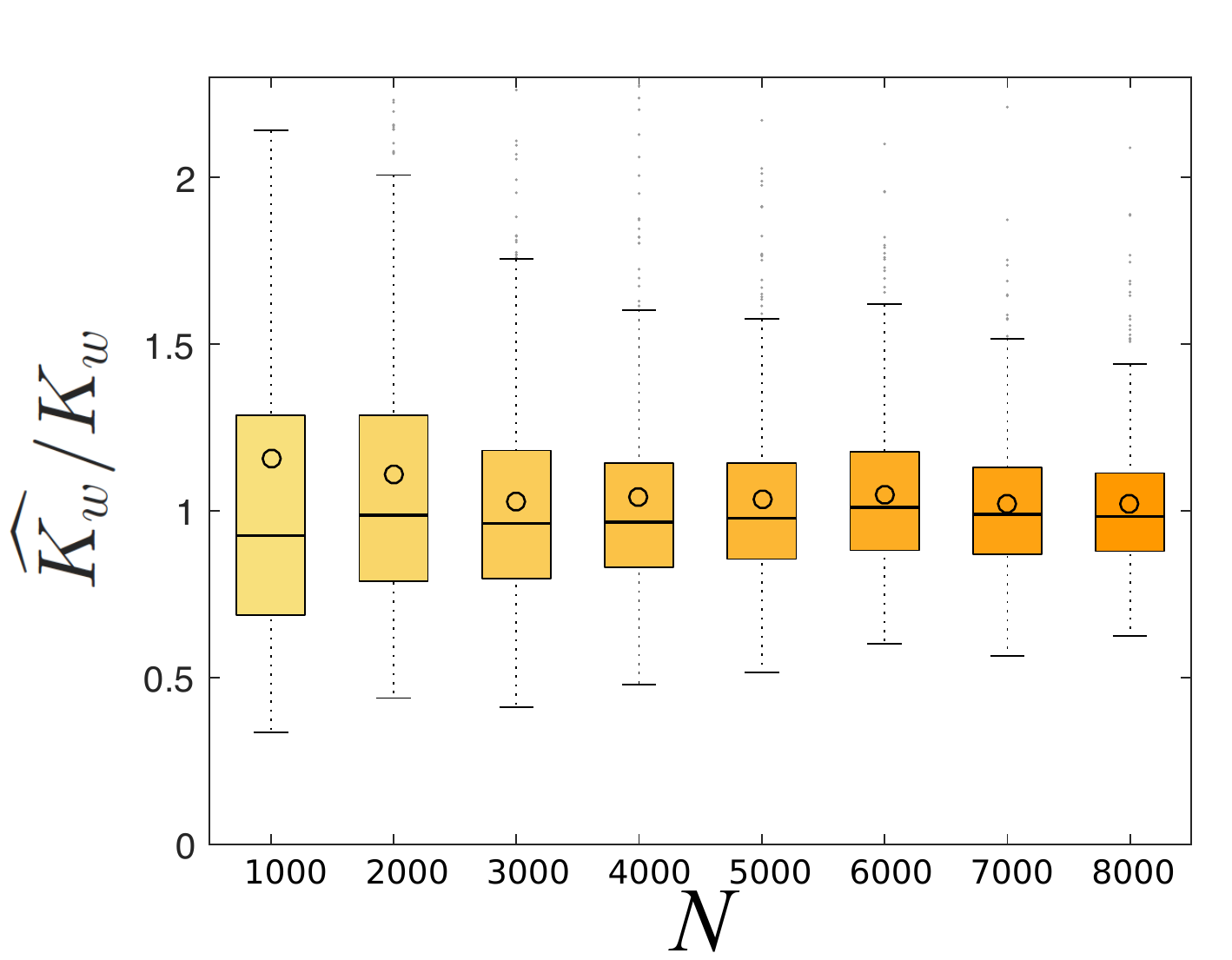}
                \caption{Average weak degree,  $q=0.1$, $B=10$}
                \label{kw_N_b10}
        \end{subfigure}%
        ~ 
        \begin{subfigure}[b]{ 0.32\textwidth}
                \includegraphics[width=\textwidth]{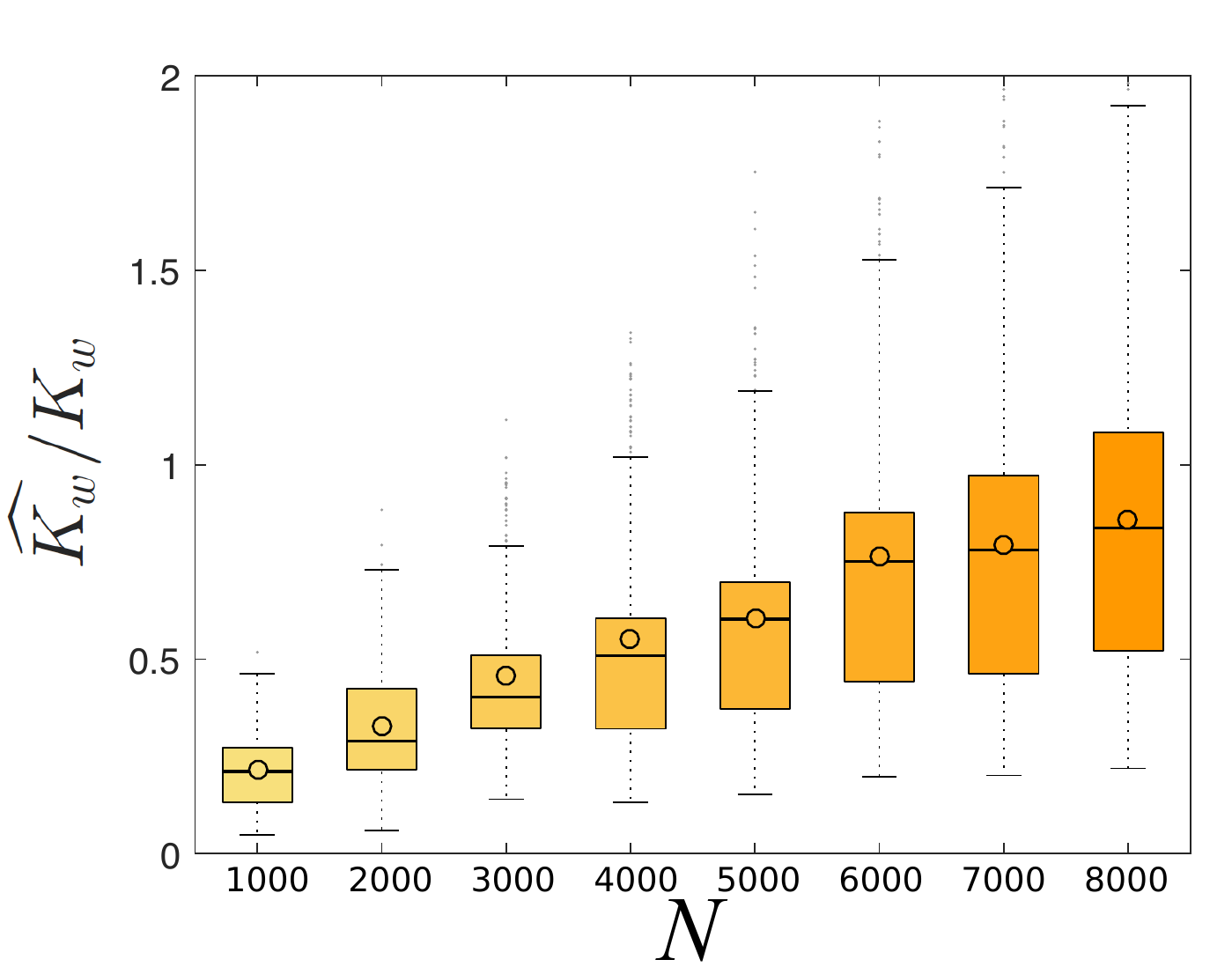}
                \caption{Average weak degree, $q=0.1$, $B=2$}
                \label{kw_N_b2}
        \end{subfigure}
        ~
        \begin{subfigure}[b]{0.32\textwidth}
                \includegraphics[width=\textwidth]{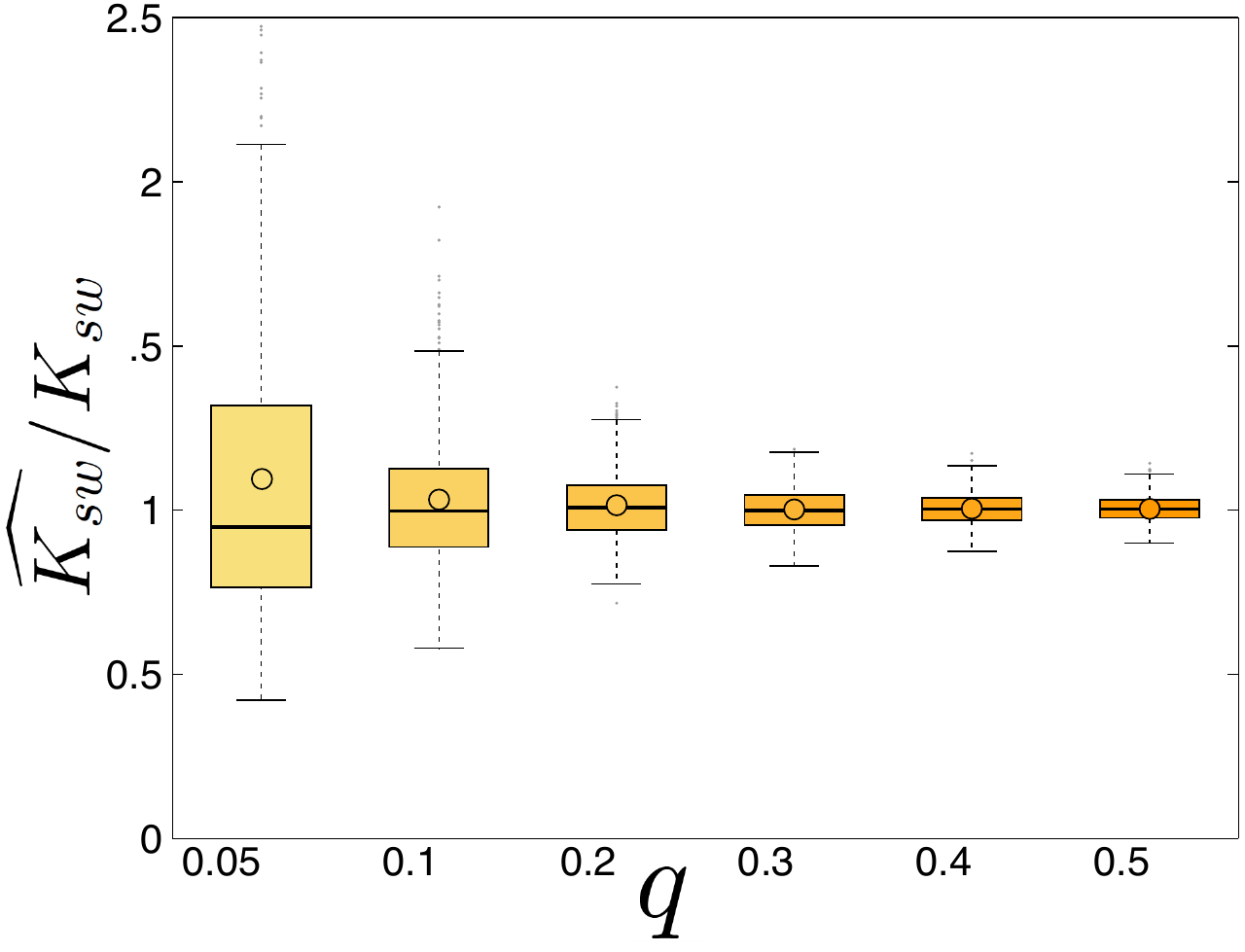}
                \caption{degrees product moment,  $N=4000$, $B=10$}
                \label{ksw_q_b10}
        \end{subfigure}%
        \\
        
        \vspace{1em}
        
        \begin{subfigure}[b]{0.32\textwidth}
                \includegraphics[width=\textwidth]{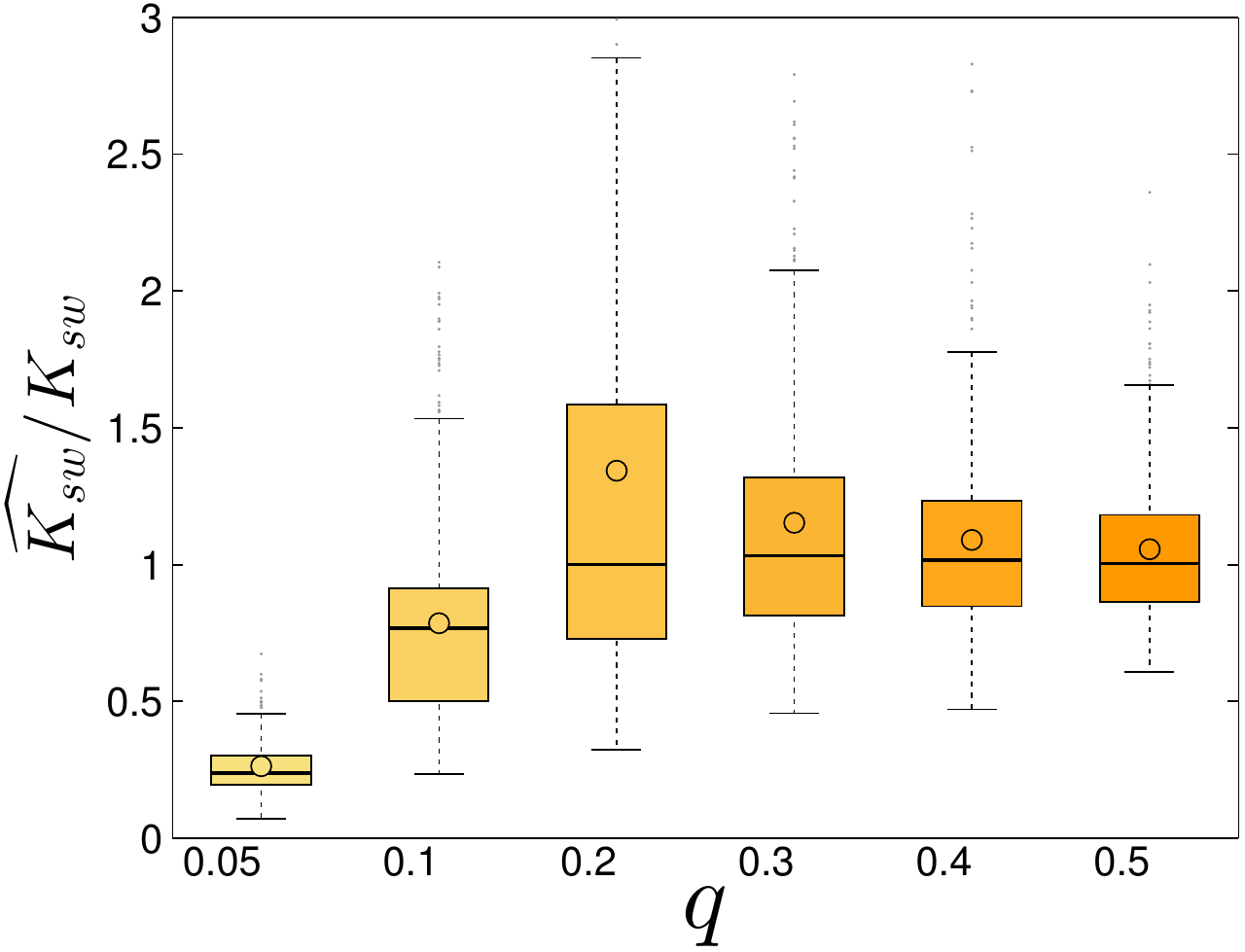}
                \caption{degrees product moment,  $N=4000$, $B=2$}
                \label{ksw_q_b2}
        \end{subfigure}
        ~
        \begin{subfigure}[b]{0.32\textwidth}
                \includegraphics[width=\textwidth]{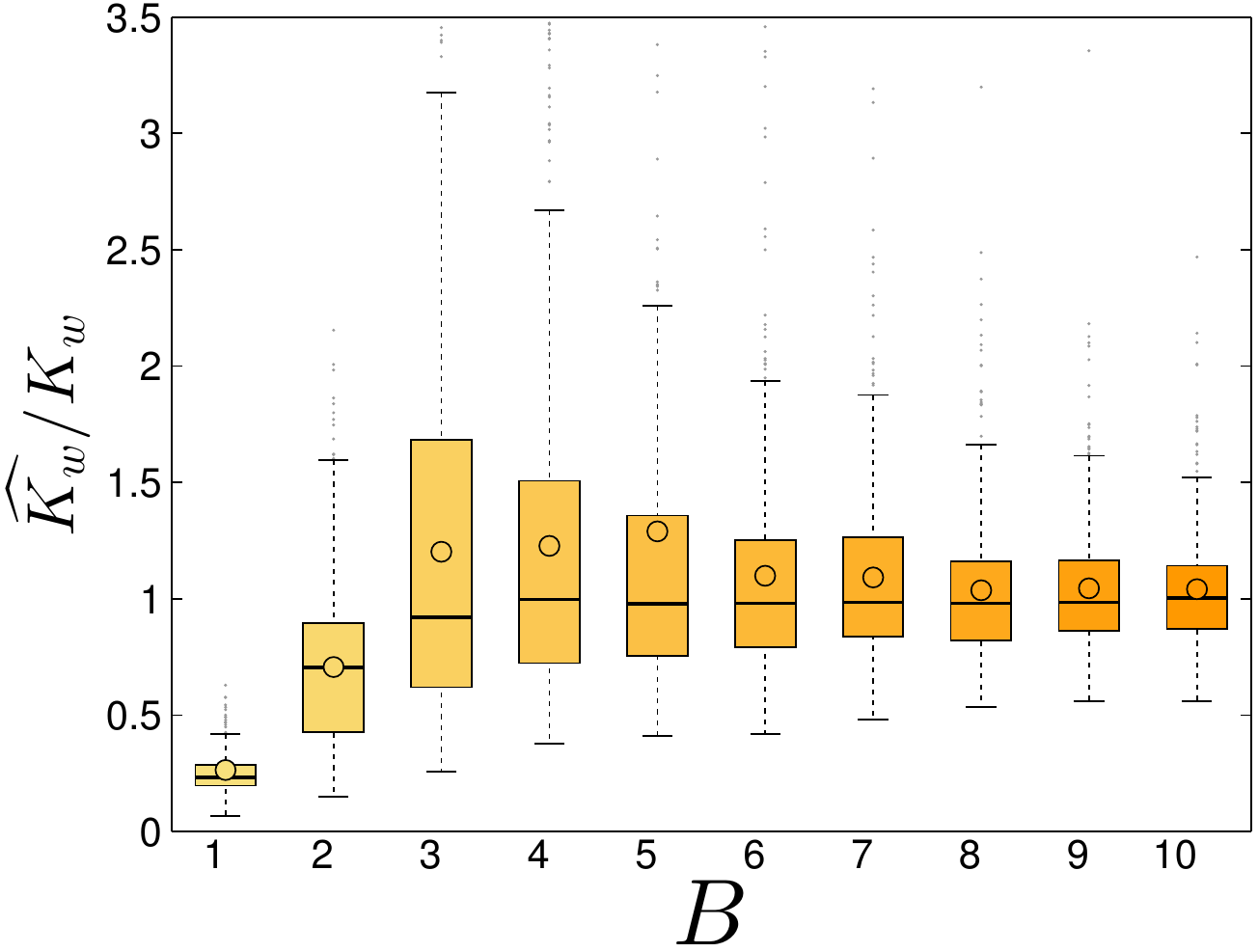}
                \caption{Average weak degree,   $N=4000$, $q=0.1$}
                \label{kw_B}
        \end{subfigure}%
        ~ 
        \begin{subfigure}[b]{0.32\textwidth}
                \includegraphics[width=\textwidth]{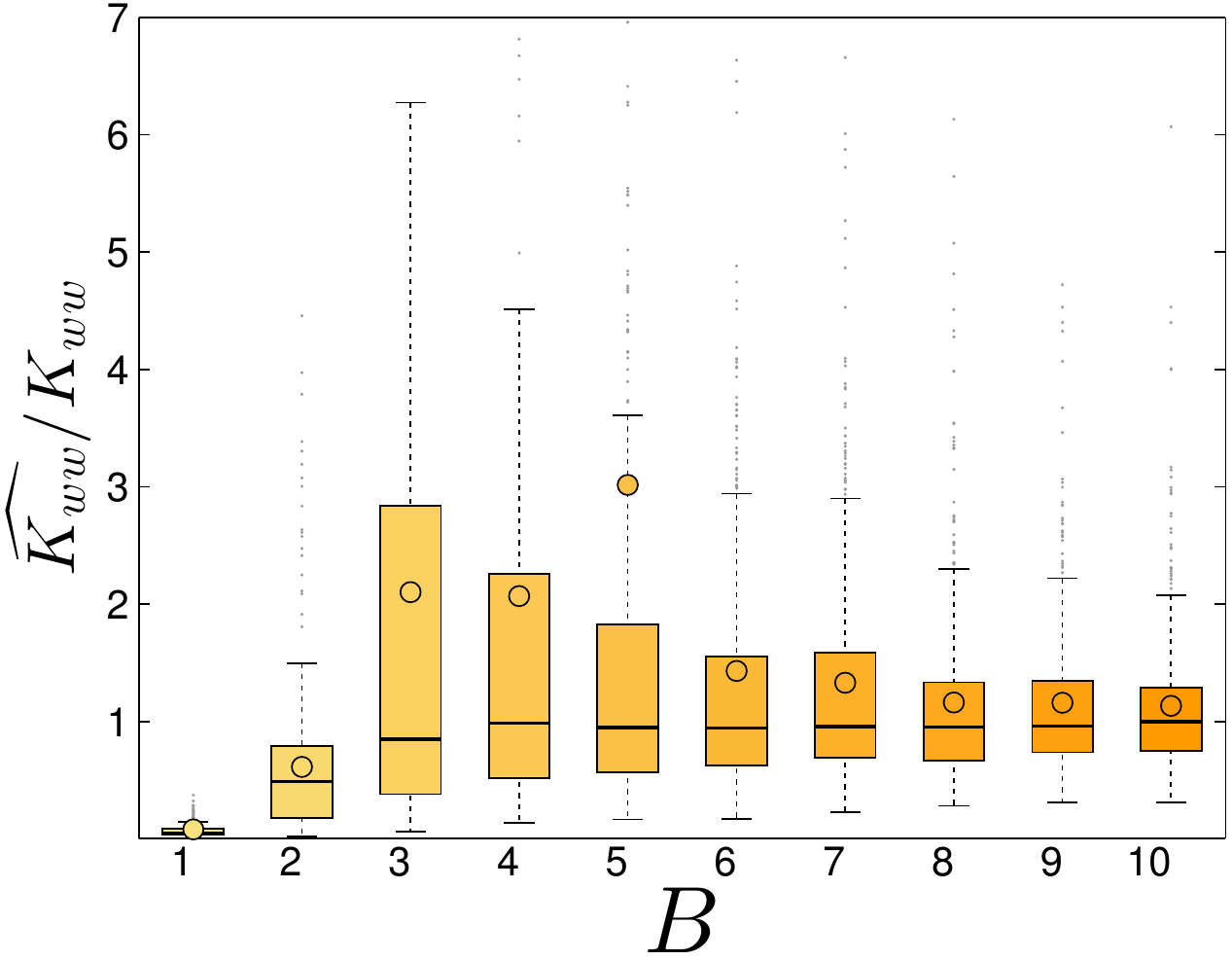}
                \caption{Second moment, weak degrees, $N=4000$,  $q=0.1$}
                \label{kw2_B}
        \end{subfigure}
        \caption{\footnotesize{Distribution of ratio of estimated values and true values.} }\label{fig:results2}
\end{figure*}

Figure~\ref{fig:results2} presents the results for $K_w$,  $K_{sw}$, and $K_{ww}$. For these estimators, the results depend on the value of $B$.  The results for $K_{sw}$ and $K_{ww}$ resemble those of $K_w$, so we omitted multiple figures. For $B=10$, the estimates are not biased for different sizes of network and samples and their variability consistently decreases as the size increases (Figures~\ref{kw_N_b10} and \ref{ksw_q_b10}). However, the behaviour of the estimators is different when $B=2$. In Figure~\ref{kw_N_b2}, the estimator has a bias for smaller values of $N$ and the bias decreases as $N$ increases. In Figure~\ref{ksw_q_b2}, the estimator shows a significant bias for $q=0.05$. As $q$ increases, first the bias and then the variability of the estimator are improved. Figures~\ref{kw_B} and \ref{kw2_B} illustrate the dependence of the performance of the estimators on $B$. It can be seen that the bias is negligible for values of $B$ as small as 4. If we  further increase $B$, the variability of estimates decreases.

To test the robustness of the results on moderate levels of skew that might be observed in offline social networks, we also tested the results on the high-clustering scale-free model of Holme and Kim~\cite{HK}, and the observed results are  reasonably accurate.  Since the results are similar to those presented, we omit them for space limitations.

\subsection{Comparing with SFC Estimators}

\begin{figure*}
        \centering
        \begin{subfigure}[b]{ 0.5\textwidth}
                \includegraphics[width=\textwidth]{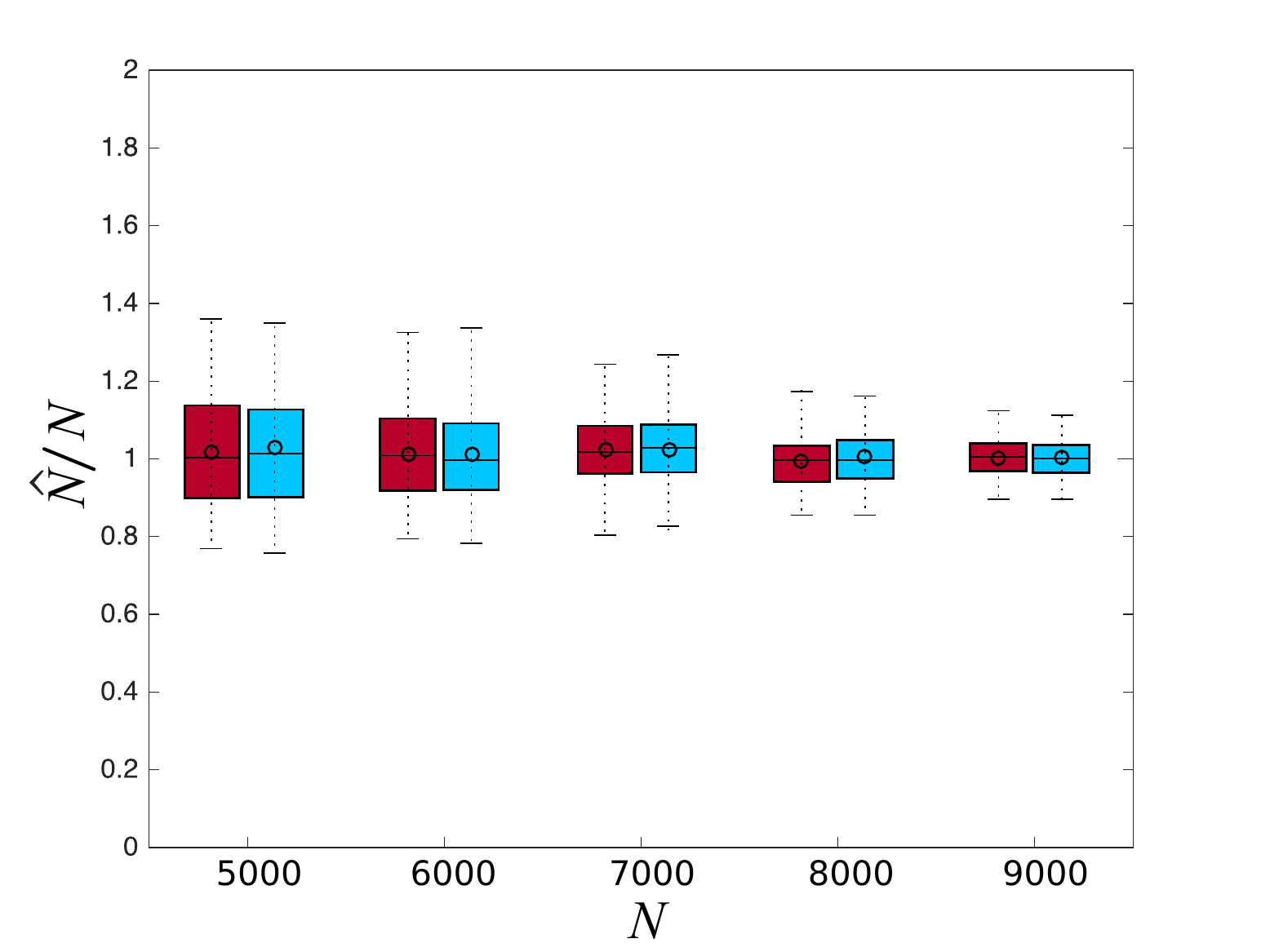}
                \caption{Number of Nodes, $q=0.1$, $B=10$}
                \label{fig:N_compare}
        \end{subfigure}%
        \begin{subfigure}[b]{ 0.5\textwidth}
                \includegraphics[width=\textwidth]{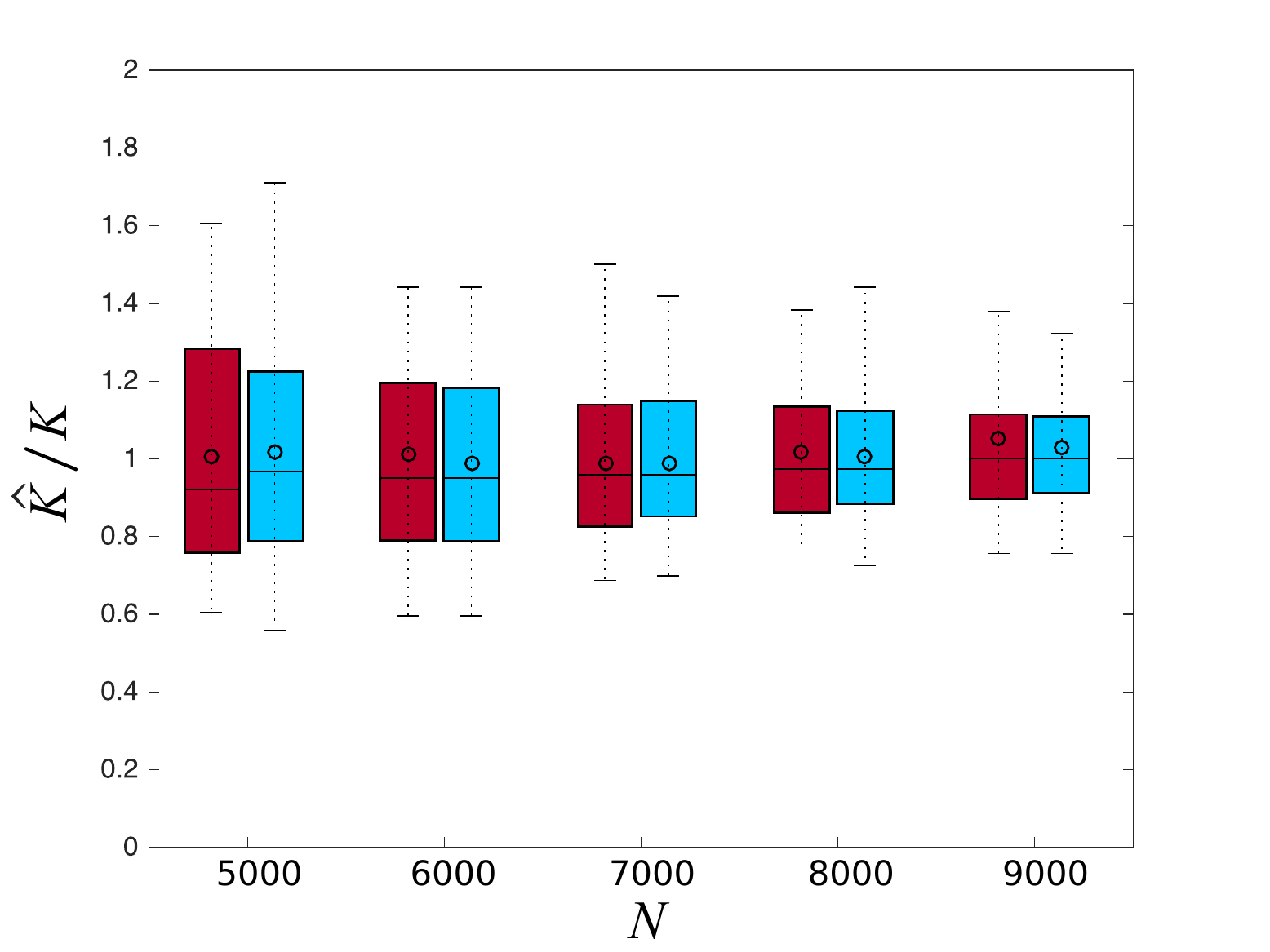}
                \caption{Average Degree, $q=0.1$, $B=10$}
                \label{fig:k_compare}
        \end{subfigure}%
        \caption{Comparing the performance of our estimators to SFC method~\cite{babaku}. Red boxes (left, within each group) show the distribution of SFC estimates and blue boxes (right, within each group) show the distribution of our estimates. (Best viewed in color.)}
\end{figure*}

Next, we compare the performance of the method proposed in this paper to the one proposed in~\cite{babaku}, which we refer to as the \emph{single fixed choice} (SFC) method, since it draws inferences about network structure based on the responses to a single fixed-choice survey question without differentiating between weak and strong ties. In order to facilitate this comparison, we use the same synthetic networks as described above. We first apply the proposed sampling and inference method. Then we collapse the two-layer network into a single network and apply the SFC sampling and inference scheme. We compare the performance of the two methods in terms of their estimate of the average degree, we take the estimates for $K_s$ and $K_w$ produced by the proposed method and compare the $K_s+K_w$ with the estimate of $K$ produced by SFC. Figures \ref{fig:N_compare} and \ref{fig:k_compare} show the performance of estimators of number of nodes and the average degrees. Although the estimates of the network size are comparable, the estimates of average degree are slightly better for the proposed approach, with it having a smaller inter-quartile range. This is not surprising, since the proposed method produces a better estimate of the average strong degree, thereby providing a more reliable estimate of the total degree.

\subsection{Comparing Results with the Crude Version} \label{sec:crude} 
  
  As discussed in Section~\ref{sec:intro}, many social network studies  of contagion use the sampled network without any inference~\cite{wellman1979community,fischer1982dwell,
behrman2002social,helleringer2007sexual,christakis2007spread,
christakis2013social}. 
 To  compare our estimators with the crude values of the sampled network, we need to choose a network statistic.  
 The effect of degree truncation in moments of the degree distribution is trivial, and it is clear that the crude values will be heavily biased, in comparison to the values produced by our estimators which appear to exhibit good performance in the experiments reported above. 
  
\begin{figure*}
        \centering
        \begin{subfigure}[b]{0.23\textwidth}
                \includegraphics[width=\textwidth]{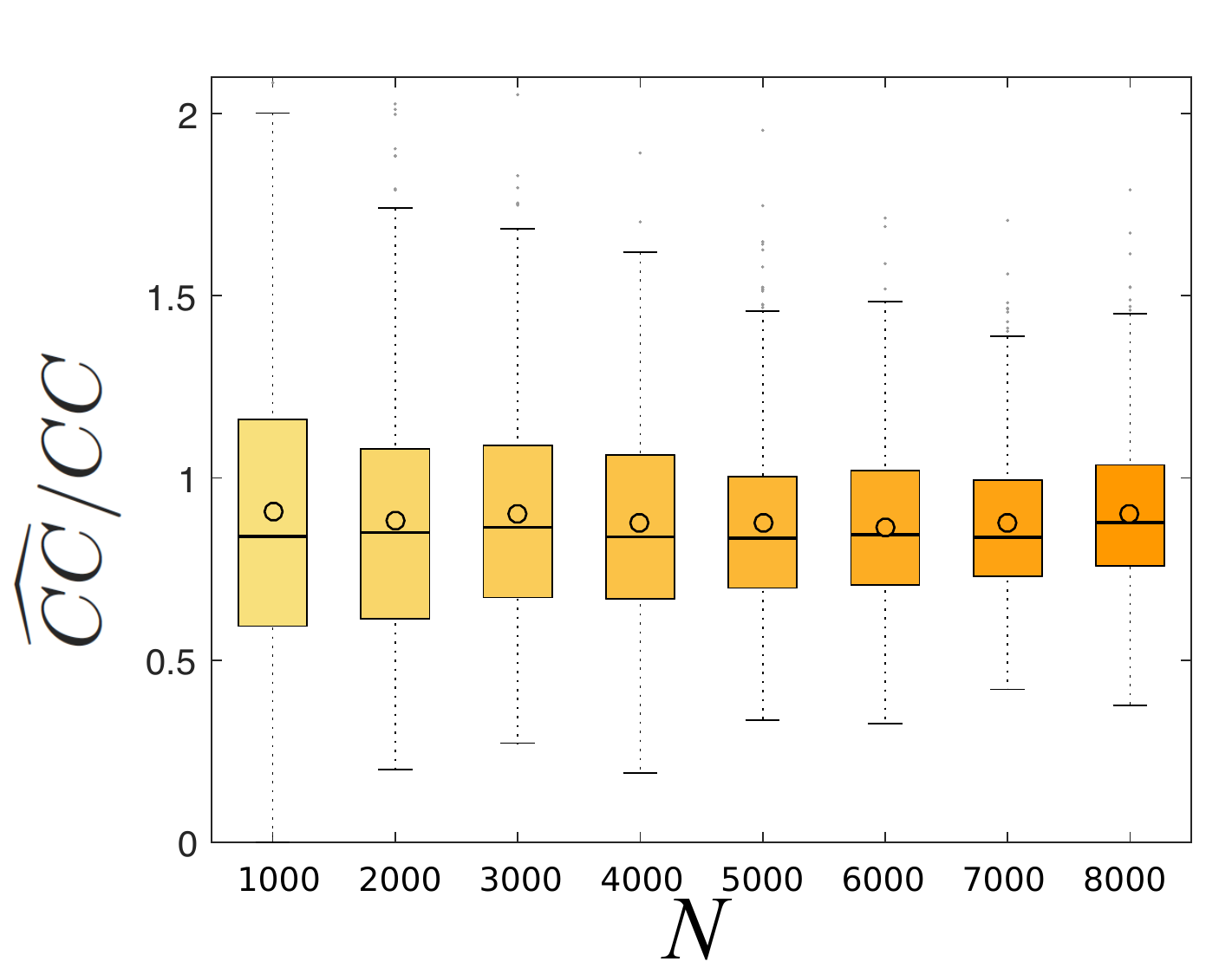}
                \caption{Proposed estimate of clustering coefficient, $q=0.1$}
                \label{cc_us_N}
        \end{subfigure}%
        ~
        \begin{subfigure}[b]{ 0.23\textwidth}
                \includegraphics[width=\textwidth]{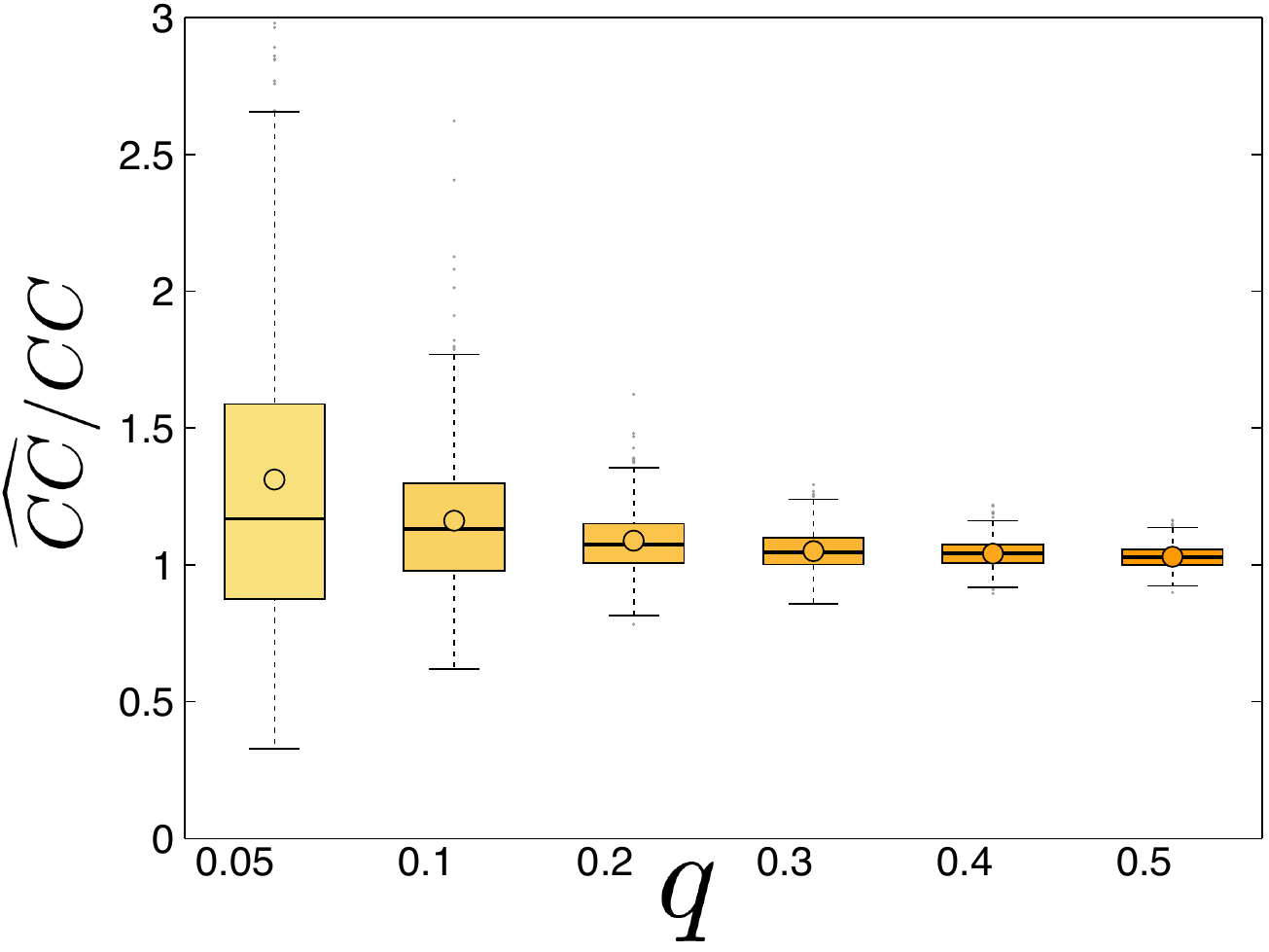}
                \caption{Proposed estimate of clustering coefficient, $N=4000$}
                \label{cc_us_q}
        \end{subfigure}
        ~
        \begin{subfigure}[b]{0.23\textwidth}
                \includegraphics[width=\textwidth]{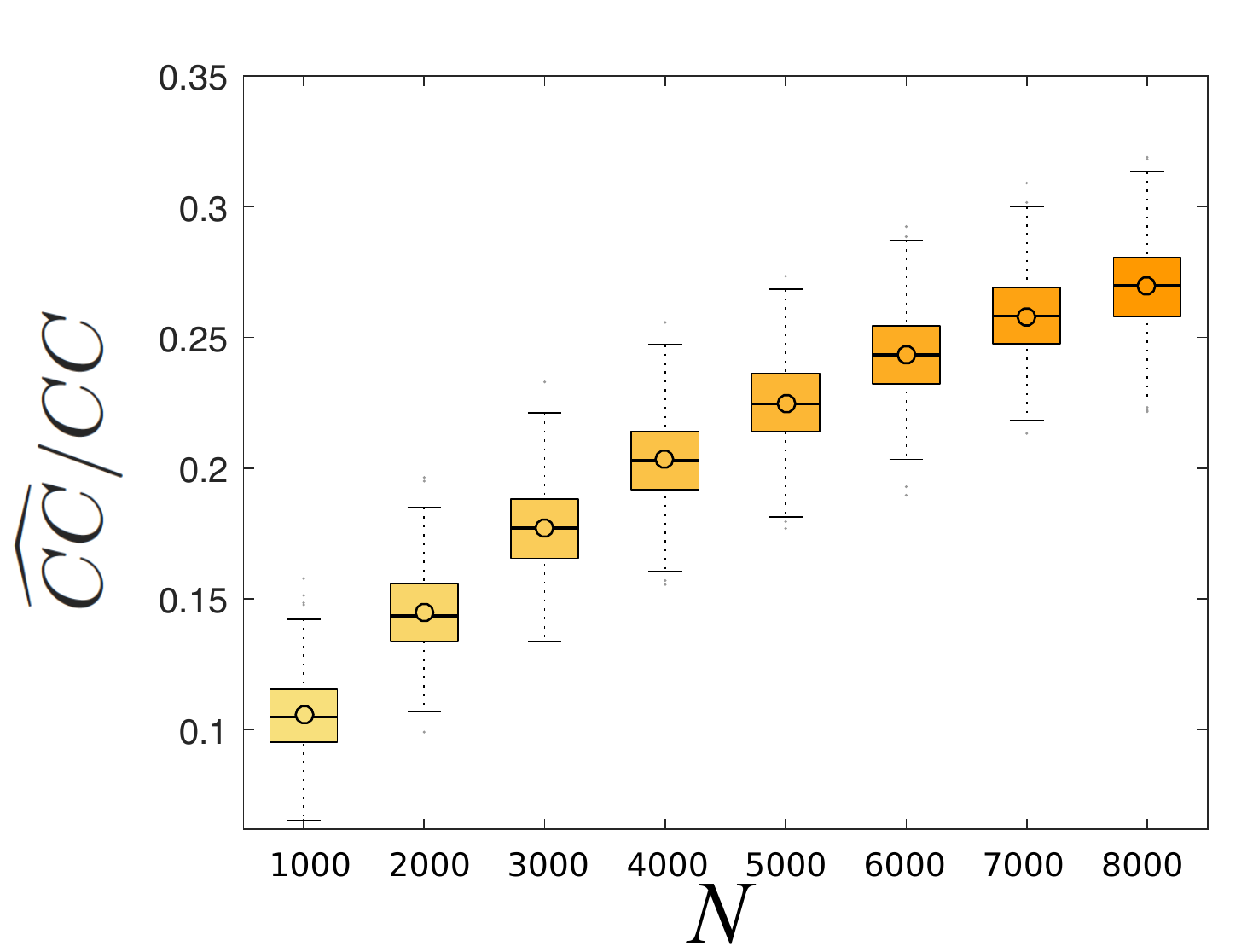}
                \caption{Crude estimate of clustering coefficient, $q=0.1$}
                \label{cc_cr_N}
        \end{subfigure}%
        ~
        \begin{subfigure}[b]{0.23\textwidth}
                \includegraphics[width=\textwidth]{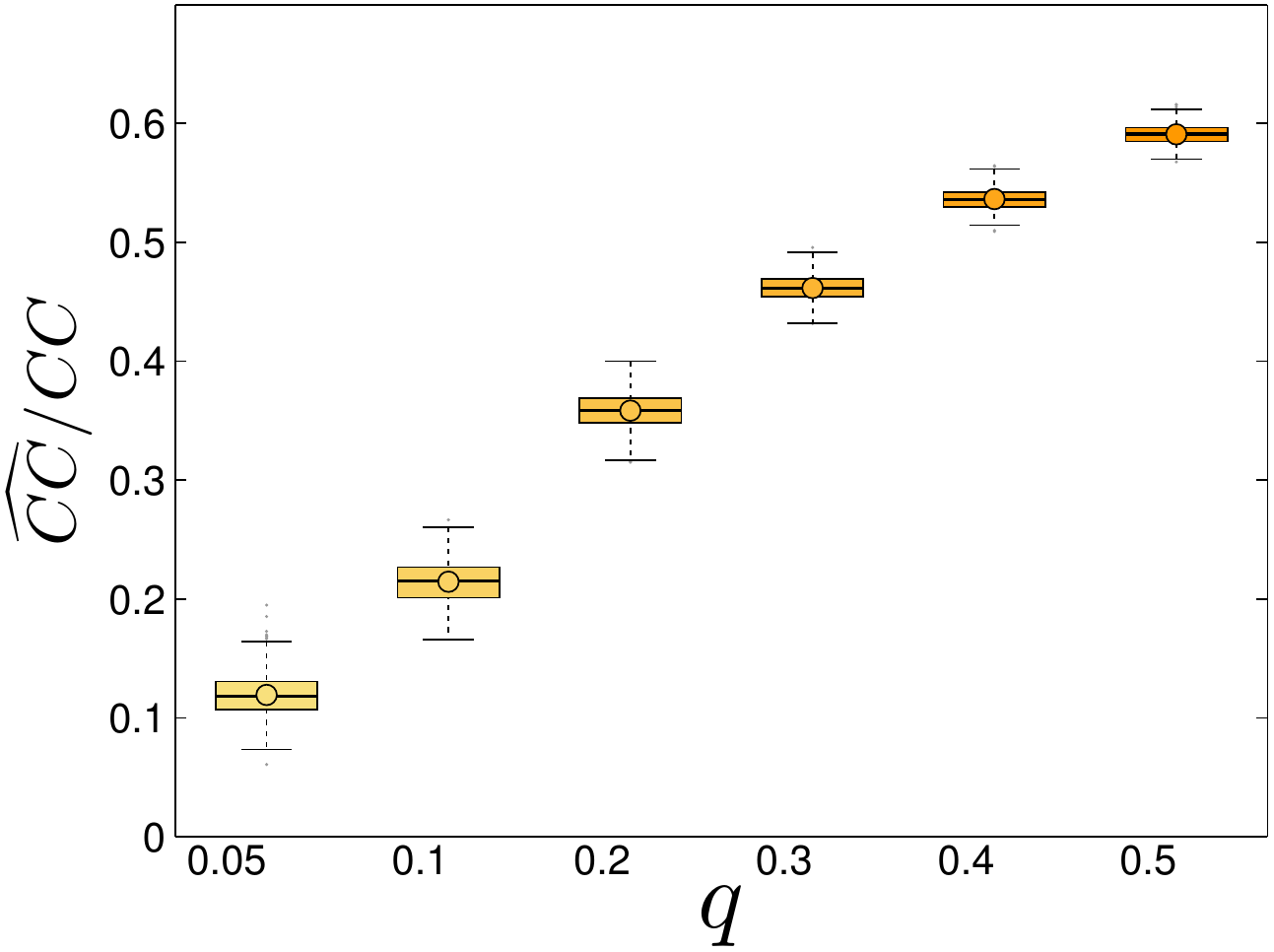}
                \caption{Crude estimate of clustering coefficient, $N=4000$}
                \label{cc_cr_q}
        \end{subfigure}
        \caption{Comparing the proposed and crude estimates of clustering coefficient for $B=10$.}\label{fig:results3}
\end{figure*}

  Instead, we consider estimating the clustering coefficient, a dimensionless quantity which is one of the most important network statistics in social network studies~\cite{wasserman1994social}. 
  It  is also desirable because it embodies all the other estimators and approximations.  
  Since we have different types of triads and triangles, we first collapse the network into one layer (with homogeneous links) and then  calculate its clustering coefficient. Figures~\ref{cc_us_N} and \ref{cc_us_q} illustrate the performance of our method in estimating the clustering coefficient.  Figures~\ref{cc_cr_N} and \ref{cc_cr_q} depict the clustering coefficient calculated directly from the crude sampled network.  It is evident that our approach outperforms the crude estimate by a large margin, and using the crude estimates results in underestimating the clustering coefficient of the network.

\subsection{Performance of Estimators on Real-world Datasets}
\begin{figure}
\centering
\includegraphics[width=0.6\textwidth]{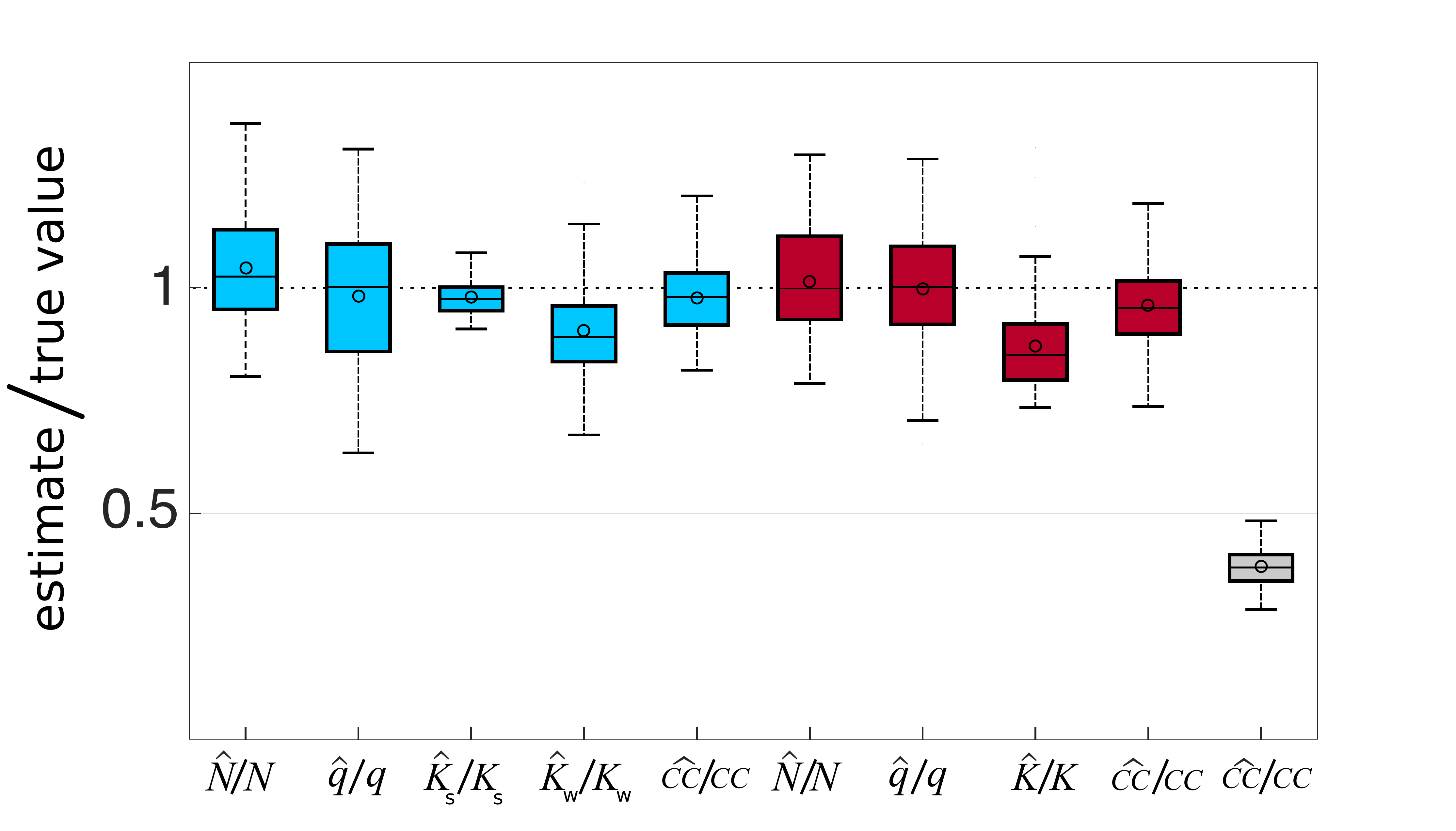}
\caption{Performance on the Villages dataset. Blue boxes correspond to the proposed estimator, red boxes correspond to the SFC estimator, and the grey box corresponds to the crude estimate of the clustering coefficient. (Best viewed in color.)}\label{fig:real}
\end{figure}

  The Villages dataset~\cite{banerjee2013} consists of surveys made in 77 villages in India. The questionnaire includes several questions used to build the social networks. To apply our method to the networks in this data set, we form the strong layer of each network based on responses to a question about relationships involving the borrowing of money, and we build the weak layer by connecting two nodes with a weak tie if they are not relatives and accompany each other when going to temple. We remove the nodes whose weak degree was less than 3 and then applied the sampling method with $B=3$. The distribution of the estimates for $N$, $q$, $K_s$, $K_w$ and the clustering coefficient (for the collapsed network) using the proposed method are presented in Fig~\ref{fig:real} in red. To compare our estimates to the SFC model, we collapse the two layers into one and estimate $N$, $q$, $K$, and clustering coefficient using SFC. The distribution of these estimates are shown in Fig~\ref{fig:real} in blue. Also, we have included the crude estimates of the clustering coefficient (calculated as explained in Section~\ref{sec:crude}) in gray. As with the simulated dataset, the clustering coefficient estimate obtained using the proposed method is significantly better than that obtained using the crude network. Moreover, the accuracy of the proposed approach in estimating a variety of network parameters provides some validation that the modeling assumptions on which this approach is based are reasonable. 

\section{Estimating the Variance of the Estimators}
\begin{figure}
        \centering
        \begin{subfigure}[b]{0.49\textwidth}
                \includegraphics[width=\textwidth]{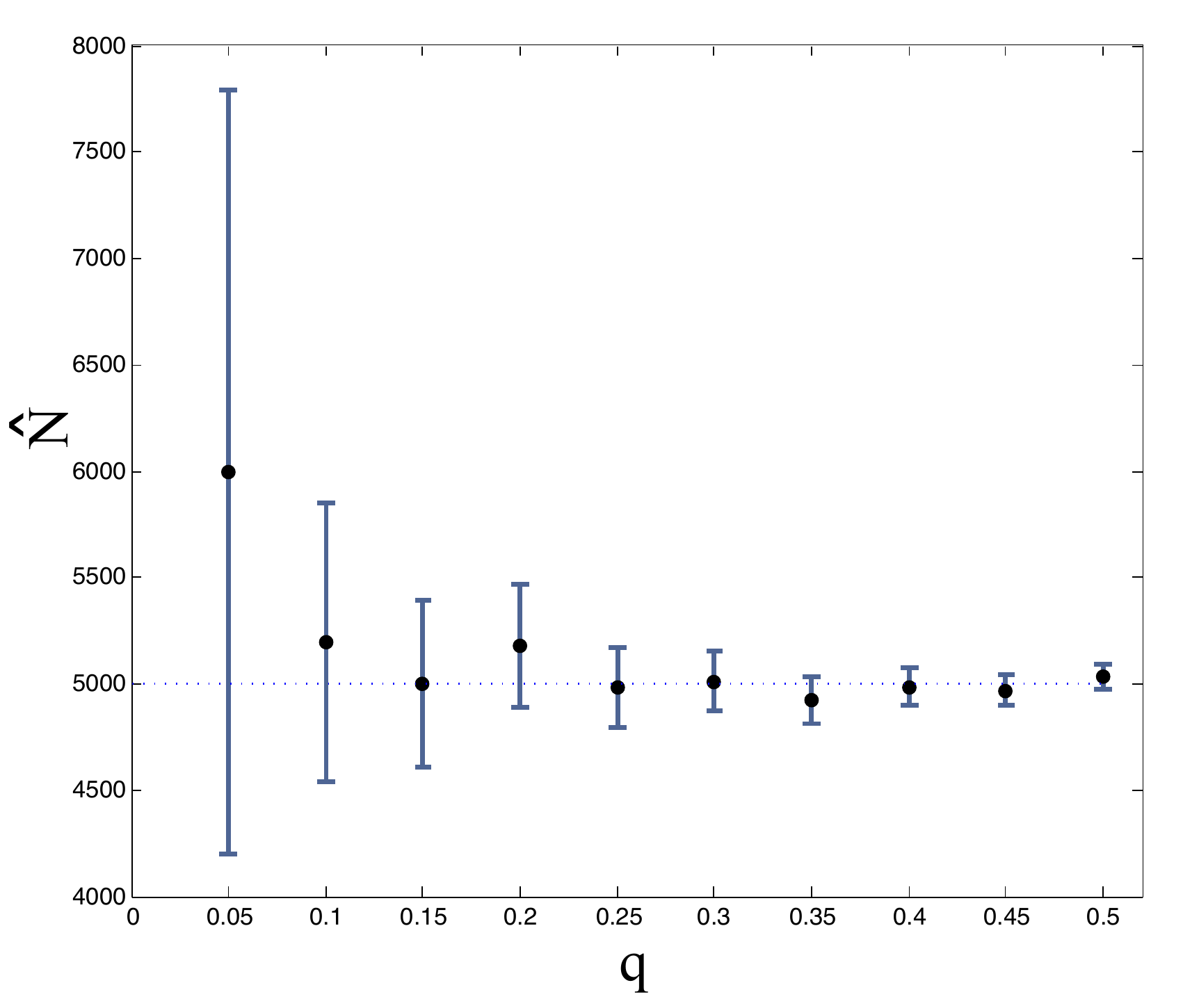}
                \caption{Estimator for $N$}
                \label{jack_N}
        \end{subfigure}%
        \begin{subfigure}[b]{0.49\textwidth}
                \includegraphics[width=\textwidth]{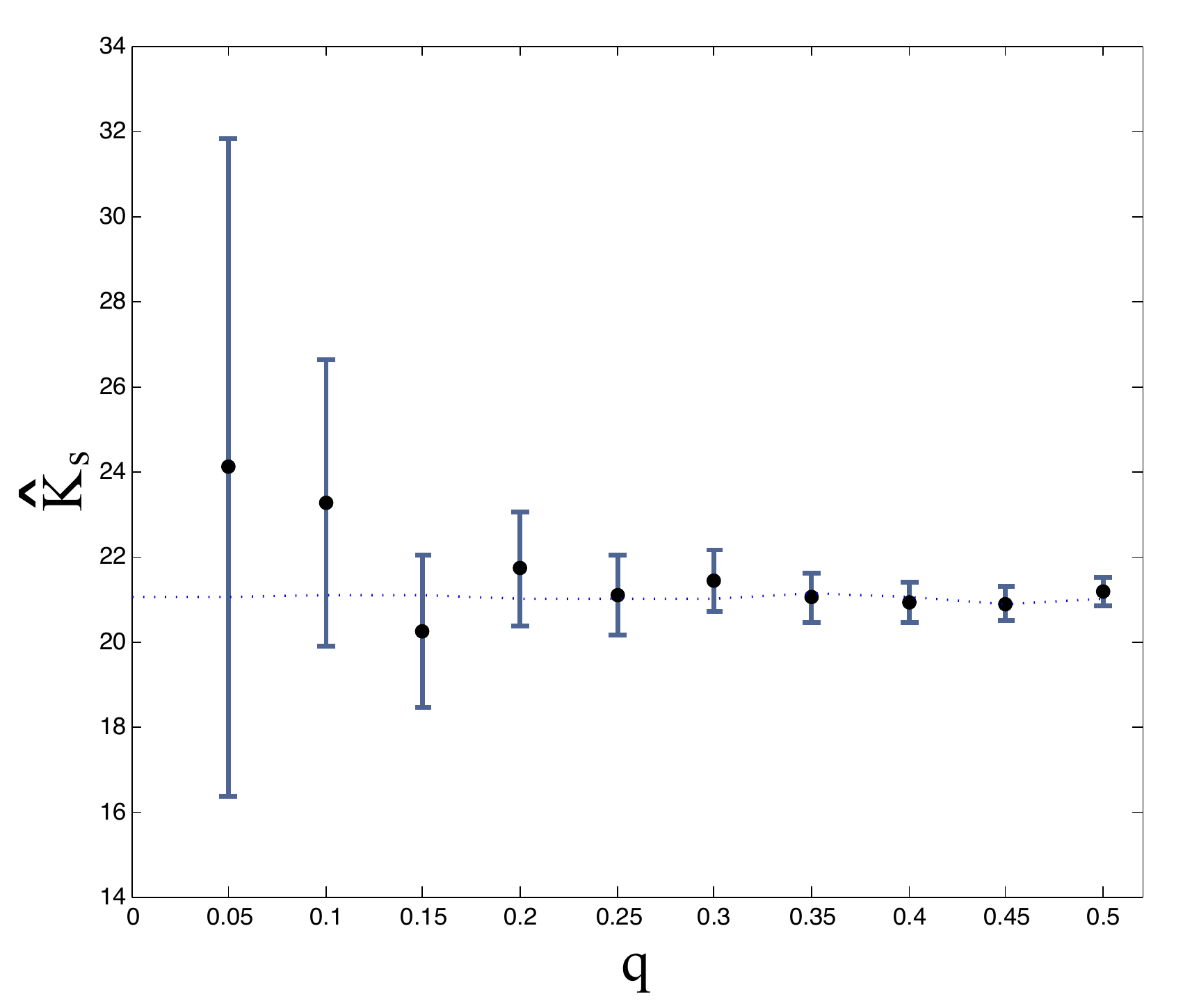}
                \caption{Estimator for $K_s$}
                \label{jack_ks}
        \end{subfigure}
        \caption{Standard deviation of two estimators. The circles represent the jackknife estimates of the mean and the error bars represent the jackknife estimate of the standard deviation. The dashed line represent the true values. }\label{jack}
\end{figure}

To estimate the variance of the estimators, we propose a variation of the Jackknife resampling method~\cite{quenouille1956notes,tukey1958bias}. In each resampling, we leave out one of the respondents   and remove all the links of that respondent in the sampled network. Then, we apply our method to estimate the desired variables in the resampled network. Variances are estimated from the distribution that is obtained by repeating this procedure for all respondents. Note that this estimated variance is different from the variance of all estimates from subsamples.  The estimated variance of an estimator for parameter $h$ in this method is equal to
\begin{equation}
Var(h)=\frac{n_0-1}{n_0}\sum_{n=1}^{n_0}(\widetilde{h}_i-\bar{h})^2,
\end{equation}
where $\widetilde{h}_i$ is the estimated value of $h$ when node $i$ is removed from the seeds and $\bar{h}$ is the average of all values of $\widetilde{h}_i$. Figure~\ref{jack} presents the results of Jackknife resampling for two of the estimators for different values of sampling probability ($N=5000$ and $B=10$ are fixed).  It can be seen that as the sample size increases, the estimated standard deviations of both estimators decrease. Moreover, for all values of $q$ we see that the true value (dashed line) falls within one standard deviation of the jackknife-estimated mean.

\section{Conclusion and Future Work}
This paper described a method for estimating characteristics of a social network topology (the network size, average number of strong and weak ties, as well as second moments of the strong and week degree distributions) from fixed choice survey data. In particular, we assumed that every respondent provides all of their strong ties and a fixed number of their weak ties. The proposed estimation methodology is based on the method of moments, under a model where respondents are sampled according to a Bernoulli process over vertices (with unknown sampling rate) and the subset of reported weak ties is sampled uniformly from all of the respondent's weak ties.

A natural extension of the work proposed is to consider surveys with a soft fixed choice design; instead of reporting exactly $B$ weak ties, each respondent may report up to $B$ weak ties. One approach to this may be to assume that each respondent $x$ samples a number $B_x \le B$ of weak ties to report with $B_x$ being independently and identically distributed with an unknown mass function over the integers from $0$ to $B$. In this case, in the context of the model developed in this paper, it turns out that it is sufficient to estimate the mean $\mathbb{E}[B]$. We are currently exploring such an estimator, as well as theoretical guarantees for the proposed inference procedure. Another possibility is to make parametric assumptions about the recollection process and to modify the assumption of seeds choosing weak ties uniformly at random.

In social health-related applications, it is commonly of interest to identify a subset of the population to be immunized, with the intention of most efficiently preventing the spread of infectious diseases, subject to a constraint on the number of individuals that can be immunized. For this reason, it would also be of interest to extend the results of this paper to estimate quantities such as the betweenness centrality (or another centrality) measure of each node, since these typically correlate highly with individuals that are well-placed (i.e., hubs) in the network.

\appendix
\section*{Appendix}
Here we present additional figures and tables of expressions used in the estimator calculations described in Section~\ref{inf}. Table~\ref{B_A} summarizes the expressions for the coefficients $a_{ij}$ and $b_{ij}$, $i,j \in \{0,1\}$, used for calculating estimates of the second moments.

Figure~\ref{all_triangles} shows all of the 42 possible ways that a triangle in $\G$ can be observed in $\GG$. Table~\ref{rho} shows the corresponding expressions $\rho_j$ for each $j=1,\dots,26$ corresponding to each example shown in Figure~\ref{all_triangles}.

Figure~\ref{all_triads} shows the 31 possible ways that an open triad in $\G$ can be observed in $\GG$, and Table~\ref{pi} provides expressions for the probability $\pi_i$, $i=1,\dots,31$ of observing each one. Table~\ref{phi} provides expressions for the probability $\phi_i$, $i=1,\dots,31$, of observing each open triad as as an open triad in $\GG$.

\begin{figure*}
\captionsetup[subfigure]{labelformat=empty}
    \begin{subfigure}{0.12\textwidth}
    \begin{tikzpicture}   
    \draw[fill=blue!5,line width=0.35mm] (0.85,1.3) circle (6pt)node[](1){} ;
    \draw[fill=blue!5,line width=0.35mm] (0,0) circle (6pt)node[](2){} ;
    \draw[fill=blue!5,line width=0.35mm] (1.7,0) circle (6pt)node[](3){} ;

    \foreach \x/\y in {}
        \path[arrows={[scale=0.8]}] (\x) edge [strong dir] (\y); 
    \foreach \x/\y in {1/2,2/1,1/3,3/1,2/3,3/2}
    	\path[arrows={[scale=0.8]}] (\x) edge [strong bidir] (\y); 
	\foreach \x/\y in {}
   		\path[arrows={[scale=0.8]}] (\x) edge [weak dir] (\y); 
    \foreach \x/\y in {}
    	\path[arrows={[scale=0.8]}] (\x) edge [weak bidir] (\y); 
    \end{tikzpicture}
    \caption{1}
  \end{subfigure}~~
    \begin{subfigure}{0.12\textwidth}
	 \begin{tikzpicture}   
    \node[draw] at (1.7,1.5) {$\times 3$};    
    \draw[fill=blue!5,line width=0.mm] (0.85,1.3) circle (6pt)node[](1){} ;
    \draw[fill=blue!5,line width=0.35mm] (0,0) circle (6pt)node[](2){} ;
    \draw[fill=black!100,line width=0.35mm] (1.7,0) circle (6pt)node[](3){} ;

    \foreach \x/\y in {2/3,1/3}
        \path[arrows={[scale=0.8]}] (\x) edge [strong dir] (\y); 
    \foreach \x/\y in {1/2,2/1}
    	\path[arrows={[scale=0.8]}] (\x) edge [strong bidir] (\y); 
	\foreach \x/\y in {}
   		\path[arrows={[scale=0.8]}] (\x) edge [weak dir] (\y); 
    \foreach \x/\y in {}
    	\path[arrows={[scale=0.8]}] (\x) edge [weak bidir] (\y); 

    \end{tikzpicture}
    \caption{2}
  \end{subfigure}~~
  \begin{subfigure}{0.12\textwidth}
   \begin{tikzpicture}   
    \node[draw] at (1.7,1.5) {$\times 2$};
    \draw[fill=blue!5,line width=0.35mm] (0.85,1.3) circle (6pt)node[](1){} ;
    \draw[fill=blue!5,line width=0.35mm] (0,0) circle (6pt)node[](2){} ;
    \draw[fill=blue!5,line width=0.35mm] (1.7,0) circle (6pt)node[](3){} ;

    \foreach \x/\y in {}
        \path[arrows={[scale=0.8]}] (\x) edge [strong dir] (\y); 
    \foreach \x/\y in {1/3,3/1,1/2,2/1}
    	\path[arrows={[scale=0.8]}] (\x) edge [strong bidir] (\y); 
	\foreach \x/\y in {2/3}
   		\path[arrows={[scale=0.8]}] (\x) edge [weak dir] (\y); 
    \foreach \x/\y in {}
    	\path[arrows={[scale=0.8]}] (\x) edge [weak bidir] (\y); 

    \end{tikzpicture}
    \caption{3}
  \end{subfigure}~~
  \begin{subfigure}{0.12\textwidth}
   \begin{tikzpicture}   
    
    \draw[fill=blue!5,line width=0.35mm] (0.85,1.3) circle (6pt)node[](1){} ;
    \draw[fill=blue!5,line width=0.35mm] (0,0) circle (6pt)node[](2){} ;
    \draw[fill=blue!5,line width=0.35mm] (1.7,0) circle (6pt)node[](3){} ;

    \foreach \x/\y in {}
        \path[arrows={[scale=0.8]}] (\x) edge [strong dir] (\y); 
    \foreach \x/\y in {1/3,3/1,1/2,2/1}
    	\path[arrows={[scale=0.8]}] (\x) edge [strong bidir] (\y); 
	\foreach \x/\y in {}
   		\path[arrows={[scale=0.8]}] (\x) edge [weak dir] (\y); 
    \foreach \x/\y in {2/3,3/2}
    	\path[arrows={[scale=0.8]}] (\x) edge [weak bidir] (\y); 

    \end{tikzpicture}
    \caption{4}
  \end{subfigure}~~
    \begin{subfigure}{0.12\textwidth}
   \begin{tikzpicture}   
    \node[draw] at (1.7,1.5) {$\times 2$};
    \draw[fill=blue!5,line width=0.35mm] (0.85,1.3) circle (6pt)node[](1){} ;
    \draw[fill=blue!5,line width=0.35mm] (0,0) circle (6pt)node[](2){} ;
    \draw[fill=black!100,line width=0.35mm] (1.7,0) circle (6pt)node[](3){} ;

    \foreach \x/\y in {1/3}
        \path[arrows={[scale=0.8]}] (\x) edge [strong dir] (\y); 
    \foreach \x/\y in {1/2,2/1}
    	\path[arrows={[scale=0.8]}] (\x) edge [strong bidir] (\y); 
	\foreach \x/\y in {2/3}
   		\path[arrows={[scale=0.8]}] (\x) edge [weak dir] (\y); 
    \foreach \x/\y in {}
    	\path[arrows={[scale=0.8]}] (\x) edge [weak bidir] (\y); 

    \end{tikzpicture}
    \caption{5}
  \end{subfigure}~~  
  \begin{subfigure}{0.12\textwidth}
   \begin{tikzpicture}   
    \node[draw] at (1.7,1.5) {$\times 2$};
    \draw[fill=black!100,line width=0.35mm] (0.85,1.3) circle (6pt)node[](1){} ;
    \draw[fill=blue!5,line width=0.35mm] (0,0) circle (6pt)node[](2){} ;
    \draw[fill=blue!5,line width=0.35mm] (1.7,0) circle (6pt)node[](3){} ;

    \foreach \x/\y in {2/1,3/1}
        \path[arrows={[scale=0.8]}] (\x) edge [strong dir] (\y); 
    \foreach \x/\y in {}
    	\path[arrows={[scale=0.8]}] (\x) edge [strong bidir] (\y); 
	\foreach \x/\y in {2/3}
   		\path[arrows={[scale=0.8]}] (\x) edge [weak dir] (\y); 
    \foreach \x/\y in {}
    	\path[arrows={[scale=0.8]}] (\x) edge [weak bidir] (\y); 

    \end{tikzpicture}
    \caption{6}
  \end{subfigure}~~
    \begin{subfigure}{0.12\textwidth}
   \begin{tikzpicture}   
    
    \draw[fill=black!100,line width=0.35mm] (0.85,1.3) circle (6pt)node[](1){} ;
    \draw[fill=blue!5,line width=0.35mm] (0,0) circle (6pt)node[](2){} ;
    \draw[fill=blue!5,line width=0.35mm] (1.7,0) circle (6pt)node[](3){} ;

    \foreach \x/\y in {2/1,3/1}
        \path[arrows={[scale=0.8]}] (\x) edge [strong dir] (\y); 
    \foreach \x/\y in {}
    	\path[arrows={[scale=0.8]}] (\x) edge [strong bidir] (\y); 
	\foreach \x/\y in {}
   		\path[arrows={[scale=0.8]}] (\x) edge [weak dir] (\y); 
    \foreach \x/\y in {2/3,3/2}
    	\path[arrows={[scale=0.8]}] (\x) edge [weak bidir] (\y); 

    \end{tikzpicture}
    \caption{7}
  \end{subfigure}~~ 
  
  \begin{subfigure}{0.12\textwidth}
    \begin{tikzpicture}   
    \node[draw] at (1.7,1.5) {$\times 2$};
    \draw[fill=blue!5,line width=0.35mm] (0.85,1.3) circle (6pt)node[](1){} ;
    \draw[fill=blue!5,line width=0.35mm] (0,0) circle (6pt)node[](2){} ;
    \draw[fill=blue!5,line width=0.35mm] (1.7,0) circle (6pt)node[](3){} ;

    \foreach \x/\y in {}
        \path[arrows={[scale=0.8]}] (\x) edge [strong dir] (\y); 
    \foreach \x/\y in {2/3,3/2}
    	\path[arrows={[scale=0.8]}] (\x) edge [strong bidir] (\y); 
	\foreach \x/\y in {1/2}
   		\path[arrows={[scale=0.8]}] (\x) edge [weak dir] (\y); 
    \foreach \x/\y in {1/3,3/1}
    	\path[arrows={[scale=0.8]}] (\x) edge [weak bidir] (\y); 

    \end{tikzpicture}
    \caption{8}
  \end{subfigure}~~
  \begin{subfigure}{0.12\textwidth}
    \begin{tikzpicture}   
    \node[draw] at (1.7,1.5) {$\times 2$};
    \draw[fill=blue!5,line width=0.35mm] (0.85,1.3) circle (6pt)node[](1){} ;
    \draw[fill=blue!5,line width=0.35mm] (0,0) circle (6pt)node[](2){} ;
    \draw[fill=blue!5,line width=0.35mm] (1.7,0) circle (6pt)node[](3){} ;

    \foreach \x/\y in {}
        \path[arrows={[scale=0.8]}] (\x) edge [strong dir] (\y); 
    \foreach \x/\y in {2/3,3/2}
    	\path[arrows={[scale=0.8]}] (\x) edge [strong bidir] (\y); 
	\foreach \x/\y in {2/1}
   		\path[arrows={[scale=0.8]}] (\x) edge [weak dir] (\y); 
    \foreach \x/\y in {1/3,3/1}
    	\path[arrows={[scale=0.8]}] (\x) edge [weak bidir] (\y); 

    \end{tikzpicture}
    \caption{9}
  \end{subfigure}~~  
  \begin{subfigure}{0.12\textwidth}
    \begin{tikzpicture}   
    \draw[fill=blue!5,line width=0.35mm] (0.85,1.3) circle (6pt)node[](1){} ;
    \draw[fill=blue!5,line width=0.35mm] (0,0) circle (6pt)node[](2){} ;
    \draw[fill=blue!5,line width=0.35mm] (1.7,0) circle (6pt)node[](3){} ;

    \foreach \x/\y in {}
        \path[arrows={[scale=0.8]}] (\x) edge [strong dir] (\y); 
    \foreach \x/\y in {2/3,3/2}
    	\path[arrows={[scale=0.8]}] (\x) edge [strong bidir] (\y); 
	\foreach \x/\y in {}
   		\path[arrows={[scale=0.8]}] (\x) edge [weak dir] (\y); 
    \foreach \x/\y in {1/2,2/1,1/3,3/1}
    	\path[arrows={[scale=0.8]}] (\x) edge [weak bidir] (\y); 

    \end{tikzpicture}
    \caption{10}
  \end{subfigure}~~
  \begin{subfigure}{0.12\textwidth}
    \begin{tikzpicture}   
    \draw[fill=blue!5,line width=0.35mm] (0.85,1.3) circle (6pt)node[](1){} ;
    \draw[fill=blue!5,line width=0.35mm] (0,0) circle (6pt)node[](2){} ;
    \draw[fill=blue!5,line width=0.35mm] (1.7,0) circle (6pt)node[](3){} ;

    \foreach \x/\y in {}
        \path[arrows={[scale=0.8]}] (\x) edge [strong dir] (\y); 
    \foreach \x/\y in {2/3,3/2}
    	\path[arrows={[scale=0.8]}] (\x) edge [strong bidir] (\y); 
	\foreach \x/\y in {2/1,3/1}
   		\path[arrows={[scale=0.8]}] (\x) edge [weak dir] (\y); 
    \foreach \x/\y in {}
    	\path[arrows={[scale=0.8]}] (\x) edge [weak bidir] (\y); 

    \end{tikzpicture}
    \caption{11}
  \end{subfigure}~~
  \begin{subfigure}{0.12\textwidth}
    \begin{tikzpicture}   
   \node[draw] at (1.7,1.5) {$\times 2$};
    \draw[fill=blue!5,line width=0.35mm] (0.85,1.3) circle (6pt)node[](1){} ;
    \draw[fill=blue!5,line width=0.35mm] (0,0) circle (6pt)node[](2){} ;
    \draw[fill=blue!5,line width=0.35mm] (1.7,0) circle (6pt)node[](3){} ;

    \foreach \x/\y in {}
        \path[arrows={[scale=0.8]}] (\x) edge [strong dir] (\y); 
    \foreach \x/\y in {2/3,3/2}
    	\path[arrows={[scale=0.8]}] (\x) edge [strong bidir] (\y); 
	\foreach \x/\y in {1/2,3/1}
   		\path[arrows={[scale=0.8]}] (\x) edge [weak dir] (\y); 
    \foreach \x/\y in {}
    	\path[arrows={[scale=0.8]}] (\x) edge [weak bidir] (\y); 

    \end{tikzpicture}
    \caption{12}
  \end{subfigure}~~
  \begin{subfigure}{0.12\textwidth}
    \begin{tikzpicture}   
    \draw[fill=blue!5,line width=0.35mm] (0.85,1.3) circle (6pt)node[](1){} ;
    \draw[fill=blue!5,line width=0.35mm] (0,0) circle (6pt)node[](2){} ;
    \draw[fill=blue!5,line width=0.35mm] (1.7,0) circle (6pt)node[](3){} ;

    \foreach \x/\y in {}
        \path[arrows={[scale=0.8]}] (\x) edge [strong dir] (\y); 
    \foreach \x/\y in {2/3,3/2}
    	\path[arrows={[scale=0.8]}] (\x) edge [strong bidir] (\y); 
	\foreach \x/\y in {1/2,1/3}
   		\path[arrows={[scale=0.8]}] (\x) edge [weak dir] (\y); 
    \foreach \x/\y in {}
    	\path[arrows={[scale=0.8]}] (\x) edge [weak bidir] (\y); 

    \end{tikzpicture}
    \caption{13}
  \end{subfigure}~~    
  \begin{subfigure}{0.12\textwidth}
    \begin{tikzpicture}   
    \draw[fill=black!100,line width=0.35mm] (0.85,1.3) circle (6pt)node[](1){} ;
    \draw[fill=blue!5,line width=0.35mm] (0,0) circle (6pt)node[](2){} ;
    \draw[fill=blue!5,line width=0.35mm] (1.7,0) circle (6pt)node[](3){} ;

    \foreach \x/\y in {}
        \path[arrows={[scale=0.8]}] (\x) edge [strong dir] (\y); 
    \foreach \x/\y in {2/3,3/2}
    	\path[arrows={[scale=0.8]}] (\x) edge [strong bidir] (\y); 
	\foreach \x/\y in {2/1,3/1}
   		\path[arrows={[scale=0.8]}] (\x) edge [weak dir] (\y); 
    \foreach \x/\y in {}
    	\path[arrows={[scale=0.8]}] (\x) edge [weak bidir] (\y); 

    \end{tikzpicture}
    \caption{14}
  \end{subfigure}~~
  
  \begin{subfigure}{0.12\textwidth}
    \begin{tikzpicture}   
    \node[draw] at (1.7,1.5) {$\times 2$};
    \draw[fill=blue!5,line width=0.35mm] (0.85,1.3) circle (6pt)node[](1){} ;
    \draw[fill=blue!5,line width=0.35mm] (0,0) circle (6pt)node[](2){} ;
    \draw[fill=black!100,line width=0.35mm] (1.7,0) circle (6pt)node[](3){} ;

    \foreach \x/\y in {2/3}
        \path[arrows={[scale=0.8]}] (\x) edge [strong dir] (\y); 
    \foreach \x/\y in {}
    	\path[arrows={[scale=0.8]}] (\x) edge [strong bidir] (\y); 
	\foreach \x/\y in {1/3,2/1}
   		\path[arrows={[scale=0.8]}] (\x) edge [weak dir] (\y); 
    \foreach \x/\y in {}
    	\path[arrows={[scale=0.8]}] (\x) edge [weak bidir] (\y); 

    \end{tikzpicture}
    \caption{15}
  \end{subfigure}~~
  \begin{subfigure}{0.12\textwidth}
    \begin{tikzpicture}   
    \node[draw] at (1.7,1.5) {$\times 2$};
    \draw[fill=blue!5,line width=0.35mm] (0.85,1.3) circle (6pt)node[](1){} ;
    \draw[fill=blue!5,line width=0.35mm] (0,0) circle (6pt)node[](2){} ;
    \draw[fill=black!100,line width=0.35mm] (1.7,0) circle (6pt)node[](3){} ;

    \foreach \x/\y in {2/3}
        \path[arrows={[scale=0.8]}] (\x) edge [strong dir] (\y); 
    \foreach \x/\y in {}
    	\path[arrows={[scale=0.8]}] (\x) edge [strong bidir] (\y); 
	\foreach \x/\y in {1/3,1/2}
   		\path[arrows={[scale=0.8]}] (\x) edge [weak dir] (\y); 
    \foreach \x/\y in {}
    	\path[arrows={[scale=0.8]}] (\x) edge [weak bidir] (\y); 

    \end{tikzpicture}
    \caption{16}
  \end{subfigure}~~
  \begin{subfigure}{0.12\textwidth}
    \begin{tikzpicture}   
    \node[draw] at (1.7,1.5) {$\times 2$};
    \draw[fill=blue!5,line width=0.35mm] (0.85,1.3) circle (6pt)node[](1){} ;
    \draw[fill=blue!5,line width=0.35mm] (0,0) circle (6pt)node[](2){} ;
    \draw[fill=black!100,line width=0.35mm] (1.7,0) circle (6pt)node[](3){} ;

    \foreach \x/\y in {2/3}
        \path[arrows={[scale=0.8]}] (\x) edge [strong dir] (\y); 
    \foreach \x/\y in {}
    	\path[arrows={[scale=0.8]}] (\x) edge [strong bidir] (\y); 
	\foreach \x/\y in {1/3}
   		\path[arrows={[scale=0.8]}] (\x) edge [weak dir] (\y); 
    \foreach \x/\y in {1/2,2/1}
    	\path[arrows={[scale=0.8]}] (\x) edge [weak bidir] (\y); 

    \end{tikzpicture}
    \caption{17}
  \end{subfigure}~~    
    \begin{subfigure}{0.12\textwidth}
    \begin{tikzpicture}   
    \node[draw] at (1.7,1.5) {$\times 6$};
    \draw[fill=black!100,line width=0.35mm] (0.85,1.3) circle (6pt)node[](1){} ;
    \draw[fill=blue!5,line width=0.35mm] (0,0) circle (6pt)node[](2){} ;
    \draw[fill=blue!5,line width=0.35mm] (1.7,0) circle (6pt)node[](3){} ;

    \foreach \x/\y in {}
        \path[arrows={[scale=0.8]}] (\x) edge [strong dir] (\y); 
    \foreach \x/\y in {}
    	\path[arrows={[scale=0.8]}] (\x) edge [strong bidir] (\y); 
	\foreach \x/\y in {2/1,3/1,2/3}
   		\path[arrows={[scale=0.8]}] (\x) edge [weak dir] (\y); 
    \foreach \x/\y in {}
    	\path[arrows={[scale=0.8]}] (\x) edge [weak bidir] (\y); 

    \end{tikzpicture}
    \caption{18}
  \end{subfigure}~~
    \begin{subfigure}{0.12\textwidth}
    \begin{tikzpicture}   
    \node[draw] at (1.7,1.5) {$\times 3$};
    \draw[fill=black!100,line width=0.35mm] (0.85,1.3) circle (6pt)node[](1){} ;
    \draw[fill=blue!5,line width=0.35mm] (0,0) circle (6pt)node[](2){} ;
    \draw[fill=blue!5,line width=0.35mm] (1.7,0) circle (6pt)node[](3){} ;

    \foreach \x/\y in {}
        \path[arrows={[scale=0.8]}] (\x) edge [strong dir] (\y); 
    \foreach \x/\y in {}
    	\path[arrows={[scale=0.8]}] (\x) edge [strong bidir] (\y); 
	\foreach \x/\y in {2/1,3/1}
   		\path[arrows={[scale=0.8]}] (\x) edge [weak dir] (\y); 
    \foreach \x/\y in {2/3,3/2}
    	\path[arrows={[scale=0.8]}] (\x) edge [weak bidir] (\y); 

    \end{tikzpicture}
    \caption{19}
  \end{subfigure}~~
  \begin{subfigure}{0.12\textwidth}
    \begin{tikzpicture}   
    \node[draw] at (1.7,1.5) {$\times 6$};
    \draw[fill=blue!5,line width=0.35mm] (0.85,1.3) circle (6pt)node[](1){} ;
    \draw[fill=blue!5,line width=0.35mm] (0,0) circle (6pt)node[](2){} ;
    \draw[fill=blue!5,line width=0.35mm] (1.7,0) circle (6pt)node[](3){} ;

    \foreach \x/\y in {}
        \path[arrows={[scale=0.8]}] (\x) edge [strong dir] (\y); 
    \foreach \x/\y in {}
    	\path[arrows={[scale=0.8]}] (\x) edge [strong bidir] (\y); 
	\foreach \x/\y in {2/1,3/1,2/3}
   		\path[arrows={[scale=0.8]}] (\x) edge [weak dir] (\y); 
    \foreach \x/\y in {}
    	\path[arrows={[scale=0.8]}] (\x) edge [weak bidir] (\y); 

    \end{tikzpicture}
    \caption{20}
  \end{subfigure}~~ 
  \begin{subfigure}{0.12\textwidth}
    \begin{tikzpicture}   
    \node[draw] at (1.7,1.5) {$\times 3$};
    \draw[fill=blue!5,line width=0.35mm] (0.85,1.3) circle (6pt)node[](1){} ;
    \draw[fill=blue!5,line width=0.35mm] (0,0) circle (6pt)node[](2){} ;
    \draw[fill=blue!5,line width=0.35mm] (1.7,0) circle (6pt)node[](3){} ;

    \foreach \x/\y in {}
        \path[arrows={[scale=0.8]}] (\x) edge [strong dir] (\y); 
    \foreach \x/\y in {}
    	\path[arrows={[scale=0.8]}] (\x) edge [strong bidir] (\y); 
	\foreach \x/\y in {2/1,3/1}
   		\path[arrows={[scale=0.8]}] (\x) edge [weak dir] (\y); 
    \foreach \x/\y in {2/3,3/2}
    	\path[arrows={[scale=0.8]}] (\x) edge [weak bidir] (\y); 

    \end{tikzpicture}
    \caption{21}
  \end{subfigure}~~ 
   
    \begin{subfigure}{0.12\textwidth}
    \begin{tikzpicture}   
	\node[draw] at (1.7,1.5) {$\times 2$};
    \draw[fill=blue!5,line width=0.35mm] (0.85,1.3) circle (6pt)node[](1){} ;
    \draw[fill=blue!5,line width=0.35mm] (0,0) circle (6pt)node[](2){} ;
    \draw[fill=blue!5,line width=0.35mm] (1.7,0) circle (6pt)node[](3){} ;

    \foreach \x/\y in {}
        \path[arrows={[scale=0.8]}] (\x) edge [strong dir] (\y); 
    \foreach \x/\y in {}
    	\path[arrows={[scale=0.8]}] (\x) edge [strong bidir] (\y); 
	\foreach \x/\y in {2/1,1/3,3/2}
   		\path[arrows={[scale=0.8]}] (\x) edge [weak dir] (\y); 
    \foreach \x/\y in {}
    	\path[arrows={[scale=0.8]}] (\x) edge [weak bidir] (\y); 

    \end{tikzpicture}
    \caption{22}
  \end{subfigure}~~      
    \begin{subfigure}{0.12\textwidth}
    \begin{tikzpicture}   
    \node[draw] at (1.7,1.5) {$\times 6$};
    \draw[fill=blue!5,line width=0.35mm] (0.85,1.3) circle (6pt)node[](1){} ;
    \draw[fill=blue!5,line width=0.35mm] (0,0) circle (6pt)node[](2){} ;
    \draw[fill=blue!5,line width=0.35mm] (1.7,0) circle (6pt)node[](3){} ;

    \foreach \x/\y in {}
        \path[arrows={[scale=0.8]}] (\x) edge [strong dir] (\y); 
    \foreach \x/\y in {}
    	\path[arrows={[scale=0.8]}] (\x) edge [strong bidir] (\y); 
	\foreach \x/\y in {2/1,1/3}
   		\path[arrows={[scale=0.8]}] (\x) edge [weak dir] (\y); 
    \foreach \x/\y in {2/3,3/2}
    	\path[arrows={[scale=0.8]}] (\x) edge [weak bidir] (\y); 

    \end{tikzpicture}
    \caption{23}
  \end{subfigure}~~
      \begin{subfigure}{0.12\textwidth}
    \begin{tikzpicture}   
    \node[draw] at (1.7,1.5) {$\times 3$};
    \draw[fill=blue!5,line width=0.35mm] (0.85,1.3) circle (6pt)node[](1){} ;
    \draw[fill=blue!5,line width=0.35mm] (0,0) circle (6pt)node[](2){} ;
    \draw[fill=blue!5,line width=0.35mm] (1.7,0) circle (6pt)node[](3){} ;

    \foreach \x/\y in {}
        \path[arrows={[scale=0.8]}] (\x) edge [strong dir] (\y); 
    \foreach \x/\y in {}
    	\path[arrows={[scale=0.8]}] (\x) edge [strong bidir] (\y); 
	\foreach \x/\y in {2/1,2/3}
   		\path[arrows={[scale=0.8]}] (\x) edge [weak dir] (\y); 
    \foreach \x/\y in {1/3,3/1}
    	\path[arrows={[scale=0.8]}] (\x) edge [weak bidir] (\y); 

    \end{tikzpicture}
    \caption{24}
  \end{subfigure}~~ 
      \begin{subfigure}{0.12\textwidth}
    \begin{tikzpicture}   
    \node[draw] at (1.7,1.5) {$\times 6$};
    \draw[fill=blue!5,line width=0.35mm] (0.85,1.3) circle (6pt)node[](1){} ;
    \draw[fill=blue!5,line width=0.35mm] (0,0) circle (6pt)node[](2){} ;
    \draw[fill=blue!5,line width=0.35mm] (1.7,0) circle (6pt)node[](3){} ;

    \foreach \x/\y in {}
        \path[arrows={[scale=0.8]}] (\x) edge [strong dir] (\y); 
    \foreach \x/\y in {}
    	\path[arrows={[scale=0.8]}] (\x) edge [strong bidir] (\y); 
	\foreach \x/\y in {2/1}
   		\path[arrows={[scale=0.8]}] (\x) edge [weak dir] (\y); 
    \foreach \x/\y in {2/3,3/2,1/3,3/1}
    	\path[arrows={[scale=0.8]}] (\x) edge [weak bidir] (\y); 

    \end{tikzpicture}
    \caption{25}
  \end{subfigure}~~ 
      \begin{subfigure}{0.12\textwidth}
    \begin{tikzpicture}   
    \draw[fill=blue!5,line width=0.35mm] (0.85,1.3) circle (6pt)node[](1){} ;
    \draw[fill=blue!5,line width=0.35mm] (0,0) circle (6pt)node[](2){} ;
    \draw[fill=blue!5,line width=0.35mm] (1.7,0) circle (6pt)node[](3){} ;

    \foreach \x/\y in {}
        \path[arrows={[scale=0.8]}] (\x) edge [strong dir] (\y); 
    \foreach \x/\y in {}
    	\path[arrows={[scale=0.8]}] (\x) edge [strong bidir] (\y); 
	\foreach \x/\y in {}
   		\path[arrows={[scale=0.8]}] (\x) edge [weak dir] (\y); 
    \foreach \x/\y in {1/2,2/1,1/3,3/1,2/3,3/2}
    	\path[arrows={[scale=0.8]}] (\x) edge [weak bidir] (\y); 

    \end{tikzpicture}
    \caption{26}
  \end{subfigure}
  \caption{All possible ways a triangle can be observed in $\GG$.  Some configurations are degenerate (due to symmetry).  For these cases, the  corresponding multiplicities are given on the top right.}\label{all_triangles}   
\end{figure*}
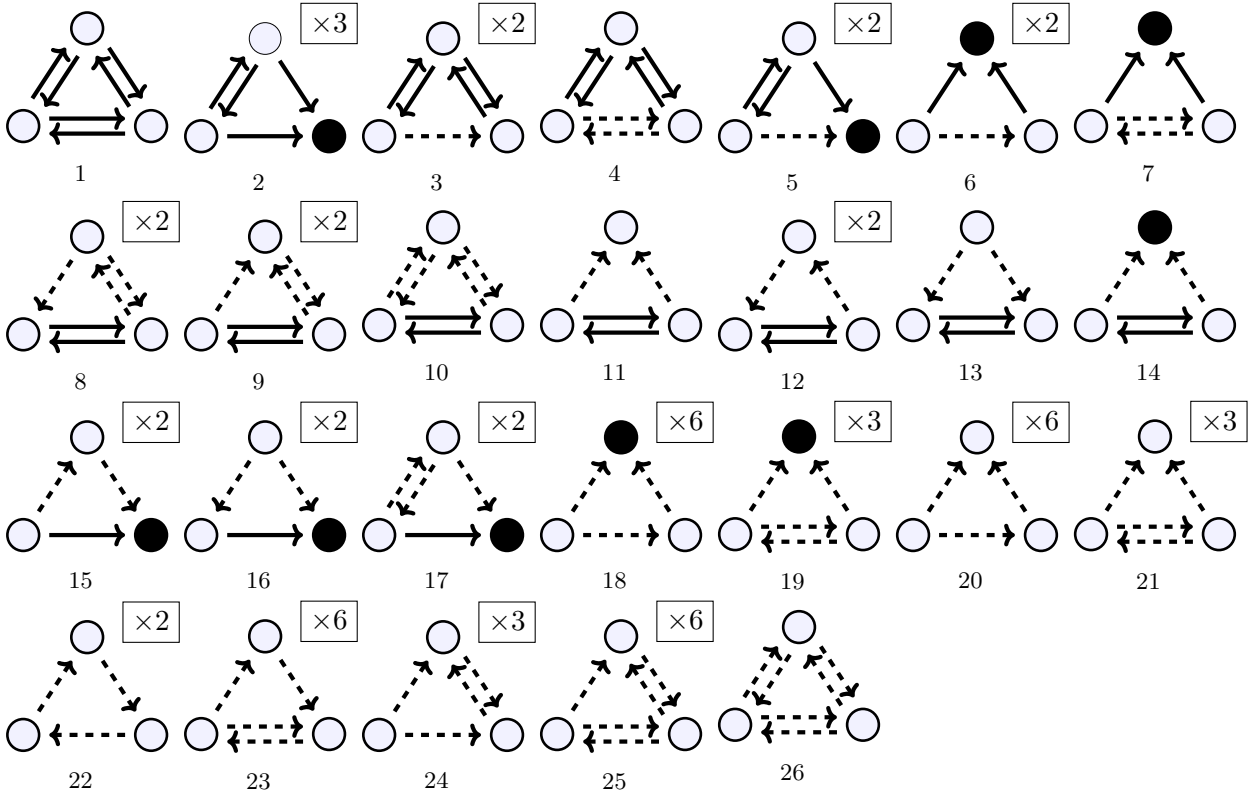

\begin{figure*}
\captionsetup[subfigure]{labelformat=empty}
  \begin{subfigure}{0.12\textwidth}
    \begin{tikzpicture}   
    \draw[fill=blue!5,line width=0.35mm] (0.85,1.3) circle (6pt)node[](1){} ;
    \draw[fill=blue!5,line width=0.35mm] (0,0) circle (6pt)node[](2){} ;
    \draw[fill=blue!5,line width=0.35mm] (1.7,0) circle (6pt)node[](3){} ;

    \foreach \x/\y in {}
        \path[arrows={[scale=0.8]}] (\x) edge [strong dir] (\y); 
    \foreach \x/\y in {1/2,2/1,1/3,3/1}
    	\path[arrows={[scale=0.8]}] (\x) edge [strong bidir] (\y); 
	\foreach \x/\y in {}
   		\path[arrows={[scale=0.8]}] (\x) edge [weak dir] (\y); 
    \foreach \x/\y in {}
    	\path[arrows={[scale=0.8]}] (\x) edge [weak bidir] (\y); 

    \end{tikzpicture}
    \caption{1}
  \end{subfigure}~~
  \begin{subfigure}{0.12\textwidth}
    \begin{tikzpicture}   
    \draw[fill=black!100,line width=0.35mm] (0.85,1.3) circle (6pt)node[](1){} ;
    \draw[fill=blue!5,line width=0.35mm] (0,0) circle (6pt)node[](2){} ;
    \draw[fill=blue!5,line width=0.35mm] (1.7,0) circle (6pt)node[](3){} ;

    \foreach \x/\y in {2/1,3/1}
        \path[arrows={[scale=0.8]}] (\x) edge [strong dir] (\y); 
    \foreach \x/\y in {}
    	\path[arrows={[scale=0.8]}] (\x) edge [strong bidir] (\y); 
	\foreach \x/\y in {}
   		\path[arrows={[scale=0.8]}] (\x) edge [weak dir] (\y); 
    \foreach \x/\y in {}
    	\path[arrows={[scale=0.8]}] (\x) edge [weak bidir] (\y); 

    \end{tikzpicture}
    \caption{2}
  \end{subfigure}~~
  \begin{subfigure}{0.12\textwidth}
    \begin{tikzpicture}   
    \draw[fill=blue!5,line width=0.35mm] (0.85,1.3) circle (6pt)node[](1){} ;
    \draw[fill=black!100,line width=0.35mm] (0,0) circle (6pt)node[](2){} ;
    \draw[fill=black!100,line width=0.35mm] (1.7,0) circle (6pt)node[](3){} ;

    \foreach \x/\y in {1/2,1/3}
        \path[arrows={[scale=0.8]}] (\x) edge [strong dir] (\y); 
    \foreach \x/\y in {}
    	\path[arrows={[scale=0.8]}] (\x) edge [strong bidir] (\y); 
	\foreach \x/\y in {}
   		\path[arrows={[scale=0.8]}] (\x) edge [weak dir] (\y); 
    \foreach \x/\y in {}
    	\path[arrows={[scale=0.8]}] (\x) edge [weak bidir] (\y); 

    \end{tikzpicture}
    \caption{3}
  \end{subfigure}~~
  \begin{subfigure}{0.12\textwidth}
    \begin{tikzpicture}   
   \node[draw] at (1.7,1.5) {$\times 2$};
    \draw[fill=blue!5,line width=0.35mm] (0.85,1.3) circle (6pt)node[](1){} ;
    \draw[fill=blue!5,line width=0.35mm] (0,0) circle (6pt)node[](2){} ;
    \draw[fill=black!100,line width=0.35mm] (1.7,0) circle (6pt)node[](3){} ;

    \foreach \x/\y in {1/3}
        \path[arrows={[scale=0.8]}] (\x) edge [strong dir] (\y); 
    \foreach \x/\y in {1/2,2/1}
    	\path[arrows={[scale=0.8]}] (\x) edge [strong bidir] (\y); 
	\foreach \x/\y in {}
   		\path[arrows={[scale=0.8]}] (\x) edge [weak dir] (\y); 
    \foreach \x/\y in {}
    	\path[arrows={[scale=0.8]}] (\x) edge [weak bidir] (\y); 

    \end{tikzpicture}
    \caption{4}
  \end{subfigure}~~    
  \begin{subfigure}{0.12\textwidth}
    \begin{tikzpicture}   
    \draw[fill=black!100,line width=0.35mm] (0.85,1.3) circle (6pt)node[](1){} ;
    \draw[fill=blue!5,line width=0.35mm] (0,0) circle (6pt)node[](2){} ;
    \draw[fill=blue!5,line width=0.35mm] (1.7,0) circle (6pt)node[](3){} ;

    \foreach \x/\y in {2/1}
        \path[arrows={[scale=0.8]}] (\x) edge [strong dir] (\y); 
    \foreach \x/\y in {}
    	\path[arrows={[scale=0.8]}] (\x) edge [strong bidir] (\y); 
	\foreach \x/\y in {3/1}
   		\path[arrows={[scale=0.8]}] (\x) edge [weak dir] (\y); 
    \foreach \x/\y in {}
    	\path[arrows={[scale=0.8]}] (\x) edge [weak bidir] (\y); 

    \end{tikzpicture}
    \caption{5}
  \end{subfigure}~~
  \begin{subfigure}{0.12\textwidth}
    \begin{tikzpicture}   
    \draw[fill=blue!5,line width=0.35mm] (0.85,1.3) circle (6pt)node[](1){} ;
    \draw[fill=black!100,line width=0.35mm] (0,0) circle (6pt)node[](2){} ;
    \draw[fill=black!100,line width=0.35mm] (1.7,0) circle (6pt)node[](3){} ;

    \foreach \x/\y in {1/2}
        \path[arrows={[scale=0.8]}] (\x) edge [strong dir] (\y); 
    \foreach \x/\y in {}
    	\path[arrows={[scale=0.8]}] (\x) edge [strong bidir] (\y); 
	\foreach \x/\y in {1/3}
   		\path[arrows={[scale=0.8]}] (\x) edge [weak dir] (\y); 
    \foreach \x/\y in {}
    	\path[arrows={[scale=0.8]}] (\x) edge [weak bidir] (\y); 

    \end{tikzpicture}
    \caption{6}
  \end{subfigure}~~
  \begin{subfigure}{0.12\textwidth}
    \begin{tikzpicture}   
    \draw[fill=blue!5,line width=0.35mm] (0.85,1.3) circle (6pt)node[](1){} ;
    \draw[fill=blue!5,line width=0.35mm] (0,0) circle (6pt)node[](2){} ;
    \draw[fill=black!100,line width=0.35mm] (1.7,0) circle (6pt)node[](3){} ;

    \foreach \x/\y in {}
        \path[arrows={[scale=0.8]}] (\x) edge [strong dir] (\y); 
    \foreach \x/\y in {1/2,2/1}
    	\path[arrows={[scale=0.8]}] (\x) edge [strong bidir] (\y); 
	\foreach \x/\y in {1/3}
   		\path[arrows={[scale=0.8]}] (\x) edge [weak dir] (\y); 
    \foreach \x/\y in {}
    	\path[arrows={[scale=0.8]}] (\x) edge [weak bidir] (\y); 

    \end{tikzpicture}
    \caption{7}
  \end{subfigure}~~
  
    \begin{subfigure}{0.12\textwidth}
    \begin{tikzpicture}   
    \draw[fill=blue!5,line width=0.35mm] (0.85,1.3) circle (6pt)node[](1){} ;
    \draw[fill=black!100,line width=0.35mm] (0,0) circle (6pt)node[](2){} ;
    \draw[fill=blue!5,line width=0.35mm] (1.7,0) circle (6pt)node[](3){} ;

    \foreach \x/\y in {1/2}
        \path[arrows={[scale=0.8]}] (\x) edge [strong dir] (\y); 
    \foreach \x/\y in {}
    	\path[arrows={[scale=0.8]}] (\x) edge [strong bidir] (\y); 
	\foreach \x/\y in {1/3}
   		\path[arrows={[scale=0.8]}] (\x) edge [weak dir] (\y); 
    \foreach \x/\y in {}
    	\path[arrows={[scale=0.8]}] (\x) edge [weak bidir] (\y); 

    \end{tikzpicture}
    \caption{8}
  \end{subfigure}~~
  \begin{subfigure}{0.12\textwidth}
    \begin{tikzpicture}   
    \draw[fill=blue!5,line width=0.35mm] (0.85,1.3) circle (6pt)node[](1){} ;
    \draw[fill=black!100,line width=0.35mm] (0,0) circle (6pt)node[](2){} ;
    \draw[fill=blue!5,line width=0.35mm] (1.7,0) circle (6pt)node[](3){} ;

    \foreach \x/\y in {1/2}
        \path[arrows={[scale=0.8]}] (\x) edge [strong dir] (\y); 
    \foreach \x/\y in {}
    	\path[arrows={[scale=0.8]}] (\x) edge [strong bidir] (\y); 
	\foreach \x/\y in {3/1}
   		\path[arrows={[scale=0.8]}] (\x) edge [weak dir] (\y); 
    \foreach \x/\y in {}
    	\path[arrows={[scale=0.8]}] (\x) edge [weak bidir] (\y); 

    \end{tikzpicture}
    \caption{9}
  \end{subfigure}~~
  \begin{subfigure}{0.12\textwidth}
    \begin{tikzpicture}   
    \draw[fill=blue!5,line width=0.35mm] (0.85,1.3) circle (6pt)node[](1){} ;
    \draw[fill=black!100,line width=0.35mm] (0,0) circle (6pt)node[](2){} ;
    \draw[fill=blue!5,line width=0.35mm] (1.7,0) circle (6pt)node[](3){} ;

    \foreach \x/\y in {1/2}
        \path[arrows={[scale=0.8]}] (\x) edge [strong dir] (\y); 
    \foreach \x/\y in {}
    	\path[arrows={[scale=0.8]}] (\x) edge [strong bidir] (\y); 
	\foreach \x/\y in {}
   		\path[arrows={[scale=0.8]}] (\x) edge [weak dir] (\y); 
    \foreach \x/\y in {1/3,3/1}
    	\path[arrows={[scale=0.8]}] (\x) edge [weak bidir] (\y); 

    \end{tikzpicture}
    \caption{10}
  \end{subfigure}~~
  \begin{subfigure}{0.12\textwidth}
    \begin{tikzpicture}   
    \draw[fill=blue!5,line width=0.35mm] (0.85,1.3) circle (6pt)node[](1){} ;
    \draw[fill=blue!5,line width=0.35mm] (0,0) circle (6pt)node[](2){} ;
    \draw[fill=blue!5,line width=0.35mm] (1.7,0) circle (6pt)node[](3){} ;

    \foreach \x/\y in {}
        \path[arrows={[scale=0.8]}] (\x) edge [strong dir] (\y); 
    \foreach \x/\y in {1/2,2/1}
    	\path[arrows={[scale=0.8]}] (\x) edge [strong bidir] (\y); 
	\foreach \x/\y in {1/3}
   		\path[arrows={[scale=0.8]}] (\x) edge [weak dir] (\y); 
    \foreach \x/\y in {}
    	\path[arrows={[scale=0.8]}] (\x) edge [weak bidir] (\y); 

    \end{tikzpicture}
    \caption{11}
  \end{subfigure}~~  
  \begin{subfigure}{0.12\textwidth}
    \begin{tikzpicture}   
    \draw[fill=blue!5,line width=0.35mm] (0.85,1.3) circle (6pt)node[](1){} ;
    \draw[fill=blue!5,line width=0.35mm] (0,0) circle (6pt)node[](2){} ;
    \draw[fill=blue!5,line width=0.35mm] (1.7,0) circle (6pt)node[](3){} ;

    \foreach \x/\y in {}
        \path[arrows={[scale=0.8]}] (\x) edge [strong dir] (\y); 
    \foreach \x/\y in {1/2,2/1}
    	\path[arrows={[scale=0.8]}] (\x) edge [strong bidir] (\y); 
	\foreach \x/\y in {3/1}
   		\path[arrows={[scale=0.8]}] (\x) edge [weak dir] (\y); 
    \foreach \x/\y in {}
    	\path[arrows={[scale=0.8]}] (\x) edge [weak bidir] (\y); 

    \end{tikzpicture}
    \caption{12}
  \end{subfigure}~~
  \begin{subfigure}{0.12\textwidth}
    \begin{tikzpicture}   
    \draw[fill=blue!5,line width=0.35mm] (0.85,1.3) circle (6pt)node[](1){} ;
    \draw[fill=blue!5,line width=0.35mm] (0,0) circle (6pt)node[](2){} ;
    \draw[fill=blue!5,line width=0.35mm] (1.7,0) circle (6pt)node[](3){} ;

    \foreach \x/\y in {}
        \path[arrows={[scale=0.8]}] (\x) edge [strong dir] (\y); 
    \foreach \x/\y in {1/2,2/1}
    	\path[arrows={[scale=0.8]}] (\x) edge [strong bidir] (\y); 
	\foreach \x/\y in {}
   		\path[arrows={[scale=0.8]}] (\x) edge [weak dir] (\y); 
    \foreach \x/\y in {1/3,3/1}
    	\path[arrows={[scale=0.8]}] (\x) edge [weak bidir] (\y); 

    \end{tikzpicture}
    \caption{13}
  \end{subfigure}~~
  \begin{subfigure}{0.12\textwidth}
    \begin{tikzpicture}   
    \draw[fill=blue!5,line width=0.35mm] (0.85,1.3) circle (6pt)node[](1){} ;
    \draw[fill=black!100,line width=0.35mm] (0,0) circle (6pt)node[](2){} ;
    \draw[fill=black!100,line width=0.35mm] (1.7,0) circle (6pt)node[](3){} ;

    \foreach \x/\y in {}
        \path[arrows={[scale=0.8]}] (\x) edge [strong dir] (\y); 
    \foreach \x/\y in {}
    	\path[arrows={[scale=0.8]}] (\x) edge [strong bidir] (\y); 
	\foreach \x/\y in {1/2,1/3}
   		\path[arrows={[scale=0.8]}] (\x) edge [weak dir] (\y); 
    \foreach \x/\y in {}
    	\path[arrows={[scale=0.8]}] (\x) edge [weak bidir] (\y); 

    \end{tikzpicture}
    \caption{14}
  \end{subfigure}~~
  
  \begin{subfigure}{0.12\textwidth}
    \begin{tikzpicture}   
    \draw[fill=black!100,line width=0.35mm] (0.85,1.3) circle (6pt)node[](1){} ;
    \draw[fill=blue!5,line width=0.35mm] (0,0) circle (6pt)node[](2){} ;
    \draw[fill=blue!5,line width=0.35mm] (1.7,0) circle (6pt)node[](3){} ;

    \foreach \x/\y in {}
        \path[arrows={[scale=0.8]}] (\x) edge [strong dir] (\y); 
    \foreach \x/\y in {}
    	\path[arrows={[scale=0.8]}] (\x) edge [strong bidir] (\y); 
	\foreach \x/\y in {3/1,2/1}
   		\path[arrows={[scale=0.8]}] (\x) edge [weak dir] (\y); 
    \foreach \x/\y in {}
    	\path[arrows={[scale=0.8]}] (\x) edge [weak bidir] (\y); 

    \end{tikzpicture}
    \caption{15}
  \end{subfigure}~~ 
    \begin{subfigure}{0.12\textwidth}
    \begin{tikzpicture}   
    \node[draw] at (1.7,1.5) {$\times 2$};
    \draw[fill=blue!5,line width=0.35mm] (0.85,1.3) circle (6pt)node[](1){} ;
    \draw[fill=black!100,line width=0.35mm] (0,0) circle (6pt)node[](2){} ;
    \draw[fill=blue!5,line width=0.35mm] (1.7,0) circle (6pt)node[](3){} ;

    \foreach \x/\y in {}
        \path[arrows={[scale=0.8]}] (\x) edge [strong dir] (\y); 
    \foreach \x/\y in {}
    	\path[arrows={[scale=0.8]}] (\x) edge [strong bidir] (\y); 
	\foreach \x/\y in {3/1,1/2}
   		\path[arrows={[scale=0.8]}] (\x) edge [weak dir] (\y); 
    \foreach \x/\y in {}
    	\path[arrows={[scale=0.8]}] (\x) edge [weak bidir] (\y); 

    \end{tikzpicture}
    \caption{16}
  \end{subfigure}~~
  \begin{subfigure}{0.12\textwidth}
    \begin{tikzpicture}   
    \node[draw] at (1.7,1.5) {$\times 2$};
    \draw[fill=blue!5,line width=0.35mm] (0.85,1.3) circle (6pt)node[](1){} ;
    \draw[fill=black!100,line width=0.35mm] (0,0) circle (6pt)node[](2){} ;
    \draw[fill=blue!5,line width=0.35mm] (1.7,0) circle (6pt)node[](3){} ;

    \foreach \x/\y in {}
        \path[arrows={[scale=0.8]}] (\x) edge [strong dir] (\y); 
    \foreach \x/\y in {}
    	\path[arrows={[scale=0.8]}] (\x) edge [strong bidir] (\y); 
	\foreach \x/\y in {1/3,1/2}
   		\path[arrows={[scale=0.8]}] (\x) edge [weak dir] (\y); 
    \foreach \x/\y in {}
    	\path[arrows={[scale=0.8]}] (\x) edge [weak bidir] (\y); 

    \end{tikzpicture}
    \caption{17}
  \end{subfigure}~~
  \begin{subfigure}{0.12\textwidth}
    \begin{tikzpicture}   
    \node[draw] at (1.7,1.5) {$\times 2$};
    \draw[fill=blue!5,line width=0.35mm] (0.85,1.3) circle (6pt)node[](1){} ;
    \draw[fill=black!100,line width=0.35mm] (0,0) circle (6pt)node[](2){} ;
    \draw[fill=blue!5,line width=0.35mm] (1.7,0) circle (6pt)node[](3){} ;

    \foreach \x/\y in {}
        \path[arrows={[scale=0.8]}] (\x) edge [strong dir] (\y); 
    \foreach \x/\y in {}
    	\path[arrows={[scale=0.8]}] (\x) edge [strong bidir] (\y); 
	\foreach \x/\y in {1/2}
   		\path[arrows={[scale=0.8]}] (\x) edge [weak dir] (\y); 
    \foreach \x/\y in {3/1,1/3}
    	\path[arrows={[scale=0.8]}] (\x) edge [weak bidir] (\y); 

    \end{tikzpicture}
    \caption{18}
  \end{subfigure}~~
  \begin{subfigure}{0.12\textwidth}
    \begin{tikzpicture}   
    \node[draw] at (1.7,1.5) {$\times 2$};
    \draw[fill=blue!5,line width=0.35mm] (0.85,1.3) circle (6pt)node[](1){} ;
    \draw[fill=blue!5,line width=0.35mm] (0,0) circle (6pt)node[](2){} ;
    \draw[fill=blue!5,line width=0.35mm] (1.7,0) circle (6pt)node[](3){} ;

    \foreach \x/\y in {}
        \path[arrows={[scale=0.8]}] (\x) edge [strong dir] (\y); 
    \foreach \x/\y in {}
    	\path[arrows={[scale=0.8]}] (\x) edge [strong bidir] (\y); 
	\foreach \x/\y in {2/1,1/3}
   		\path[arrows={[scale=0.8]}] (\x) edge [weak dir] (\y); 
    \foreach \x/\y in {}
    	\path[arrows={[scale=0.8]}] (\x) edge [weak bidir] (\y); 

    \end{tikzpicture}
    \caption{19}
  \end{subfigure}~~
    \begin{subfigure}{0.12\textwidth}
    \begin{tikzpicture}   
    \node[draw] at (1.7,1.5) {$\times 2$};
    \draw[fill=blue!5,line width=0.35mm] (0.85,1.3) circle (6pt)node[](1){} ;
    \draw[fill=blue!5,line width=0.35mm] (0,0) circle (6pt)node[](2){} ;
    \draw[fill=blue!5,line width=0.35mm] (1.7,0) circle (6pt)node[](3){} ;

    \foreach \x/\y in {}
        \path[arrows={[scale=0.8]}] (\x) edge [strong dir] (\y); 
    \foreach \x/\y in {}
    	\path[arrows={[scale=0.8]}] (\x) edge [strong bidir] (\y); 
	\foreach \x/\y in {2/1,3/1}
   		\path[arrows={[scale=0.8]}] (\x) edge [weak dir] (\y); 
    \foreach \x/\y in {}
    	\path[arrows={[scale=0.8]}] (\x) edge [weak bidir] (\y); 

    \end{tikzpicture}
    \caption{20}
  \end{subfigure}~~  
    \begin{subfigure}{0.12\textwidth}
    \begin{tikzpicture}   
   \node[draw] at (1.7,1.5) {$\times 2$};
    \draw[fill=blue!5,line width=0.35mm] (0.85,1.3) circle (6pt)node[](1){} ;
    \draw[fill=blue!5,line width=0.35mm] (0,0) circle (6pt)node[](2){} ;
    \draw[fill=blue!5,line width=0.35mm] (1.7,0) circle (6pt)node[](3){} ;

    \foreach \x/\y in {}
        \path[arrows={[scale=0.8]}] (\x) edge [strong dir] (\y); 
    \foreach \x/\y in {}
    	\path[arrows={[scale=0.8]}] (\x) edge [strong bidir] (\y); 
	\foreach \x/\y in {2/1}
   		\path[arrows={[scale=0.8]}] (\x) edge [weak dir] (\y); 
    \foreach \x/\y in {3/1,1/3}
    	\path[arrows={[scale=0.8]}] (\x) edge [weak bidir] (\y); 

    \end{tikzpicture}
    \caption{21}
  \end{subfigure}~~
  
    \begin{subfigure}{0.12\textwidth}
    \begin{tikzpicture}   
    \draw[fill=blue!5,line width=0.35mm] (0.85,1.3) circle (6pt)node[](1){} ;
    \draw[fill=blue!5,line width=0.35mm] (0,0) circle (6pt)node[](2){} ;
    \draw[fill=blue!5,line width=0.35mm] (1.7,0) circle (6pt)node[](3){} ;

    \foreach \x/\y in {}
        \path[arrows={[scale=0.8]}] (\x) edge [strong dir] (\y); 
    \foreach \x/\y in {}
    	\path[arrows={[scale=0.8]}] (\x) edge [strong bidir] (\y); 
	\foreach \x/\y in {1/2,1/3}
   		\path[arrows={[scale=0.8]}] (\x) edge [weak dir] (\y); 
    \foreach \x/\y in {}
    	\path[arrows={[scale=0.8]}] (\x) edge [weak bidir] (\y); 

    \end{tikzpicture}
    \caption{22}
  \end{subfigure}~~
    \begin{subfigure}{0.12\textwidth}
    \begin{tikzpicture}   
    \node[draw] at (1.7,1.5) {$\times 2$};
    \draw[fill=blue!5,line width=0.35mm] (0.85,1.3) circle (6pt)node[](1){} ;
    \draw[fill=blue!5,line width=0.35mm] (0,0) circle (6pt)node[](2){} ;
    \draw[fill=blue!5,line width=0.35mm] (1.7,0) circle (6pt)node[](3){} ;

    \foreach \x/\y in {}
        \path[arrows={[scale=0.8]}] (\x) edge [strong dir] (\y); 
    \foreach \x/\y in {}
    	\path[arrows={[scale=0.8]}] (\x) edge [strong bidir] (\y); 
	\foreach \x/\y in {1/2}
   		\path[arrows={[scale=0.8]}] (\x) edge [weak dir] (\y); 
    \foreach \x/\y in {1/3,3/1}
    	\path[arrows={[scale=0.8]}] (\x) edge [weak bidir] (\y); 

    \end{tikzpicture}
    \caption{23}
  \end{subfigure}~~  
    \begin{subfigure}{0.12\textwidth}
    \begin{tikzpicture}   
    \draw[fill=blue!5,line width=0.35mm] (0.85,1.3) circle (6pt)node[](1){} ;
    \draw[fill=blue!5,line width=0.35mm] (0,0) circle (6pt)node[](2){} ;
    \draw[fill=blue!5,line width=0.35mm] (1.7,0) circle (6pt)node[](3){} ;

    \foreach \x/\y in {}
        \path[arrows={[scale=0.8]}] (\x) edge [strong dir] (\y); 
    \foreach \x/\y in {}
    	\path[arrows={[scale=0.8]}] (\x) edge [strong bidir] (\y); 
	\foreach \x/\y in {}
   		\path[arrows={[scale=0.8]}] (\x) edge [weak dir] (\y); 
    \foreach \x/\y in {2/1,1/3,1/2,3/1}
    	\path[arrows={[scale=0.8]}] (\x) edge [weak bidir] (\y); 

    \end{tikzpicture}
    \caption{24}
  \end{subfigure}~~
  \caption{All possible ways an open triad can be obserevd in $\GG$.}\label{all_triads} 
\end{figure*}
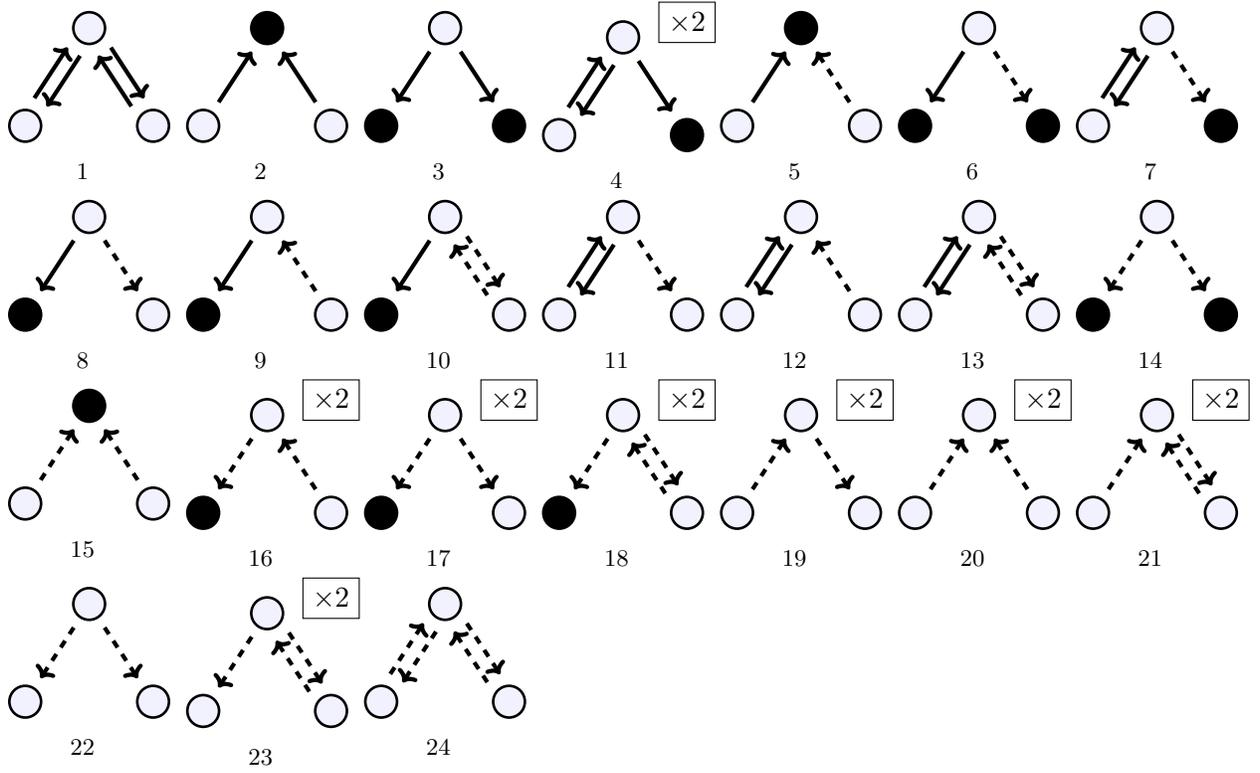

\begin{table*}[]
\centering
\caption{Approximate probabilities corresponding to the outcome of the sampling process for seeds which are part of triangles and open triads}
\resizebox{0.8\textwidth}{!}{%
\begin{tabular}{|llllllllll|} \hline & & & & & & & & &  \\ 
$ b_{00}\simeq \frac{(K_w-B)(K_w-B-1)}{K_w(K_w-1)}$& & & 
$b_{01}\simeq\frac{B(K_w-B)}{K_w(K_w-1)}$& & & 
$b_{11}\simeq \frac{B}{K_w} $& & & 
$ $\\
$ b_{02}\simeq \frac{B(B-1)}{K_w(K_w-1)}$& & & 
$b_{11}\simeq \frac{B}{K_w} $& & & 
$b_{20}=1 $& & & 
$ $\\
$a_{00}=\simeq 1-\frac{B}{K_w} $& & & 
$a_{01}\simeq \frac{B}{K_w} $& & & 
$ a_{10}=1$& & & 
$ $
\\ & & & & & & & & &  \\ \hline
\end{tabular}
}\label{B_A}
\end{table*}

\begin{table*}[]
\centering
\caption{Probabilities of observing different triangles}
\resizebox{\textwidth}{!}{%
\begin{tabular}{|llllllllll|} \hline & & & & & & & & &  \\ 
$\rho_1=q^3b_{20}^3$ & & & 
$\rho_2=3q^2(1-q)b_{20}^2$& & & 
$\rho_3=2q^3b_{20}b_{11}b_{10}$& & & 
$\rho_4=q^3b_{20}b_{11}^2$\\
$\rho_5=2q^2(1-q)b_{20}b_{11}$& & & 
$\rho_{6}=2q^2(1-q)b_{11}b_{10}$& & & 
$\rho_{7}=q^2(1-q)b_{11}^2$& & & 
$\rho_{8}=2q^3b_{02}b_{10}b_{11}$\\
$\rho_{9}=2q^3b_{01}b_{11}^2$& & & 
$\rho_{10}=q^3b_{02}b_{11}^2$& & & 
$\rho_{11}=q^3b_{00}b_{11}^2$& & & 
$\rho_{12}=2q^3b_{01}b_{11}b_{10}$\\
$\rho_{13}=q^3b_{02}b_{10}^2$& & & 
$\rho_{14}=q^2(1-q)b_{11}^2$& & & 
$\rho_{15}=2q^2(1-q)b_{01}b_{11}$& & & 
$\rho_{16}=2q^2(1-q)b_{02}b_{10}$\\
$\rho_{17}=2q^2(1-q)b_{02}b_{11}$& & & 
$\rho_{29}=6q^2(1-q)b_{02}b_{01}$& & &  
$\rho_{19}=3q^2(1-q)b_{02}^2$& & & 
$\rho_{20}=6q^3b_{00}b_{01}b_{02}$\\
$\rho_{21}=3q^3b_{00}b_{02}^2$& & & 
$\rho_{22}=2q^3b_{01}^3$& & & 
$\rho_{23}=6q^3b_{01}^2b_{02}$& & & 
$\rho_{24}=3q^3b_{01}^2b_{02}$\\
$\rho_{25}=3q^3b_{01}b_{02}^2$& & & 
$\rho_{26}=q^3b_{02}^3$& & & & & & \\ & & & & & & & & &  \\ \hline
\end{tabular}
}\label{rho}
\end{table*}

\begin{table*}[]
\centering
\caption{Probabilities of observing different open triads originated from triangles}
\resizebox{\textwidth}{!}{%
\begin{tabular}{|llllllllll|} \hline & & & & & & & & &  \\
$\pi_1=q^3b_{20}b_{10}^2$& & & 
$\pi_2=q^2(1-q)b_{10}^2$& & & 
$\pi_3=q(1-q)^2b_{20}$& & & 
$\pi_4=2q^2(1-q)b_{20}b_{10}$\\
$\pi_5=q^2(1-q)b_{10}b_{01}$& & & 
$\pi_6=q(1-q)^2b_{11}$& & & 
$\pi_7=q^2(1-q)b_{11}b_{10}$& & & 
$\pi_8=q^2(1-q)b_{11}b_{00}$\\ 
$\pi_{9}=q^2(1-q)b_{10}b_{01}$& & & 
$\pi_{10}=q^2(1-q)b_{11}b_{01}$& & & 
$\pi_{11}=q^3b_{10}b_{11}b_{00}$& & & 
$\pi_{12}=q^3b_{10}^2b_{01}$\\ 
$\pi_{13}=q^3b_{10}b_{11}b_{01}$& & & 
$\pi_{14}=q(1-q)^2b_{02}$& & & 
$\pi_{15}=q^2(1-q)b_{01}^2$& & &
$\pi_{16}=2q^2(1-q)b_{01}^2$\\
$\pi_{17}=2q^2(1-q)b_{00}b_{02}$& & & 
$\pi_{18}=2q^2(1-q)b_{02}b_{01}$& & & 
$\pi_{19}=2q^3b_{01}^2b_{00}$& & & 
$\pi_{20}=q^3b_{00}b_{01}^2$\\
$\pi_{21}=2q^3b_{01}^3$& & & 
$\pi_{22}=q^3b_{02}b_{00}^2$& & & 
$\pi_{23}=2q^3b_{02}b_{00}b_{01}$& & & 
$\pi_{24}=q^3b_{02}b_{01}^2$\\ & & & & & & & & &  \\ \hline
\end{tabular}
}\label{pi}
\end{table*}

\begin{table*}[]
\centering
\caption{Probabilities of observing different open triads originated from open triads}
\resizebox{\textwidth}{!}{%
\begin{tabular}{|llllllllll|} \hline & & & & & & & & &  \\

$\phi_1=q^3b_{20}a_{10}^2$& & & 
$\phi_2=q^2(1-q)a_{10}^2$& & & 
$\phi_3=q(1-q)^2b_{20}$& & & 
$\phi_4=2q^2(1-q)b_{20}a_{10}$\\
$\phi_5=q^2(1-q)a_{10}a_{01}$& & & 
$\phi_6=q(1-q)^2b_{11}$& & & 
$\phi_7=q^2(1-q)b_{11}a_{10}$& & & 
$\phi_8=q^2(1-q)b_{11}a_{00}$\\
$\phi_{9}=q^2(1-q)b_{10}a_{01}$& & & 
$\phi_{10}=q^2(1-q)b_{11}a_{01}$& & & 
$\phi_{11}=q^3b_{11}a_{10}a_{00}$& & & 
$\phi_{12}=q^3b_{10}a_{10}a_{01}$\\
$\phi_{13}=q^3b_{11}a_{10}a_{01}$& & & 
$\phi_{14}=q(1-q)^2b_{02}$& & & 
$\phi_{15}=q^2(1-q)a_{01}^2$& & & 
$\phi_{16}=2q^2(1-q)b_{01}a_{01}$\\
$\phi_{17}=2q^2(1-q)b_{02}a_{00}$& & & 
$\phi_{18}=2q^2(1-q)b{02}a_{01}$& & & 
$\phi_{19}=2q^3b_{01}a_{01}a_{00}$& & & 
$\phi_{20}=q^3b_{00}a_{01}^2$\\
$\phi_{21}=2q^3b_{01}a_{01}^2$& & &  
$\phi_{22}=q^3b_{02}a_{00}^2$& & & 
$\phi_{23}=2q^3b_{02}a_{00}a_{01}$& & & 
$\phi_{24}=q^3b_{02}a_{01}^2$\\ & & & & & & & & &  \\ \hline
\end{tabular}
}\label{phi}
\end{table*}

\bibliographystyle{abbrv}
\bibliography{myref}

\end{document}